\documentclass[]{revtex4}
\usepackage[]{graphicx}
\usepackage[]{graphics}
\usepackage{latexsym}
\usepackage{amssymb}

\def\lsim{\lower.5ex\hbox{$\; \buildrel < \over \sim \;$}}
\def\gsim{\lower.5ex\hbox{$\; \buildrel > \over \sim \;$}}
\def\pmb#1{\setbox0=\hbox{$#1$}%
\kern-.025em\copy0\kern-\wd0
\kern.05em\copy0\kern-\wd0
\kern-.025em\raise.0433em\box0}
\def\lsim{\lower.5ex\hbox{$\; \buildrel < \over \sim \;$}}
\def\gsim{\lower.5ex\hbox{$\; \buildrel > \over \sim \;$}}
\def\AHR{analogue Hawking radiation}
\def\AH{acoustic horizon}
\def\vh{\bf {{\vert_{(r=r_h)}}}}
\def\vc{\bf {{\vert_{(r=r_c)}}}}

\def\egam{$\left\{{\cal E},{\gamma}\right\}$}
\def\eker{$\left[{\cal E},\lambda,\gamma,a\right]$}

\def\ptw{\left[{\cal E},\lambda\right]}

\begin{document}
\title{Astrophysical Accretion as an Analogue Gravity Phenomena}
\date{\today}
\author{Tapas Kumar Das$^{1,2}$} 
\email{tapas@mri.ernet.in}
\homepage{http://www.mri.ernet.in/~tapas}
\affiliation{$^1$ Theoretical Institute for Advanced Research in Astrophysics, Hsinchu, Taiwan}
\affiliation{$^2$Harish Chandra Research
%{\ad $^1$Harish Chandra Research
Institute, 
%Dept. of Atomic Energy, Govt. of India, 
Jhunsi, Allahabad 211 019, India (Permanent Affiliation).}

\begin{abstract}
\noindent
Inspite of the remarkable
resemblance in between a black hole and an ordinary thermodynamic system,
black holes never radiate according to the classical laws of physics.
The introduction of quantum effects radically changes the scenario. Black holes radiate
due to quantum effects. Such radiation is known as Hawking radiation and the corresponding
radiation temperature is referred as the Hawking temperature.
Observational
manifestation of Hawking effect for 
astrophysical black holes is beyond the scope of present day's experimental
techniques. Also, Hawking quanta may posses trans-Planckian frequencies, and 
physics beyond the
Planck scale is not well understood. The
above mentioned difficulties with Hawking effect were the motivations to search for
an analogous version of Hawking radiation, and the theory of acoustic/analogue black holes
were thus introduced.

Classical black hole analogues (alternatively, the analogue systems) are fluid dynamical
analogue of general relativistic black holes. Such analogue effects may be  observed when
the
acoustic perturbation (sound waves) propagates through a classical dissipation-less
transonic fluid. The acoustic horizon, which resembles the actual black hole event horizon
in many ways,
may be generated at the transonic point in the fluid flow. Acoustic horizon emits
acoustic radiation with quasi thermal phonon spectra, which is analogous to the actual Hawking radiation.

Transonic accretion onto astrophysical black holes is a very interesting example of
classical analogue system found naturally in the Universe. An accreting black hole system
as a classical analogue is unique in the sense that {\it only} for such a system,
{\it both} kind of horizons, the electromagnetic and the acoustic (generated due to
transonicity of accreting fluid) are {\it simultaneously} present in the {\it same} system.
Hence accreting astrophysical black holes are the most ideal candidate to study theoretically
and to compare the properties of these two different kind of horizons. Such a
system is also unique in the aspect that  
accretion onto the black holes represents the {\it only}
classical analogue system found in the nature so far,
where the analogue Hawking temperature
may {\it exceed} the actual Hawking temperature.
In this review article, it will be demonstrated that, in general,
the transonic accretion in astrophysics
can be considered as an example of
the classical analogue gravity model.
\end{abstract}
\maketitle
\noindent
\section{Black Holes}
\noindent
Black holes 
are the vacuum solutions of Einstein's field equations in general
relativity. Classically, a black hole is conceived as a singularity in space
time, censored from the rest of the Universe by a mathematically defined one way
surface, the event horizon. 
Black holes are completely characterized only by three externally 
observable parameters, the mass of the black hole $M_{BH}$, the rotation (spin) $J_{BH}$
and charge $Q_{BH}$. All other informations about the matter which formed the 
black hole or is falling into it, disappear behind the event horizon, are therefore
permanently inaccessible to the external observer.
Thus the space time metric defining the vacuum exterior
of a classical black hole is characterized by 
$M_{BH},J_{BH}$ and $Q_{BH}$ only. The 
most general family of 
black hole solutions have non zero values of 
$M_{BH},J_{BH}$ and $Q_{BH}$ (rotating charged black holes), and are 
known as the Kerr-Newman black holes.
The following table classifies various categories of black hole solutions
according to the value of $M_{BH},J_{BH}$ and $Q_{BH}$.
%\hskip -4.0truecm
\begin{center}
{
\begin{tabular}{|l|l|l|l|r|r|} \hline\hline
{\bf Types of Black Hole} & {\bf Mass} & {\bf Angular Momentum} & {\bf Charge} \\ \hline\hline
{{\large\it Kerr-Newman}}&&&\\
{(Newman et. al. 1965)}
& {${\bf M}_{\bf BH}>{\bf 0}$}
&{{${\bf J}_{\bf BH}{\ne}{\bf 0}$}}
&{${\bf Q}_{\bf BH}{\ne}{\bf 0}$}\\ \hline
{\large\it Kerr} (Kerr 1963) &
{${\bf M}_{\bf BH}>{\bf 0}$}
&{{${\bf J}_{\bf BH}{\ne}{\bf 0}$}}
&{${\bf Q}_{\bf BH}{=}{\bf 0}$}\\ \hline
{\large\it Reissner-Nordstr$\ddot{\rm o}$m} &&&\\
{(Reissner 1916; Weyl 1917; Nordstr$\ddot{o}$m 1918)} & {${\bf M}_{\bf BH}>{\bf 0}$}
&{{${\bf J}_{\bf BH}{=}{\bf 0}$}}
&{${\bf Q}_{\bf BH}{\ne}{\bf 0}$}\\ \hline
{\large\it Schwarzschild}&&& \\
{(Schwarzschild 1916)} &
{${\bf M}_{\bf BH}>{\bf 0}$}
&{{${\bf J}_{\bf BH}{=}{\bf 0}$}}
&{${\bf Q}_{\bf BH}{=}{\bf 0}$}\\ \hline\hline 
\end{tabular}
\vskip 0.5truecm
{\bf Table 1:} Classification of black holes according to the value 
of its Mass, angular momentum and charge.
}
\end{center}

%For $J=q=0$, one obtains a Schwarzschild black hole, and for $q=0$,
%one obtains a Kerr black hole. These two kind of black holes 
%(Kerr and Schwarzschild) are important in
%astrophysics. 
The Israle-Carter-Robinson theorem (Israle 1967; Carter 1971; Robinson 1975),
when coupled with Price's conjecture (Price 1972), ensures that any object with 
event horizon must rapidly settles down to the Kerr metric, radiating away all its
irregularities and distortions which may deviate them from the black hole solutions
exactly described by the Kerr metric. 

In astrophysics, black holes are the end point of gravitational
collapse of massive celestial objects. 
The Kerr-Newman and the Reissner-Nordstr$\ddot{\rm o}$m black hole
solutions usually do not play any significant role in astrophysical context.
Typical astrophysical black holes are supposed to be immersed in an 
charged plasma environment. Any net charge $Q_{BH}$ will thus rapidly be 
neutrilized by teh ambient magnetic field. The time scale for such charge 
relaxation would be roughly of the order of $\left(M_{BH}/M_{\odot}\right){\mu}{\rm sec}$
($M_{\odot}$ being the mass of the Sun,
see, e.g., Hughes 2005 for further details), which is obviously far shorter compared 
to the rather long timescale relevant to {\it observing} most of the 
properties of the astrophysical black holes. Hence the Kerr solution provides the
complete description of most stable astrophysical black holes. However, the study of 
Schwarzschild black holes, although less general compared to the Kerr type holes,
is still greatly relevant in astrophysics.

Astrophysical black holes may be broadly
classified into two categories, the stellar mass ($M_{BH}{\sim}$ a few
$M_{\odot}$),
and super massive ($M_{BH}{\ge}{10^6}M_{\odot}$)
black holes.
While the birth history of the stellar mass black holes is theoretically known
with almost absolute certainty (they are the endpoint of the gravitational
collapse of massive stars), the formation scenario of the supermassive black
hole is not unanimously understood.
A super massive black hole may form through the
monolithic collapse of early proto-spheroid gaseous mass originated at the time
of galaxy formation. Or a number of stellar/intermediate mass
%($M_{BH}{\sim}{10^{3-4}}M_{\odot}$)
black holes may merge to form it. Also the
runaway growth of a seed black hole by accretion in a specially favoured high-density
environment may lead to the formation of super massive black holes. However, it is yet to be well understood
exactly which of the above mentioned processes routes toward the formation of
super massive black holes;
see, e.g., Rees 1984, 2002; Haiman \& Quataert 2004; and Volonteri 2006,
%\cite{rees}
for comprehensive
review on the formation and evolution of super massive black holes.

Both kind of astrophysical black holes,
the stellar mass and super massive black holes,
however, accrete matter from the surroundings. Depending on the intrinsic
angular momentum content of accreting material, either spherically symmetric
(zero angular momentum flow of matter), or
axisymmetric (matter flow with non-zero finite angular momentum)
flow geometry is invoked to study an accreting black hole system (see 
the excellent monographs by Frank, King \& Raine 1992, and
Kato, Fukue \& Mineshige 1998, for details about the astrophysical 
accretion processes).
We will get back to the accretion process in greater detail in subsequent 
sections.
%\cite{frank}.\\
\section{Black Hole Thermodynamics}
\noindent
Within the framework of purely classical physics, black holes in any diffeomorphism covariant
theory of gravity (where the field equations directly follow from the
diffeomorphism covariant Lagrangian) and in general relativity, mathematically
resembles some aspects of classical thermodynamic systems 
(Wald 1984, 1994, 2001; Keifer 1998; Brown 1995,
%(\cite{wald84}\cite{wald94}\cite{wald2001}\cite{kiefer98}, \cite{brown95}
and references therein). 
In early seventies, a
series of influential works (Bekenstein 1972, 1972a, 1973, 1975; Israel 1976; 
Bardeen, Carter \& Hawking 1973, see also Bekenstein 1980 for a review)
%(\cite{beken72}\cite{beken72a}\cite{beken73}\cite{beken75}\cite{israel76}\cite{bardeen}, 
%see also [14] 
%\cite{beken80} 
%for a review)
revealed the idea that classical black
holes in general relativity, obey certain laws which bear remarkable analogy to the
ordinary laws of classical thermodynamics. Such analogy between black hole
mechanics and ordinary thermodynamics (`The Generalized Second Law', as it is
customarily called) leads to the idea of the `surface gravity' of black
hole,\footnote {The surface gravity may be defined as the acceleration measured
by red-shift of light rays passing close to the horizon (see, 
e.g., Helfer 2003, and references therein for further details.)}
%(\cite{helfer2003})} 
$\kappa$, which can be obtained by computing the norm of the gradient of
the norms of the Killing fields evaluated at the stationary black hole
horizon, and is found to be constant on the horizon (analogous to the constancy of
temperature T on a body in thermal equilibrium - the `Zeroth Law' of classical
thermodynamics). Also, $\kappa=0$ can not be accomplished by performing finite number of
operations (analogous to the `weak version' of the third law of classical
thermodynamics where temperature of a system cannot be made to reach at absolute
zero, see discussions in Keifer 1998).
%\cite{kiefer98}). 
It was found by analogy via black
hole uniqueness theorem (see, e.g., Heusler 1996,
%\cite{heusler96} 
and references therein) 
that the role of entropy in
classical thermodynamic system is played by a constant multiple of the surface
area of a classical black hole.
\noindent
\section{Hawking Radiation}
\noindent
The resemblance between the laws of ordinary
thermodynamics to those of black hole mechanics were, however, initially
regarded as purely formal. This is because, the physical temperature of a black
hole is absolute zero (see, e.g. Wald 2001).
%\cite{wald2001}). 
Hence physical relationship between
the surface gravity of the black hole and the temperature of a classical 
thermodynamic system can not be conceived. This further
indicates that a classical black hole can never radiate. However, introduction
of quantum effects might bring a radical change to the situation. In an epoch
making paper published in 1975, Hawking (Hawking 1975)
%\cite{hawking75} 
used quantum field theoretic
calculation on curved spacetime to show that the physical
temperature and entropy of black hole {\it does} have finite non-zero value
(see Page 2004 and Padmanabhan 2005
%\cite{page2004} 
for intelligible reviews of black hole thermodynamics and 
Hawking radiation). A
classical space time describing gravitational collapse leading to the
formation of a Schwarzschild black
hole was assumed to be the dynamical back ground, and a linear quantum field,
initially in it's vacuum state prior to the collapse, was considered to propagate
against this background. The vacuum expectation value of the energy momentum
tensor of this field turned out to be negative near the horizon. This phenomenon
leads to the flux of negative energy into the hole.  Such negative energy flux would
decrease the mass of the black hole and would lead to the fact that the quantum
state of the outgoing mode of the field would contain particles.\footnote {For
a lucid description of the physical interpretation of Hawking radiation,
see, e.g., Wald 1994; Keifer 1998; Helfer 2003; Page 2004 and
Padmanabhan 2005.} 
%\cite{wald94}, \cite{kiefer98}, 
%\cite{helfer2003}.}. 
The expected number of such
particles would correspond to radiation from a perfect black body of finite size.
Hence the spectrum of such radiation is thermal in nature, 
and the temperature of such
radiation, the Hawking temperature $T_H$ from a Schwarzschild black 
hole, can be computed as 
\begin{equation}
T_H=\frac{{\hbar}c^3}{8{\pi}k_BGM_{BH}}
\label{eq1}
\end{equation}
where $G$ is the 
universal gravitational constant,
$c,{\hbar}$ and $k_B$ are the velocity of light in vacuum, the Dirac's 
constant and the Boltzmann's
constant, respectively.

The semi classical description for
Hawking radiation treats the gravitational field classically and the
quantized radiation field satisfies the d'Alembert equation. At any time, black
hole evaporation is an adiabatic process if the residual mass of the hole at
that time remains larger than the Planck mass.
%\section{Toward an Analogy of Hawking Effect: The motivation}
\section{Toward an Analogy of Hawking Effect}
\noindent
Substituting the values of the fundamental constants in Eq. (1), one can rewrite
$T_H$ for a Schwarzschild black hole as (Helfer 2003):
%(\cite{helfer2003}):
\begin{equation}
T_H ~{\sim}~ 6.2{\times}10^{-8}
\left(\frac{M_{\odot}}{M_{BH}}\right) {\rm Degree~~Kelvin}
\label{eq2}
\end{equation}
It is evident from the above equation that for one solar mass black hole, the
value of the Hawking temperature would be too small to be experimentally
detected. A rough estimate shows that $T_{H}$ for stellar mass black holes would
be around $10^{7}$ times colder than the cosmic microwave background radiation.
The situation for super massive black hole will be much more worse, as 
$T_H {\propto} {1}/{M_{BH}}$. Hence $T_{H}$ would be a measurable quantity
only for primordial black holes with very small size and mass, if such black holes
really exist, and if instruments can be fabricated to detect them. The lower
bound of mass for such black holes may be estimated analytically. The time-scale
${\cal T}$ (in years) over which the mass of the black hole changes significantly due to
the Hawking's process may be obtained as (Helfer 2003):
%(\cite{helfer2003})
\begin{equation}
{\cal T}{\sim}\left(\frac{M_{BH}}{M_{\odot}}\right)^3 10^{65}~~{\rm Years}
\label{eq3}
\end{equation}
As the above time scale is a measure of the lifetime of the hole itself, the
lower bound for a primordial hole may be obtained by setting $ {\cal T}$ equal
to the present age of the Universe. Hence the lower bound
for the mass of the primordial black holes
comes out to be around $10^{15}$ gm. The size of such a black hole
would be of the order of $10^{-13}$ cm and the corresponding $T_{H}$ would be
about
$10^{11}{\rm K}$, which is comparable with the macroscopic 
fluid temperature of the freely falling matter (spherically symmetric accretion) onto
an one solar mass isolated Schwarzschild black hole
(see section 12.1 for further details). However, present day
instrumental technique is far from efficient to detect these primordial black
holes with such an extremely small dimension, if such holes exist at all in
first place. Hence, the observational manifestation of Hawking radiation seems
to be practically impossible. 

On the other hand, due to the infinite redshift caused by the event horizon, the
initial configuration of the emergent Hawking Quanta is supposed to possess
trans-Planckian frequencies and the corresponding wave lengths are beyond the Planck scale.
Hence, low energy effective theories cannot self consistently deal with the
Hawking radiation (see, e.g., Parentani 2002 for further
details).
%\cite{parentani2002}. 
Also, the nature of the fundamental degrees
of freedom and the physics of such ultra short distance is yet to be well
understood. Hence, some of the fundamental issues like the statistical meaning of the
black hole entropy, or the exact physical origin of the out going mode of the
quantum field, remains unresolved (Wald 2001).
%\cite{wald2001}.

Perhaps the above mentioned difficulties associated with the theory of
Hawking radiation served as the principal motivation to launch
a theory, analogous to the Hawking's one, effects of which would be possible
to comprehend through relatively more perceivable physical systems.
The theory of {\AHR}  opens up the possibility to experimentally verify some
basic features of black hole physics by creating the sonic horizons in the
laboratory. A number of works have been carried out to formulate the
condensed matter or optical analogue of event horizons
\footnote{Literature on study of analogue systems in condensed matter or optics 
are quite large in numbers. Condensed matter or optical analogue systems
deserve the right to be discussed as separate review articles on its own. In
this article, we, by no means, 
are able to provide the complete list of references for theoretical or
experimental works on such systems. However, to have an idea on 
the analogue effects in condensed matter or optical 
systems, readers are refereed to the monograph by Novello, Visser \& Volovik (2002),
the most comprehensive review article by Barcelo, Liberati \& Visser 2005,
%\cite{novelo}
for review, a greatly enjoyable popular science article 
published in the Scientific American by 
Jacobson \& Parentani 2005, 
and to some of the representative papers 
like Jacobson \& Volovik 1998; Volovik 1999, 2000, 2001;
Garay, Anglin, Cirac \& Zoller 2000, 2001; 
Reznik 2000; Brevik \& Halnes 2002; Sch$\ddot{\rm u}$tzhold
\& Unruh 2002; 
Sch$\ddot{\rm u}$tzhold, G$\ddot{\rm u}$nter \&
Gerhard 2002;  Leonhardt 2002, 2003; de Lorenci,
Klippert \& Obukhov 2003 and Novello,
Perez Bergliaffa, Salim, de Lorenci \& Klippert 2003. As already 
mentioned, this list 
of references, however, is by no means complete.}.
%\cite{fulllist}.}.
The theory of analogue 
Hawking radiation may find important uses in the fields of 
investigation of quasi-normal modes (Berti, Cardoso \& Lemos 2004; Cardoso, 
Lemos \& Yoshida 2004),
acoustic super-radiance
(Basak \& Majumdar 2003; Basak 2005;
Lepe \& Saavedra 2005; Slatyer, \& Savage 2005;
Cherubini, Federici \& Succi 2005; Kim, Son, \& Yoon 2005; 
Choy, Kruk, Carrington, Fugleberg, Zahn, Kobes, Kunstatter
\& Pickering 2005; Federici, Cherubini, Succi \& Tosi 2005),
%(\cite{berti}\cite{cardoso}),
FRW cosmology (Barcelo, Liberati \& Visser 2003)
%\cite{barcelo}, 
inflationary
models, quantum gravity and sub-Planckian models of string theory (Parentani 2002).
%\cite{parentani2002}. 

For space limitation, in this article, we will, however, mainly describe the formalism behind the 
{\it classical} analogue systems.
By `classical analogue systems' we refer to the examples where 
the analogue effects are studied in classical systems (fluids), and not in 
quantum fluids.
In the following sections, we discuss the basic features of a classical analogue
system. 
\noindent
\section{Analogue Gravity Model and the Black Hole Analogue}
\noindent
In recent years, strong analogies have been established between the physics
of acoustic perturbations in an inhomogeneous dynamical fluid system, and
some kinematic features of space-time in general relativity. An effective
metric, referred to as the `acoustic metric', which
describes the geometry of the manifold in which  acoustic perturbations
propagate, can be constructed. This effective geometry can capture the properties of curved
space-time in general relativity. Physical models constructed utilizing such
analogies are called `analogue gravity models' (for details on analogue
gravity models, see, e.g. the review articles by Barcelo, Liberati \& Visser (2005) and
Cardoso (2005), 
and the monograph by Novello, Visser \& Volovik (2002)).
\par One of the most significant effects of analogue gravity is the
`classical black hole analogue'. Classical black hole analogue effects may be  observed when acoustic perturbations (sound waves)
propagate through a classical, dissipation-less, inhomogeneous transonic
fluid. Any acoustic perturbation, dragged by a supersonically
moving fluid, can never escape upstream by penetrating the `sonic surface'.
Such a sonic surface is a collection of transonic points in space-time,
and can act as a `trapping' surface for outgoing {\it phonons}. Hence, the
sonic surface is actually an {\it acoustic horizon}, which resembles
 a black hole event horizon in
many ways and is generated at the transonic
point in the fluid flow. The acoustic horizon is essentially a null hyper
surface, generators of which are the {\it acoustic} null geodesics, i.e.
the phonons. The acoustic horizon emits acoustic radiation with quasi thermal
phonon spectra, which is analogous to the actual Hawking radiation. The
temperature of the radiation emitted from the acoustic horizon is referred
to as the analogue Hawking temperature.

Hereafter,
we shall use $T_{{AH}}$
to denote the analogue Hawking temperature, and $T_H$ to denote the
the actual Hawking temperature as defined in (\ref{eq1}).
We shall also use the words `analogue', `acoustic' and `sonic' synonymously
in describing the horizons or black holes. Also the phrases `analogue (acoustic) Hawking
radiation/effect/temperature' should be taken as identical in meaning with the phrase
`analogue (acoustic) radiation/effect/temperature'. A system manifesting the
effects of analogue radiation, will be termed as analogue system.

In a pioneering work, Unruh (1981)
%\cite{unruh81} 
showed that a classical system,
relatively  more clearly perceivable than a quantum black hole system, does
exist, which resembles the black hole as far as the quantum thermal radiation is
concerned. The behaviour of a linear quantum field in a classical gravitational 
field was simulated by the propagation of acoustic disturbance in a convergent
fluid flow. In such a system, it is possible to study the effect of the reaction
of the quantum field on it's own mode of propagation and to contemplate the
experimental investigation of the thermal emission mechanism. Considering 
the equation of motion for a transonic barotropic irrotational fluid, Unruh (1981)
showed 
%\cite{unruh81} 
that the scaler field representing the acoustic perturbation
(i.e, the propagation of sound wave) satisfies a differential equation which is
analogous to the equation of a massless scaler field propagating in a metric.
Such a metric
closely resembles the Schwarzschild metric near the horizon. Thus acoustic
propagation through a supersonic fluid forms an analogue of event horizon, as
 the `acoustic horizon' at the transonic point. The behaviour of the normal modes near
the acoustic horizon indicates that the acoustic wave with a quasi-thermal
spectrum will be emitted from the acoustic horizon and the temperature of such
acoustic emission may be calculated as (Unruh 1981):
%\cite{unruh81}:
\begin{equation}
T_{AH}=\frac{\hbar}{4{\pi}k_B}\left[\frac{1}{c_s}\frac{{\partial}{u_{\perp}^2}}{\partial{\eta}}
\right]_{\rm r_h}
\label{eq4}
\end{equation}
Where $r_h$ represents the location of the acoustic horizon,
$c_s$ is the sound speed,  $u_{\perp}$ is the component of the
dynamical flow velocity normal to the acoustic horizon, and
$ {\partial}/{{\partial}{\eta}}$
represents derivative in the direction normal to the 
acoustic horizon.

Equation (\ref{eq4}) has clear resemblance with (\ref{eq1}) and hence 
$T_{AH}$ is designated as analogue Hawking temperature and such quasi-thermal
radiation from acoustic (analogue) black hole is known as the analogue Hawking
radiation. Note that the sound speed $c_s$ in Unruh's original
treatment (the above equation) was assumed to be constant in space,
i.e., an isothermal equation of state had been invoked to describe the
fluid.

Unruh's work was followed by other important papers
(Jacobson 1991, 1999; Unruh 1995; Visser 1998; Bili$\acute{c}$ 1999)
%\cite{jacob,95unruh,visser,99jacob,bilic}. 
A more general treatment of the
classical analogue radiation for Newtonian fluid was discussed by Visser
(1998), who considered a general barotropic, inviscid fluid. The
acoustic metric for a point sink was shown to be conformally related to the
Painlev\'e-Gullstrand-Lema\^\i{}tre  form of the Schwarzschild metric 
(Painlev\'e 1921; Gullstrand 1922; Lema\^\i{}tre 1933)
and a more general
expression for analogue temperature was obtained, where unlike  Unruh's original
expression (\ref{eq4}), the speed of sound  was allowed to depend on space coordinates.
%In order to determine the analogue Hawking temperature of a classical analogue
%system, one needs to know the location of the acoustic horizon, the velocity of the
%fluid and the speed of sound and their space gradients at the acoustic horizon.
\par In the analogue gravity systems discussed above, the fluid flow is
non-relativistic in flat Minkowski space, whereas the sound wave propagating
 through
the non-relativistic fluid is coupled to a curved pseudo-Riemannian metric. This
approach has been extended to relativistic fluids (Bili$\acute{c}$ 1999) by incorporating
the general relativistic fluid dynamics. 
%Since the introduction of
%viscosity may destroy  Lorenz invariance, the acoustic analogue is best
%studied in a vorticity free dissipation-less fluid.

In subsequent sections, we will pedagogically develop the concept of the acoustic geometry
and related quantities, like the acoustic surface gravity and 
the acoustic Hawking temperature.
\section{Curved Acoustic Geometry in a Flat Space-time}
\noindent
Let  $\psi$ denote
 the velocity potential describing the fluid flow in Newtonian space-time, i.e.
 let
${\vec{u}}=-{\nabla}\psi$, where ${\vec{u}}$ is the velocity vector describing
the dynamics of a Newtonian fluid. The specific enthalpy
$h$ of a barotropic Newtonian fluid satisfies ${\nabla}h=(1/{\rho}){\nabla}p$,
where $\rho$ and $p$ are the density and the pressure of the fluid.
One then writes the Euler equation as
\begin{equation}
-{\partial}_t{\psi}+h+\frac{1}{2}\left({\nabla}{\psi}\right)^2+\Phi=0,
\label{eq5}
\end{equation}
where $\Phi$ represents the  potential  associated with any external driving force.
Assuming small fluctuations around some steady background
 $\rho_0, p_0$ and $\psi_0$, one
can linearize the continuity and the Euler equations and
obtain a wave equation
(see Landau \& Lifshitz 1959, and
Visser 1998, for further detail). 

%We will follow Visser (1998) to provide the following derivation.
%This derivation is based in part on references
%\cite{landau,visser}.
The continuity and Euler's equations may be
expressed as:
\begin{equation}
\frac{\partial{\rho}}{\partial{t}}+{\nabla}\cdot \left(\rho{\vec{u}}\right)
\label{eq6}
\end{equation}
\begin{equation}
\rho\frac{d{\vec {u}}}{dt}{\equiv}
\rho\left[\frac{\partial{\vec {u}}}{\partial{t}}
+\left({\vec {u}}\cdot {\nabla}\right){\vec {u}}\right]
=-\nabla{p}+{\vec {F}}
\label{eq7}
\end{equation}
with ${\vec {F}}$ being the sum of all external forces acting on the 
fluid which may be expressed in terms of a potential
\begin{equation}
{\vec {F}}=-{\rho}{\nabla}{\Phi},
\label{eq8}
\end{equation}
Euler's equation may now be recast in the form
\begin{equation}
\frac{\partial{\vec {u}}}{\partial{t}}=
{\vec {u}} \times ( \nabla \times {\vec {u}})  - {1\over\rho} \nabla p
- \nabla\left( {{1\over2}} u^2 + \Phi\right)
\label{eq9}
\end{equation}
Next we assume the fluid to be inviscid, irrotational, and
barotropic. Introducing the specific enthalpy $h$, such that
\begin{equation}
{\nabla}h=\frac{\nabla{p}}{\rho}
\label{eq10}
\end{equation}
and  the velocity potential $\psi$ for which 
${\vec {u}}=-{\nabla}\psi$, 
Eq. (\ref{eq9}) may be written as
\begin{equation}
-\frac{\partial{\psi}}{\partial{t}} + h + {1\over2} (\nabla\psi)^2
 + \Phi = 0
\label{eq11}
\end{equation}

One now linearizes the continuity and  Euler's equation around some 
unperturbed background flow variables $\rho_0$, $p_0$, $\psi_0$.
Introducing
\begin{eqnarray}
\rho=\rho_0+\epsilon \rho_1 +{\cal O}(\epsilon^2), & & p=p_0+\epsilon p_1 +{\cal O}(\epsilon^2) ,
\nonumber \\
\psi=\psi_0+\epsilon \psi_1 +{\cal O}(\epsilon^2),
 & & h=h_0+\epsilon h_1,
\label{eq12}
\end{eqnarray} 
from the continuity equation we obtain
\begin{equation}
\frac{\partial\rho_0}{\partial{t}}+
\nabla\cdot(\rho_0 \; {\vec {u}}_0) = 0; \;\;\;\;\;
\frac{\partial\rho_1}{\partial{t}}+
\nabla\cdot(\rho_1 \; {\vec {u}}_0 + \rho_0 \; {\vec {u}}_1) = 0.
\label{eq13}
\end{equation}
Equation (\ref{eq10}) implies
\begin{equation}
h_1=p_1\frac{dh}{dp}=
\frac{p_1}{\rho_0}\, .
\label{eq14}
\end{equation}
Using this the linearized Euler equation reads
\begin{equation}
-\frac{\partial{\psi_0}}{\partial{t}} + h_0 + {1\over2} (\nabla\psi_0)^2
 + \Phi = 0; \;\;\;\;\;
-\frac{\partial{\psi_0}}{\partial{t}}
+ {p_1\over\rho_1} - {\vec {u}}_0 \cdot \nabla\psi_1 = 0.
%\eqno{(A-7)}
\label{eq15}
\end{equation}
Re-arrangement of the last equation 
together  with the barotropic assumption yields
\begin{equation}
\rho_1 =
{\partial \rho\over\partial p} \; p_1 =
{\partial \rho\over\partial p} \; \rho_0  \;
( \partial_t \psi_1 + {\vec {u}}_0 \cdot \nabla\psi_1 )\, .
%\eqno{(A-8)}
\label{eq16}
\end{equation}
Substitution of this  into the linearized continuity equation 
gives the sound wave equation 
\begin{equation}
-\frac{\partial}{\partial{t}}
\left[ {\partial\rho\over\partial p} \; \rho_0 \;
            \left(\frac{\partial\psi_1}
{\partial{t}} + {\vec {u}}_0 \cdot \nabla\psi_1\right)
     \right]+
\nabla \cdot
     \left[ \rho_0 \; \nabla\psi_1
            - {\partial\rho\over\partial p} \; \rho_0 \; {\vec {v}}_0 \;
              \left( \frac{\partial\psi_1}{\partial{t}} 
+ {\vec {v}}_0 \cdot \nabla\psi_1\right)
     \right]
=0 .
\label{eq17}
\end{equation}
Next, we define the local speed of sound by 
\begin{equation}
c_s^2={\partial{p}}/{\partial\rho},
%\eqno{(A-10)}
\label{eq18}
\end{equation}
where the partial derivative is taken at constant specific entropy.
With help of the $4\times{4}$ matrix 
\begin{equation}
f^{\mu\nu} \equiv
{\rho^0}
\left[ \matrix{-{I}&\vdots&-{\vec {u}}\cr
               \cdots\cdots&\cdot&\cdots\cdots\cdots\cdots\cr
               -{\vec {u}}&\vdots&( c_s^2 - {{u}^2} )\cr }
\right]
\label{eq19}
\end{equation}
where ${I}$ is the $3\times3$ identity matrix,
 one can put  Eq. (\ref{eq17}) to the form 
 \begin{equation}
{\partial}_{\mu}\left(f^{\mu{\nu}}{\partial}_{\nu}{\psi^1}\right)=0.
%\label{eq59}
\label{eq20}
\end{equation}
Equation (\ref{eq20}) describes  the propagation 
of the linearized scalar potential $\psi^1$. The function $\psi^1$
represents the low amplitude fluctuations 
around the steady background 
 $(\rho_0,p_0,\psi_0)$ and 
thus describes the propagation  
of acoustic perturbation, .i.e. the propagation of 
sound waves.

The form of Eq. (\ref{eq20}) suggests that it may be 
regarded as  a d'Alembert  equation in curved space-time
geometry.
In any 
pseudo-Riemannian manifold the  d'Alembertian operator can be expressed as
(Misner, Thorne \& Wheeler 1973)
\begin{equation}
\Box =\frac{1}{\sqrt{-\left|g_{\mu{\nu}}\right|}}
{\partial}_{\mu}
{\sqrt{-\left|g_{\mu{\nu}}\right|}}
g^{\mu{\nu}}
{\partial}_\nu
\, ,
\label{eq21}
\end{equation}
where  $\left|g_{\mu{\nu}}\right|$ is the 
determinant  and 
 $g^{\mu{\nu}}$ is the  inverse of the metric
 $g_{\mu{\nu}}$.
Next, if one identifies
\begin{equation}
f^{\mu{\nu}}=\sqrt{-\left|g_{\mu{\nu}}\right|}g^{{\mu}\nu} ,
\label{eq22}
\end{equation}
 one can recast the acoustic wave equation 
in the form (Visser 1998)
\begin{equation}
\frac{1}{\sqrt{-\left|G_{\mu{\nu}}\right|}}
{\partial}_\mu
\left(
{\sqrt{-\left|G_{\mu{\nu}}\right|}}
G^{\mu{\nu}}
\right){\partial_\nu}\psi^1=0 ,
\label{eq23}
\end{equation}
where $G_{\mu{\nu}}$ is the acoustic metric tensor for the 
Newtonian fluid. The explicit form of ${G}_{\mu{\nu}}$
is obtained as
\begin{equation}
G_{\mu\nu}
\equiv {\rho_0}
\left[ \begin{matrix}{-(c_s^2-u^2)&\vdots&-{{\vec u}}\cr
               \cdots\cdots\cdots\cdots&\cdot&\cdots\cdots\cr
               -{\vec u}&\vdots& {I}\cr } \end{matrix}\right]
\label{eq24}
\end{equation}
The Lorentzian metric described by (\ref{eq24})
has an associated non-zero acoustic Riemann tensor for 
non-homogeneous, flowing fluids.

Thus, the propagation of acoustic perturbation, or the sound 
wave, embedded in a barotropic, irrotational, non-dissipative Newtonian fluid 
flow may be described by a scalar d'Alembert equation in a {\it curved} acoustic 
geometry. The corresponding  acoustic metric tensor is a matrix that 
depends on dynamical and thermodynamic variables parameterizing the fluid flow.

For analogue systems discussed above, the fluid particles are coupled to
the {\it flat} metric of Mankowski's space (because the governing equation for fluid
dynamics in the above treatment is completely Newtonian), whereas the sound wave
propagating through the non-relativistic fluid is coupled to the {\it curved}
pseudo-Riemannian metric. Phonons (quanta of acoustic perturbations) are the
null geodesics, which generate the null surface, i.e., the acoustic horizon.
Introduction of viscosity may destroy the Lorentzian invariance and hence
the acoustic analogue is best observed in a vorticity free completely
dissipation-less fluid (Visser 1998,
%(\cite{visser98},
and references therein). That is why, the
Fermi superfluids and the Bose-Einstein condensates are ideal
to simulate the analogue effects.
                                                                                                     
The most important issue emerging out of the above
discussions is that (see Visser 1998 and 
Barcelo, Liberati \& Visser 2005
%\cite{visser98}
for further details):
Even if the governing equation for fluid flow is completely non-relativistic
(Newtonian), the acoustic fluctuations embedded into it are described by a curved
pseudo-Riemannian geometry. This information is useful to portray the immense
importance of the study of the acoustic black holes, i.e. the black hole
analogue, or simply, the analogue systems.

The acoustic metric  (\ref{eq24}) in many aspects resembles a black hole type geometry
in general relativity. For example, the notions such as
`ergo region' and
  `horizon' may be introduced in full analogy with those of general relativistic
 black holes.
For a stationary flow, the time translation Killing vector
$\xi\equiv\partial/\partial t$
leads to the concept of
{\em acoustic ergo sphere} as a surface at which
$G_{\mu\nu}\xi^\mu\xi^\nu$ changes its sign.
 The acoustic ergo sphere is  the envelop of the {\em acoustic ergo region}
where $\xi^{\mu}$ is space-like with respect to the acoustic metric.
Through the equation
$G_{\mu\nu}\xi^\mu\xi^\nu =g_{tt}=u^2-c_s^2$,
it is obvious that
inside the ergo region  the fluid is
supersonic.
The `acoustic horizon'
can be defined  as the boundary of a region from which acoustic null geodesics
or phonons, cannot escape. Alternatively, the acoustic horizon is defined as a
time like hypersurface defined by the equation
\begin{equation}
c_s^2-u_{\perp}^2=0 ,
\label{eq25}
\end{equation}
where $u_{\perp}$ is the component of the fluid velocity perpendicular to the acoustic horizon. Hence, any steady supersonic flow described in a stationary geometry by a time independent velocity vector field forms an ergo-region, inside
which the acoustic horizon is generated at those points
where the normal component of the fluid
velocity is equal to the speed of sound.
                                                                                                     
In analogy to general relativity, one also defines the surface gravity
and the corresponding Hawking temperature
associated with the acoustic horizon.
The acoustic surface gravity may be obtained (Wald 1984) by computing
the gradient
of the norm of the Killing field which becomes null
vector field at the acoustic horizon.
The acoustic surface gravity $\kappa$ for a Newtonian fluid is then
given by (Visser 1998)
\begin{equation}
\kappa=\frac{1}{2c_s}
\left|\frac{\partial}{\partial\eta}\left(c_s^2-u_{\perp}^2\right)\right|
\, .
\label{eq26}
\end{equation}
The corresponding  Hawking temperature
 is then defined as usual:
\begin{equation}
T_{AH}=\frac{\kappa}{2\pi\kappa_B}\, .
\label{eq27}
\end{equation}
\section{Curved Acoustic Geometry in a Curved Space-time}
\noindent
The above formalism may be extended to relativistic fluids in curved space-time
background (Bili\'c 1999). The propagation of acoustic disturbance in a
perfect relativistic inviscid irrotational fluid
is  also described  by the wave equation of the form
(\ref{eq23})
in which
 the acoustic metric tensor and its inverse are defined as
(Bili\'c 1999; Abraham, Bili\'c \& Das 2006; Das, Bili\'c \& Dasgupta 2006)
\begin{equation}
{G}_{\mu\nu} =\frac{\rho}{h c_s}
\left[g_{\mu\nu}+(1-c_s^2)v_{\mu}v_{\nu}\right];
\;\;\;\;
G^{\mu\nu} =\frac{h c_s}{\rho}
\left[g^{\mu\nu}+(1-\frac{1}{c_s^2})v^{\mu}v^{\mu}
\right] ,
\label{eq28}
\end{equation}
where $\rho$ and $h$ are, respectively,
the rest-mass density and the specific enthalpy of the relativistic fluid,
$v^{\mu}$ is the four-velocity,
and $g_{\mu\nu}$ the background space-time metric.
A ($-,+++$) signature has been used to derive (\ref{eq28}).
The ergo region is again defined as the region where the stationary
Killing vector $\xi$ becomes spacelike and the acoustic horizon
 as a timelike hypersurface the wave velocity of which
  equals the speed of sound at every point. The defining equation
for the acoustic horizon is again of the form  (\ref{eq25})
in which the three-velocity component  perpendicular to the horizon
is given by
\begin{equation}
u_{\perp}=\frac
{
\left(\eta^\mu v_\mu\right)^2}
{
\left(\eta^\mu{v_\mu}\right)^2 +\eta^\mu\eta_\mu},
\label{eq29}
\end{equation}
where $\eta^\mu$ is the unit normal to the horizon.
For further details about the propagation of the acoustic 
perturbation, see Abraham, Bili\'c \& Das 2006.
                                                                                                     
It may be shown that, the discriminant of the acoustic metric
for an axisymmetric flow
\begin{equation}
{\cal D}=G_{t\phi}^2-G_{tt}G_{\phi\phi},
\label{eq30}
\end{equation}
vanishes at the acoustic horizon.
A supersonic flow
is characterized by the condition ${\cal D}>0$, whereas for a subsonic flow,
${\cal D}<0$ (Abraham, Bili\'c \& Das 2006). According to the classification of Bercelo,
Liberati, Sonego \& Visser (2004),
a transition from a subsonic (${\cal D}<0$)
to a supersonic (${\cal D}>0$) flow
is an acoustic {\em black hole},
whereas  a transition from a
supersonic to a subsonic flow is an acoustic
{\em white hole}.

For a stationary configuration, the surface gravity can be computed in terms
of the Killing vector
\begin{equation}
\chi^{\mu}=\xi^{\mu}+\Omega\phi^{\mu}
\label{eq31}
\end{equation}
that is null at the acoustic horizon.
Following the standard procedure (Wald 1984; Bili\'c 1999) one finds that
the expression
\begin{equation}
\kappa\chi^\mu=
\frac{1}{2}
G^{\mu\nu}\eta_\nu\frac{\partial}{\partial{\eta}}
(G_{\alpha\beta}\chi^{\alpha}\chi{\beta})
\label{eq32}
\end{equation}
holds at the acoustic horizon,
where the constant $\kappa$ is  the surface gravity.
From this expression one 
deduces the magnitude of the surface gravity  as
(see
Bili\'c 1999; Abraham, Bili\'c \& Das 2006; Das, Bili\'c \& Dasgupta 2006
for further details)
%\cite{bilic}
\begin{equation}
\kappa=
\left|\frac{\sqrt{-{\chi}^{\nu}{\chi}_{\nu}}}{1-c_s^2}
\frac{\partial}{\partial{\eta}}
\left(u-c_s\right)\right|_{\rm r=r_h}
\label{eq33}
\end{equation}
\section{Quantization of Phonons and the Hawking Effect}
%\label{subsec:quant}}
%\label{quantization}
The purpose of this section (has been adopted from 
Das, Bili\'c \& Dasgupta 2006) is to demonstrate how the
quantization of phonons in the  presence of
the acoustic horizon yields
acoustic Hawking radiation.
The acoustic perturbations considered here are classical
sound waves or {\em phonons} that
satisfy the massless wave 
equation 
%(\ref{eq028})
in curved background, i.e.  
the general relativistic analogue of
(\ref{eq23}),
with the metric $G_{\mu\nu}$
given by (\ref{eq28}).
Irrespective of the underlying microscopic structure,
acoustic perturbations are quantized.
A precise quantization scheme for an analogue gravity
system may be rather involved 
(Unruh \& Sch$\ddot{\rm u}$tzhold 2003).
However, at the scales larger than
the atomic scales below which a perfect fluid description breaks down,
the atomic substructure may be neglected and
the field may be considered elementary. Hence,
the quantization proceeds in the same way as in the case of a
scalar field in curved space (Birrell \& Davies 1982)
with a suitable UV cutoff for the scales below a typical
atomic size of a few \AA. 

For our purpose, the most convenient
quantization prescription is the Euclidean path integral
formulation.
Consider a 2+1-dimensional axisymmetric geometry
describing the fluid flow (since we are going to apply this 
on the equatorial plane of the axisymmetric black hole accretion disc,
see section 13 for further details).
The equation of motion (\ref{eq23}) with (\ref{eq28}) 
follows from the  variational principle applied to
the action functional 
\begin{equation}
S[\varphi]=\int dtdrd\phi\, 
\sqrt{-G}\,
G^{\mu\nu}
\partial_{\mu}\varphi
\partial_{\nu}\varphi\, .
\label{eq34}
\end{equation}
We define the functional integral 
\begin{equation}
Z= \int {\cal D}\varphi e^{-S_{\rm E}[\varphi]} ,
\label{eq35}
\end{equation}
where $S_{\rm E}$ is the Euclidean  action 
obtained from (\ref{eq34}) by setting
%\begin{equation}
$t=i\tau$
%\label{eq232}
%\end{equation}
and continuing the Euclidean time $\tau$ from imaginary to real values.
For a field theory at zero temperature, the integral
over $\tau$ extends up to infinity.
Here,
 owing to the presence of the acoustic horizon,
 the integral over $\tau$ 
 will be cut at the inverse Hawking temperature $2\pi/\kappa$
 where $\kappa$ denotes the analogue surface gravity.
 To illustrate how this happens, consider, for simplicity, a non-rotating 
 fluid ($v_\phi=0$) in the Schwarzschild  space-time.
 It may be easily shown that the acoustic metric takes the form
 \begin{equation}
ds^2=g_{tt}\frac{c_s^2-u^2}{1-u^2}dt^2 -2u\frac{1- c_s^2}{1-u^2}
drdt-\frac{1}{g_{tt}}\frac{2-c_s^2u^2}{1-u^2}dr^2 +r^2d\phi^2\, ,
\label{eq36}
\end{equation}
where $g_{tt}=-(1-2/r)$, $u=|v_r|/\sqrt{-g_{tt}}$,
and we have omitted the irrelevant conformal factor $\rho/(hc_s)$.
Using the coordinate transformation
\begin{equation}
dt\rightarrow dt+\frac{u}{g_{tt}}\frac{1- c_s^2}{c_s^2-u^2}dr
\label{eq37}
\end{equation}
we remove the off-diagonal part from (\ref{eq36}) and obtain
\begin{equation}
ds^2=g_{tt}\frac{c_s^2-u^2}{1-u^2}dt^2 -
\frac{1}{g_{tt}}\left[\frac{2-c_s^2u^2}{1-u^2}+
\frac{u^2(1- c_s^2)^2}{(c_s^2-u^2)(1-u^2)}\right]dr^2 
 +r^2d\phi^2.
\label{eq38}
\end{equation}
Next, we evaluate the metric near the acoustic horizon at
$r=r_{\rm h}$ using the expansion in $r-r_{\rm h}$ at first order
 \begin{equation}
c_s^2-u^2\approx 2 c_s \left. \frac{\partial}{\partial r}(c_s-u)
\right|_{r_{\rm h}}(r-r_{\rm h})
\label{eq39}
\end{equation}
and making the substitution
\begin{equation}
r-r_{\rm h}=\frac{-g_{tt}}{2c_s (1-c_s^2)}
\left. \frac{\partial}{\partial r}(c_s-u)\right|_{r_{\rm h}} R^2,
\label{eq40}
\end{equation}
where $R$ denotes a new radial variable.
Neglecting the first term in the square brackets in (\ref{eq38})
and setting $t=i\tau$, we obtain the Euclidean metric in the form
\begin{equation}
ds_{\rm E}^2=\kappa^2 R^2 d\tau^2 +dR^2 +r_{\rm h}^2 d\phi^2\, ,
\label{eq41}
\end{equation}
where 
\begin{equation}
\kappa=\frac{-g_{tt}}{1-c_s^2}
\left|\frac{\partial}{\partial r}(u-c_s)\right|_{r_{\rm h}}\, .
\label{eq42}
\end{equation}
Hence, the metric near $r=r_{\rm h}$ is the product of the metric on S$^1$
and the Euclidean Rindler space-time 
\begin{equation}
ds_{\rm E}^2=dR^2 + R^2 d(\kappa \tau)^2 .
\label{eq43}
\end{equation}
With the periodic identification 
$\tau\equiv \tau+2\pi/\kappa$, the metric (\ref{eq43})
describes $\Re^2$ in plane polar coordinates. 

Furthermore, making the substitutions
$R=e^{\kappa x}/\kappa$ and $\phi=y/r_{\rm h}+\pi$,
the Euclidean action takes the form of the 
2+1-dimensional free scalar field action
at non-zero temperature
\begin{equation}
S_{\rm E}[\varphi]=\int_0^{2\pi/\kappa} d\tau 
\int_{-\infty}^{\infty}dx
\int_{-\infty}^{\infty} dy \frac{1}{2} (\partial_{\mu} \varphi)^2,
 \label{eq44}
\end{equation} 
where we have set
the upper and lower bounds of the integral over $dy$
to $+\infty$ and $-\infty$, respectively,
assuming  that $r_{\rm h}$ is sufficiently large.
Hence, the functional integral $Z$ in (\ref{eq35}) 
is evaluated over the fields $\varphi(x,y,\tau)$ that are periodic in
$\tau$ with period  $2\pi/\kappa$.
In this way, the functional $Z$ is just the
 partition function for a grand-canonical ensemble of free  bosons
  at the 
Hawking temperature
$T_{\rm H}=\kappa/(2\pi\kappa_B)$.
However, the radiation spectrum will not be exactly thermal
since we have to cut off the scales below the atomic scale
(Unruh 1995). The choice of the cutoff and the deviation of 
the acoustic radiation spectrum from the thermal spectrum is
closely related to the so-called {\em transplanckian problem}
of Hawking radiation 
(Jacobson 1999a, 1992; Corley \& Jacobson 1996).

In the  Newtonian approximation, 
(\ref{eq42}) reduces to the usual
non-relativistic expression for the acoustic surface gravity 
represented by (\ref{eq26}).
%\begin{equation}
%\kappa=
%\left|\frac{\partial}{\partial r}(u-c_s)\right|_{r_{\rm h}}\, .
%\label{eq45}
%\end{equation}
\section{Salient Features of Acoustic Black Holes and its
Connection to Astrophysics}
\noindent
In summary, analogue (acoustic)
black holes (or systems) are fluid-dynamic analogue of general relativistic black
holes. Analogue black holes possess analogue (acoustic) event horizons at local
transonic points. Analogue black holes emit
analogue Hawking radiation, the temperature of which is termed as analogue
Hawking temperature, which may be computed using Newtonian
description of fluid flow. Black hole analogues are important to study because
it may be possible to create them experimentally in laboratories to study some
properties of the black hole event horizon, and to study the experimental
manifestation of Hawking radiation.

According to the discussion presented in previous sections,
it is now obvious that, to calculate the analogue surface gravity
$\kappa$ and the analogue Hawking temperature $T_{AH}$ for a classical
analogue gravity system, one {\it does} need to know the {\it exact} location
(the radial length scale) of the acoustic horizon $r_h$, the dynamical and
the acoustic velocity corresponding to the flowing fluid at the
acoustic horizon, and its space derivatives, respectively. Hence an
astrophysical fluid system, for which the above mentioned quantities
can be calculated, can be shown to represent an classical
analogue gravity model.

For acoustic black holes, in general, the ergo-sphere and the acoustic horizon do not coincide. However,
for some specific stationary geometry they do. This is the case, e.g.
 in the following two
examples:
                                                                                                     
\begin{enumerate}
\item
Stationary spherically symmetric configuration where
fluid is radially falling into a pointlike drain at the origin. Since
$u=u_{\perp}$ {\em everywhere}, there will  be no
 distinction between the ergo-sphere and the acoustic horizon.
An astrophysical example of such a situation is the
stationary
spherically symmetric Bondi-type accretion (Bondi 1952)
onto a
Schwarzschild  black hole, or onto other non
rotating compact astrophysical objects in general,
see section 10.2 for further details on spherically symmetric 
astrophysical accretion.
\item
Two-dimensional axisymmetric configuration, where the fluid is
radially
moving towards a drain placed at the origin.  Since only
the radial component of the  velocity is non-zero,
$u=u_{\perp}$ everywhere. Hence, for this system, the acoustic horizon will
coincide with the ergo region.
An astrophysical example is an axially symmetric
accretion with zero angular momentum onto a Schwarzschild black
hole or onto a non-rotating neutron star,
see section 10.3 for further details of axisymmetric accretion.
\end{enumerate}

In subsequent sections, we thus concentrate on transonic black hole
accretion in astrophysics. We will first review various kind of
astrophysical accretion, emphasizing mostly on the black
hole accretion processes. We will then show that sonic points may
form in such accretion and the sonic surface is essentially an
acoustic horizon. We will provide the formalism using which
one can calculate the exact location of the acoustic horizon
(sonic points) $r_h$, the dynamical accretion velocity $u$ and
the acoustic velocity $c_s$ at $r_h$, and the space gradient of
those velocities $(du/dr)$ and $(dc_s/dr)$ at $r_h$, respectively.
Using those quantities, we will then calculate $\kappa$ and $T_{AH}$
for an accreting black hole system. Such calculation will
ensure that accretion processes in astrophysics can be regarded
as a natural example of classical analogue gravity model.
\section{Transonic Black Hole Accretion in Astrophysics}
\subsection{A General Overview}
\noindent
Gravitational capture of surrounding fluid by massive
astrophysical objects is known as accretion. There remains a major difference between black hole
accretion and accretion onto other cosmic objects including neutron stars and
white dwarfs. For celestial bodies other than black holes, infall of matter terminates
either by a direct collision with the hard surface of the accretor or with the outer boundary of the
magneto-sphere, resulting the luminosity (through energy release)
from the surface. Whereas for black hole accretion, matter ultimately
dives through the event horizon from where radiation is prohibited to escape according
to the rule of classical general relativity, and the  emergence of luminosity occurs
{\it on the way} towards the black hole event horizon. The efficiency
of accretion process may be thought as a measure of the
fractional conversion of gravitational binding energy of matter
to the emergent radiation, and is considerably high for black
hole accretion  compared to accretion onto any other
astrophysical objects. Hence accretion onto classical astrophysical black holes
has
been recognized as a fundamental phenomena of increasing
importance in relativistic and high energy astrophysics.
The
extraction of gravitational energy from the black hole accretion is believed to
power the energy generation mechanism of
X-ray  binaries and of the most luminous objects of the
Universe, the Quasars and active galactic nuclei (Frank,
King \& Raine 1992).
%\cite{frank}.
The
black hole accretion is, thus, the most appealing way through which the
all pervading power of gravity is explicitly manifested.

As it is absolutely impossible to provide a detail discussion of a topic
as vast and diverse as accretion onto various astrophysical objects 
in such a small span, this section will mention only a few topic
and will concentrate on fewer still, related mostly to accretion onto black hole.
For details of various aspects of accretion processes onto compact objects,
recent reviews like Pringle 1981; Chakrabarti 1996a; Wiita 1998; Lin \& Papaloizou 1996;
Blandford 1999; Rees 1997; Bisnovayati-Kogan 1998; Abramowicz et al
1998;  and the monographs by 
Frank, King \& Raine 1992, and
Kato, Fukue \& Mineshige 1998,
will be of great help.

Accretion processes onto black holes may be broadly classified into two
different categories. When accreting material does not have any intrinsic
angular momentum, flow is spherically symmetric and any parameters
governing the accretion will be a function of radial distance only. On the
other hand, for matter accreting with considerable intrinsic angular
momentum,
\footnote{It happens when matter falling onto the black holes comes
from the neighbouring stellar companion in the binary, or when the matter
appears as a result of a tidal disruption of stars whose trajectory
approaches sufficiently close to the hole so that self-gravity could
be overcome. The first situation is observed in many galactic X-ray
sources containing a stellar mass black hole and the second one happens in
Quasars and AGNs if the central supermassive hole is
surrounded by a dense stellar cluster.}
 flow geometry is not that trivial. 
In this situation, before the infalling matter plunges through
 the event horizon, accreting fluid will be thrown into circular orbits
 around the hole, moving inward usually when viscous stress in the fluid helps to transport
 away the excess amount of angular momentum. This outward viscous transport
 of angular momentum of the accreting matter leads to the formation
 of accretion disc around the hole. The structure and radiation spectrum
 of these discs depends on various physical parameters governing the flow
 and on specific boundary conditions.

If the instantaneous dynamical velocity and local acoustic velocity
 of the accreting fluid, moving along a space curve parameterized by $r$, are
$u(r)$ and $c_s(r)$, respectively, then the local Mach number $M(r)$ of the
 fluid can be defined as $M(r)={u(r)}/{c_s(r)}$.
The flow will be locally
 subsonic or supersonic according to $M(r) < 1$ or $ >1$, i.e., according to
 $u(r)<c_s(r)$ or $u(r)>c_s(r)$. The flow is transonic if at any moment
 it crosses $M=1$. This happens when a subsonic to supersonic or supersonic to
 subsonic transition takes place either continuously or discontinuously.
The point(s) where such crossing
takes place continuously is (are) called sonic point(s),
 and where such crossing takes place discontinuously are called shocks
 or discontinuities.
At a distance far away from the black hole, accreting material almost 
always remains subsonic (except for the supersonic 
stellar wind fed accretion) since it possesses negligible dynamical 
flow velocity. On the other hand, the flow velocity will approach 
the velocity of light ($c$) while crossing the event horizon, while the maximum 
possible value of sound speed (even for the steepest possible equation 
of state) would be $c/\sqrt{3}$, resulting $M>1$ close to the 
event horizon.
In order to
satisfy such inner boundary condition imposed by the
event horizon, accretion onto black holes
exhibit transonic properties in general.
%\begin{figure}
%\vbox{
%\vskip -1.0cm
%\centerline{
%\psfig{file=figure1.ps,height=11cm,width=11cm,angle=0.0}}
%\psfig{file=f1.ps,height=11cm,width=11cm,angle=0.0}}
%{\bf Figure 1:} Spherically symmetric transonic black hole 
%accretion with acoustic horizon.}
%\end{figure}
%
\begin{center}
\begin{figure}[h]
\includegraphics[scale=0.6,angle=0.0]{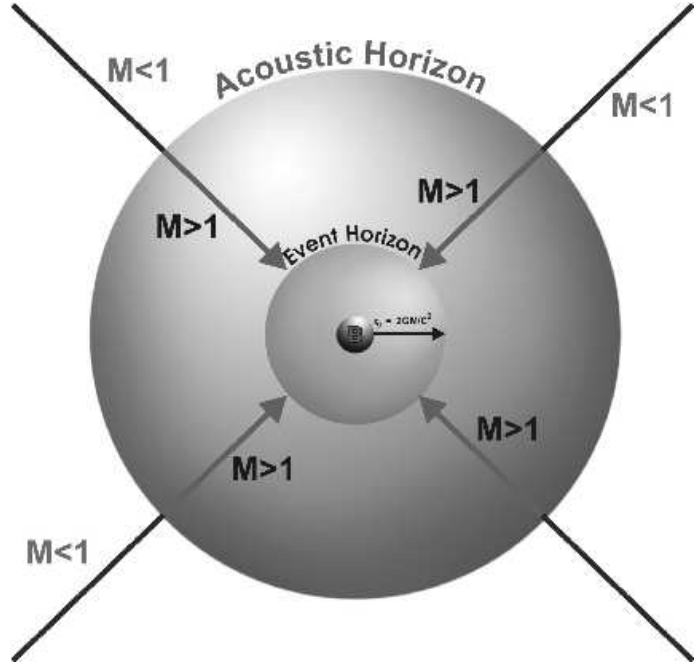}
\caption[]{Spherically symmetric transonic black hole
accretion with acoustic horizon.}
\label{fig1}
\end{figure}
\end{center}
%\vskip 0.25truecm
%\begin{center}
%{\bf Figure 1:} Spherically symmetric transonic black hole 
%accretion with acoustic horizon.\\[0.25cm]
%\end{center}
\subsection{Mono-transonic Spherical Accretion}
\noindent
Investigation of accretion processes onto celestial objects
was initiated by Hoyle \& Lyttleton (1939)
%\cite{hoyle39}
by computing the rate at which
pressure-less matter would be captured by a moving star. Subsequently,
theory of
stationary, spherically symmetric and transonic hydrodynamic accretion of
adiabatic fluid onto a gravitating astrophysical object at rest was
formulated in a seminal paper by Bondi (1952)
%\cite{bondi}
using purely Newtonian potential
and by including the pressure effect of the accreting material.
Later
on, Michel (1972)
%\cite{mitchel}
discussed fully general relativistic polytropic accretion on
to a Schwarzschild black hole by formulating the governing equations for steady
spherical flow of perfect fluid in Schwarzschild metric. Following
Michel's relativistic generalization of Bondi's treatment,
Begelman (1978)
%\cite{begel}
and Moncrief (1980)
%\cite{moncrief}
discussed some aspects of the sonic points of the
flow for such an accretion.
Spherical accretion and wind in general relativity have also been considered
using equations of state other than the polytropic one and
by incorporating various radiative processes (Shapiro 1973, 1973a; 
Blumenthal \& Mathews 1976; Brinkmann 1980).
%(\cite{shapiro73a}\cite{shapiro73b}\cite{blumenthol}\cite{brinkman}).
Malec (1999)
%(\cite{malec})
provided
the solution for general relativistic  spherical accretion with and
without back reaction, and showed that relativistic effects enhance mass
accretion when back reaction  is neglected. The exact values of dynamical
and thermodynamic accretion variables on the sonic surface,
and at extreme close vicinity of the black hole event horizons, have recently
been calculated using complete general relativistic (Das 2002)
%\cite{das2002}
as well as pseudo general relativistic (Das \& Sarkar 2001)
%\cite{dassarkar}
treatments.

Figure 1 pictorially illustrates the generation of the
acoustic horizon for spherical transonic accretion.
Let us assume that an isolated black hole at rest accretes
matter. The black hole (denoted by {\bf B} in the figure)
is assumed to be of Schwarzschild type, and is embedded
by an gravitational event horizon of radius $2GM_{BH}/c^2$. Infalling
matter is assumed not to possess any intrinsic angular momentum, and
hence, falls freely on to the black hole radially. Such an accreting
system possesses spherical symmetry. Far away from the black hole
the dynamical fluid velocity is negligible and hence the matter is
subsonic, which is demonstrated in the figure by {\bf M $<$ 1}.
In course of its motion toward the event horizon, accreting material
acquires sufficiently large dynamical velocity due to the black hole's
strong gravitational attraction. Consequently, at a certain radial
distance, the Mach number becomes unity. The particular value
of $r$, for which {\bf M=1}, is referred as the
transonic point or the sonic point, and is denoted by 
$r_h$, as mentioned in the
above section. For $r<r_h$, matter becomes supersonic and any
acoustic signal created in that region is {\it bound} to be dragged
toward the black hole, and can not escape to the region
$r>r_h$. In other words, any co-moving observer from 
$r{\le}r_h$ can not communicate with another observer at
$r>r_h$ by sending any signal traveling with velocity
$v_{\rm signal}{\le}c_s$.Hence the spherical surface through
$r_h$ is actually an acoustic horizon for stationary
configuration, which is generated when accreting fluid makes a
transition from subsonic ({\bf M $<$ 1}) to the supersonic
({\bf M $>$ 1}) state. In subsequent sections, we will demonstrate
how one can determine the location of $r_h$ and how the
surface gravity and the analogue Hawking temperature
corresponding to such $r_h$ can be computed. Note, however, that for
spherically symmetric accretion, {\it only one} acoustic
horizon may form for a given set of initial boundary configuration
characterizing the stationary configuration. For matter accreting
with non-zero intrinsic angular momentum, {\it multiple}
acoustic horizons can be obtained. Details of such
configurations will be discussed in subsequent sections.
%\begin{figure}
%\vbox{
%\vskip -1.0cm
%\centerline{
%\psfig{file=black-hole.ps,height=11cm,width=11cm,angle=0.0}}}
%psfig{file=redgiant.ps,angle=270.0}}}
%\psfig{file=arrowhead.eps}}}
%\end{figure}

%\section{Shock in Spherical Accretion}
It is perhaps relevant to mention that spherical black hole accretion can 
allow standing shock formation.
Perturbations of various kinds may produce discontinuities in
an astrophysical fluid flow.
By  {\em discontinuity} at a surface in a fluid flow we understand
any discontinuous change of
a dynamical or a thermodynamic quantity across  the
surface. The corresponding surface is called a {\em surface of discontinuity}.
Certain boundary conditions must be satisfied across such surfaces and
according to
these conditions, surfaces of discontinuities are classified into various categories.
The most important such discontinuities
are  {\em shock waves} or {\em shocks}.

While the possibility of the formation of a standing
spherical shock around compact objects was first
conceived long ago (Bisnovatyi-Kogan,
Zel`Dovich, \& Sunyaev 1971), most of the works on
shock formation in spherical accretion share more or
less the same philosophy that one should incorporate
shock formation to increase the efficiency of directed
radial infall in order to explain the high luminosity
of AGNs and QSOs and to model their broad band
spectrum (Jones \& Ellison 1991).
Considerable work has been done in this direction
where several authors have investigated the formation
and dynamics of standing shock in spherical accretion
(M\'esz\'aros \& Ostriker 1983; Protheros \& Kazanas 1983;
Chang \& Osttriker 1985; Kazanas \& Ellision 1986;
Babul, Ostriker \& M\'esz\'aros 1989; Park 1990, 1990a).

Study of spherically symmetric black hole accretion leads to
the discovery of related
interesting problems like
entropic-acoustic or various
other instabilities in spherical accretion (Foglizzo \& Tagger 2000;
Blondin \& Ellison 2001; Lai \& Goldreich 2000; Foglizzo 2001;
Kovalenko \& Eremin 1998),
the realizability and the stability properties of 
Bondi solutions (Ray \& Bhattacharjee 2002),
production of high energy cosmic rays from AGNs
(Protheroe \& Szabo  1992), study of the hadronic model of AGNs
(Blondin  \& Konigl 1987; Contopoulos \& Kazanas 1995),
high energetic emission from relativistic
particles in our galactic centre (Markoff, Melia \& Sarcevic 1999),
explanation of high lithium abundances in the
late-type, low-mass companions of the soft X-ray
transient, (Guessoum \& Kazanas 1999), study of accretion powered
spherical winds emanating from galactic and extra
galactic black hole environments (Das 2001).
\subsection{Breaking the Spherical Symmetry: Accretion Disc}
\subsubsection{A General Overview}
\noindent
In sixties, possible disc-like structures around one of the binary components
were found (Kraft, 1963) and some tentative suggestions that matter should
accrete in the form of discs were put forward
(Pendergest \& Burbidge 1968; Lynden-Bell 1969).
Meanwhile, it was understood that
for spherically symmetric accretion discussed above, the (radial) infall velocity is
very high, hence emission from such a rapidly falling matter was not found to
be strong enough to explain the high luminosity of Quasars and AGNs.
Introducing the idea of magnetic dissipation, efforts were made to improve
the luminosity (Shvartsman 1971, 1971a; Shapiro 1973, 1973a).

Theoretically, accretion discs around black holes were first envisaged
to occur within a binary stellar system where one of the components is
compact object (i.e., white dwarfs, neutron stars or a black hole) and the
secondary would feed matter onto the primary either through an wind
or through Roche lobe overflow. In either case, the accreted matter
would clearly possesses substantial intrinsic angular momentum with
respect to the compact object (a black hole, for our discussion). A flow
with that much angular momentum will have much smaller infall velocity
and much higher density compared to the spherical accretion. The infall
time being higher, viscosity within the fluid, presumably produced
by turbulence or magnetic field, would have time to dissipate angular
momentum (except in regions close to the black holes, since large 
radial velocity close the event horizon leads to the
typical value of dynamical time scale much smaller compared to 
the viscous time scale)
and energy. As matter loses angular
momentum, it sinks deeper into the gravitational potential well and radiate
more efficiently. The flow encircles the compact accretor and forms a
quasi-stationary disc like structure around the compact object and preferably
in the orbital plane of it. Clear evidences for such accretion discs around
white dwarfs in binaries was provided by analysis of Cataclysmic variable
(Robinson 1976).

Accretion forming a Keplarian disc 
\footnote{The
`Keplerian' angular momentum refers  to the value of angular
momentum of a rotating fluid for which the centrifugal force exactly compensates for the
gravitational attraction. If the angular momentum distribution is
sub-Keplerian, accretion flow will possess non-zero advective velocity.}
around a Schwarzschild black hole produces
efficiency $\eta$ (the fraction of 
gravitational energy released) of the 
order of $\eta\sim0.057$ and accretion onto a maximally rotating
Kerr black hole is even more efficient, yielding $\eta\sim0.42$.
However, the actual efficiencies depends on quantities such as viscosity
parameters and the cooling process inside the disc
(see Wiita 1998 and references therein). This energy is released
in the entire electromagnetic spectrum and the success of a disc model depends
on its ability to describe the way this energy is distributed in various
frequency band.

In case of binary systems, where one of the components is a compact
object like white dwarfs, neutron star or a black hole, the companion
is stripped off its matter due to the tidal effects. The stripped off
matter, with angular momentum equal to that of the companion, gradually
falls towards the central compact object
as the angular momentum is removed by viscosity. As the flow possesses a
considerable angular momentum to begin with, it is reasonable to assume
that the disc will form and the viscosity would transport angular momentum
from inner part of the disc radially to the outer part which allows matter
to further fall onto the compact body. This situation could be described
properly by standard thin accretion disc, which may be Keplarian in nature. On the other
hand, in the case of active galaxies and quasars, the situation could be
somewhat different. The  supermassive
($M_{BH} \gsim 10^6 M_{\odot}$) central black hole
is immersed in the
intergalactic matter. In absence of any binary companion, matter is
supplied to the central black hole very intermittently, and the angular
momentum of the accreting matter at the outer edge of the disc may be
sub-Keplarian. This low angular momentum flow departs the disc from
Keplarian in nature and a `thick disc' is more appropriate to
describe the behaviour instead of standard thin, Keplarian Shakura
Sunyaev (Shakura \& Sunyaev 1973) disc.
\subsubsection{Thin Disc Model}
\noindent
In standard thin disc model
(Shakura \& Sunyaev 1973; Novikov \& Thorne 1973), originally conceived to describe Roche lobe
accretion in a binary system, the local height $H(r)$ of the disc  is assumed
to be small enough compared to the local radius of the disc $r$, i.e., the
`thinness'
 condition is dictated by the fact that $H(r) << r$.
Pressure is neglected so that the radial
force balance equations dictates the specific angular momentum distribution to
become Keplarian and the radial velocity is negligible compared
to the azimuthal velocity ($v_r << v_\phi$). Unlike the spherical
accretion, temperature
distribution is far below than virial.
Under the above mentioned set of assumptions, radial
equations of steady state disc structure could be decoupled from the
vertical ones and could be solved independently. The complete solutions
describing the steady state disc structure can be obtained by solving
four relativistic conservation equations, namely; the conservation of rest
mass, specific angular momentum, specific energy and vertical momentum balance
condition. In addition, a viscosity law may be specified which may
transport angular momentum outwards allowing matter to fall in. On the top of
it, in standard thin disc model, the shear is approximated as
proportional to the pressure of the disc with proportionality constant
$\alpha$, $\alpha$ being the viscosity parameter having 
numerical value less than unity.

High uncertainty remains in investigating the exact nature of the viscosity
inside a thin accretion disc (see Wiita 1998 and references therein).
One of the major problems is to explain the
origin of sufficiently large viscosity that seems to be present inside
accretion discs in the binary system. Unfortunately, under nearly all
astrophysically relevant circumstances, all of the well understood
microscopic transverse momentum transport mechanism such as ionic, molecular
and radiative viscosity are extremely small. Observations with direct relevance
to the nature and strength of the macroscopic viscosity mechanism are very
difficult to make; the only fairly direct observational evidence for the strength
of disc viscosity comes from the dwarf novae system. For a black hole as compact
accretor, such observational evidences is far from reality till date. Therefore
advances in understanding the disc  viscosity is largely based on theoretical
analysis and numerical techniques. Usually accepted view is that the
viscosity may be due to magnetic transport of angular momentum or due to
small scale turbulent dissipation. Over the past several years an
explanation of viscosity in terms of Velikhov-Chandrasekhar-Balbus-Hawley
instability (linear magnetic instability) has been investigated;
see, e.g., Balbus \& Hawle 1998 for further details.
\subsubsection{Thick Disc Model}
\noindent
The assumptions implying accretion discs are always thin can break down in the
innermost region. Careful consideration of the effects of general
relativity show that the flow must go supersonically through a cusp.
For considerably high accretion rate,
 radiation emitted
by the in-falling matter exerts a significant pressure on the gas. The radiation
pressure inflates the disc, and make it geometrically thick
($H(r) \sim r$, at least for the inner $10-100 r_g$), which is often
otherwise known as `accretion torus'.
This considerable amount of radiation
pressure must be incorporated to find the dynamical structure of the disc
and in determining the thermodynamical quantities inside the disc.
Incorporation of the radiation pressure term in Euler equation dictates
the angular momentum deviation  from that of the Keplarian. The angular momentum
distribution becomes super (sub) Keplarian if the pressure gradient is positive
(negative).

Introducing a post-Newtonian
(these `pseudo' potentials are widely used to 
mimic the space time around the Schwarzschild or the Kerr metric very nicely,
see section 14 for details)
$\Phi = - \frac {GM_{BH}}{(r - 2r_g)}$
in lieu of the usual $\Phi_{Newtonian} = - \frac{GM_{BH}}{r}$
(where $r_g$ is the `gravitational' radius),
Paczy\'nski and Wiita
(1980)
provided the first thick disc model which joins with the standard thin disc
at large radius without any discontinuity. They pointed out several
important features of these configuration.
It has been shown that the structure
of thick disc in inner region is nearly independent of the viscosity
and efficiency of accretion drops dramatically. More sophisticated model
of radiation supported thick disc including self-gravity of the disc
with full general relativistic treatment was introduced later
(Wiita 1982; Lanza 1992).
\subsubsection{Further Developments}
\noindent
Despite having a couple of interesting features,
standard thick accretion disc model
suffers from some limitations for which its study fell from favour
in the late '80s. Firstly, the strong anisotropic nature of the emission
properties of the disc has been a major disadvantage. Secondly, a non-accreting
thick disc is found to be dynamically and globally unstable to
non-axisymmetric perturbations. However, an ideal `classical thick
disc', if modified to incorporate high accretion rates involving both
low angular momentum and considerable radial infall velocity self-consistently,
may remain viable. 
Also, it had been 
realized that neither the Bondi (1952) flow nor the
standard thin disc model could individually fit the bill
completely. Accretion disc theorists were
convinced about the necessity of having an intermediate model which could
bridge the gap between purely spherical flow (Bondi type) and purely rotating flow
(standard thin disc).
Such modification could be accomplished by 
incorporating
a self-consistent `advection' term which could take care of finite radial
velocity
of accreting material (for the black hole candidates which may gradually
approaches the velocity of light to satisfy the inner boundary condition
on event horizon) along with its rotational velocity and generalized
heating and cooling terms (Hoshi \& Shibazaki 1977;
Liang \& Thompson 1980; Ichimaru 1977;
Paczy\'nski \& Bisnobhatyi-Kogan 1981; Abramowicz \& Zurek 1981;
Muchotrzeb \& Paczy\'nski 1982; Muchotrzeb 1983;
Fukue 1987; Abramowicz et al. 1988; Narayan \& Yi 1994;
Chakrabarti 1989, 1996).

%These goals could be achieved by self-consistently
%introducing the shock waves and a sub-Keplarian inner disc in an
%initially thin disc. Inside the shock wave, flow is likely to be hotter
%and to puff up and resembles the standard thick disc model. This is the
%subject of the `advective accretion disc model'.

\subsection{Multi-transonic Accretion Disc}
%As discussed in the previous section,
%for accretion with non-zero angular momentum density, spherical
%symmetry is broken and the disc accretion phenomena is studied employing
%axisymmetric configuration.
%Accreting matter is thrown into circular orbits around the central accretor,
%leading to the formation of the accretion discs. 
For certain values of the intrinsic angular
momentum density of accreting material, the number of sonic point, unlike spherical
accretion, may {\it exceed} one, and accretion is called `multi-transonic'. Study of
such multi-transonicity was initiated by Abramowicz \& Zurek (1981).
%\cite{zurek}.
Subsequently, multi-transonic accretion disc has been
studied in a number of works (Fukue 1987; Chakrabarti 1990, 1996;
Kafatos \& Yang 1994;
Yang \& Kafatos 1995; 
Pariev 1996; Peitz \& Appl 1997; Lasota \& Abramowicz 1997; 
Lu, Yu, Yuan \& Young 1997; Das 2004; Barai, Das \& Wiita 2004;
Abraham, Bili\'c \& Das 2006; Das, Bili\'c \& Dasgupta 2006).
%(\cite{fukue}\cite{chak90}\cite{kafatosyang}\cite{yangkafatos}
%\cite{pariev}\cite{peitz}\cite{lu}\cite{das2004}\cite{barai}).
All the above works, except Barai, Das \& Wiita 2004,
%\cite{barai},
usually deal with low angular
momentum sub-Keplerian {\it inviscid} flow around a Schwarzschild 
black hole or a prograde flow around a Kerr black hole. 
Barai, Das \& Wiita 2004 
%, Das \& Wiita \cite{barai}
studied the retrograde
flows as well and showed that a higher angular momentum (as high as
Keplerian) retrograde flow can also produce multi-transonicity.
Sub-Keplerian
weakly rotating flows
are exhibited in
various physical situations, such as detached binary systems
fed by accretion from OB stellar winds (Illarionov \&
Sunyaev 1975; Liang \& Nolan 1984),
%\cite{ilario}\cite{liang},
semi-detached low-mass non-magnetic binaries (Bisikalo et al.\ 1998),
%\cite{bisikalo},
and super-massive black holes fed
by accretion from slowly rotating central stellar clusters
(Illarionov 1988; Ho 1999
%(\cite{ilario1}\cite{ho}
and references therein). Even for a standard Keplerian
accretion disc, turbulence may produce such low angular momentum flow
(see, e.g., Igumenshchev
\& Abramowicz 1999,
%\cite{igu},
and references therein). 
\subsection{Non-axisymmetric Accretion Disc}
All the above mentioned works deals with `axisymmetric' accretion,
for which the orbital angular momentum of the entire disc plane
remains aligned with the spin angular momentum of the compact object
of our consideration. In a strongly coupled binary system (with
a compact object as one of the components), accretion may experience
a non-axisymmetric potential because the secondary donor star may
exert non-axisymmetric tidal force on the accretion disc around the
compact primary. In general, non-axisymmetric tilted disc may form if
the accretion takes place out of the symmetry plane of the spinning
compact object. Matter in such misaligned disc will experience a
torque due to the general relativistic Lense-Thirring effect
(Lense \& Thirring 1918), leading to the precession of the inner disc
plane. The differential precession with radius may cause stress and
dissipative effects in the disc. If the torque remains strong enough
compared to the internal viscous force, the inner region of the initially
tilted disc may be forced to realigned itself with the spin angular
momentum (symmetry plane) of the central accretor. This phenomena of
partial re-alignment (out to a certain radial distance known as the
`transition radius' or the `alignment radius') of the initially
non-axisymmetric disc is known as the `Bardeen-Petterson effect'
(Bardeen \& Petterson 1975). Such a transition radius can be obtained
by balancing the precession and the inward drift or the viscous time
scale.
                                                                                
Astrophysical accretion disc subjected to the Bardeen-Petterson effect
becomes `twisted' or `warped'. A large scale warp (twist) in the disc
may modify the emergent spectrum and can influence the direction of the
Quasar and micro-quasar jets emanating out from the inner region of
the accretion disc (see, e.g., Maccarone 2002; Lu \& Zhou 2005,
and references therein).
                                                                                
Such a twisted disc may be thought as an ensemble of annuli of increasing
radii, for which the variation of the direction of the orbital angular
momentum occurs smoothly while crossing the alignment radius. System
of equations describing such twisted disc have been formulated by several
authors (see, e.g., Peterson 1977; Kumar 1988; Demianski \& Ivanov 1997;
and references therein), and the time scale required for a Kerr black hole
to align its spin angular momentum with that of the initially
misaligned accretion disc, has also been estimated (Scheuer \&
Feiler 1996). Numerical simulation using three dimensional Newtonian
Smooth Particle Hydrodynamics
(SPH) code (Nelson \& Papaloizou 2000) as well as using fully general
relativistic framework (Fragile \& Anninos 2005) reveal the geometric
structure of such discs.

We would, however, not like to explore the non-axisymmetric accretion further
in this 
review. One of the main reasons for which is, as long as the acoustic 
horizon forms at a radial
length scale smaller than that of the alignment radius (typically
100 $r_g$ - 1000 $r_g$, according to the original estimation of
Bardeen \& Petterson 1975), one need not implement the non-axisymmetric
geometry to study the analogue effects.
\subsection{Angular Momentum Supported Shock in Multi-transonic Accretion Disc}
In an adiabatic flow of the Newtonian fluid, the shocks obey the following
conditions  (Landau \& Lifshitz 1959)
\begin{equation}
\left[\left[{\rho}u\right]\right]=0,~
\left[\left[p+{\rho}u^2\right]\right]=0,~
\left[\left[\frac{u^2}{2}+h\right]\right]=0,
\label{eq46}
\end{equation}
where $[[f]]$ denotes the discontinuity of $f$ across the surface of discontinuity, i.e.
\begin{equation}
\left[\left[f\right]\right]=f_2 -f_1,
\label{eq47}
\end{equation}
with $f_2$ and $f_1$ being the boundary values
of the quantity $f$ on the two sides of
the surface.
 Such shock waves
are quite often generated in
various kinds of supersonic astrophysical flows having
intrinsic angular momentum, resulting
in a flow which
becomes subsonic. This is because the repulsive centrifugal potential barrier
experienced by such flows is sufficiently strong to brake the infalling
motion and a stationary solution
could be introduced only through a shock. Rotating, transonic astrophysical fluid
flows are thus believed to be `prone' to the shock formation phenomena.
                                                                                                     
One also
expects that a shock formation in black-hole accretion discs
might be a general phenomenon because shock waves
in rotating astrophysical flows potentially
provide an important and efficient mechanism
for conversion of a significant amount of the
gravitational energy  into
radiation by randomizing the directed infall motion of
the accreting fluid. Hence, the shocks  play an
important role in governing the overall dynamical and
radiative processes taking place in astrophysical fluids and
plasma accreting
onto black holes.
The study of steady, standing, stationary shock waves produced in black
hole accretion has acquired an important status, 
and a 
number of works studied the 
shock 
formation in black hole accretion discs
(Fukue 1983; Hawley, Wilson \& Smarr 1984; Ferrari et al.
1985; Sawada, Matsuda \& Hachisu 1986; Spruit 1987;
Chakrabarti 1989; Abramowicz \& Chakrabarti 1990;
%Chakrabarti \& Molteni 1993; 
Yang \& Kafatos 1995;
Chakrabarti 1996a;
Lu, Yu, Yuan \& Young 1997;
Caditz \& Tsuruta 1998; T\'oth, Keppens
\& Botchev 1998;
Das 2002; 
Takahashi, Rillet, Fukumura \& Tsuruta 2002;
Das, Pendharkar \& Mitra 2003; 
Das 2004; Chakrabarti \& Das 2004;
Fukumura \& Tsuruta 2004;
Abraham, Bili\'c \& Das 2006;
Das, Bili\'c \& Dasgupta 2006)
For more  details
and for a more exhaustive list of references
see, e.g., Chakrabarti 1996c and Das 2002.

Generally,
the issue of the formation of steady, standing shock waves in black-hole accretion discs is
addressed in
two different ways.
First, one can study the formation of Rankine-Hugoniot shock waves in a
polytropic flow. Radiative cooling in this type of shock is quite inefficient. No energy is
dissipated at the shock and the total specific energy of the accreting material is a shock-conserved
quantity. Entropy is generated at the shock and the post-shock flow possesses
 a higher entropy accretion rate
than its pre-shock counterpart. The flow changes its temperature permanently at the shock. Higher
post-shock temperature puffs up the post-shock flow and a quasi-spherical,
 quasi-toroidal centrifugal
pressure supported region is formed in the inner region of the accretion disc
(see Das 2002, and references therein for further detail) which 
locally mimics a thick accretion flow.

Another class of the shock studies concentrates on
the shock formation in isothermal black-hole accretion
discs. The characteristic features of such shocks are quite different from the
non-dissipative shocks discussed
above. In isothermal shocks, the
accretion flow  dissipates a part of its
energy and entropy at
the shock surface to keep the post-shock temperature equal to its pre-shock value.
This maintains the vertical
thickness of the flow exactly the
same just before and just after the shock is formed. Simultaneous jumps in
energy and entropy join the pre-shock supersonic flow to its post-shock
subsonic counterpart.
For detailed
discussion
and references
see, e.g., Das, Pendharkar \& Mitra 2003, and Fukumura \& Tsuruta 2004.

In section 13.5, we will construct and solve the equations governing the general 
relativistic Rankine-Hugoniot shock. The shocked accretion flow in 
general relativity and in post-Newtonian pseudo-Schwarzschild potentials
will be discussed in the section 13.5 - 13.8 and 16.2 respectively.
\section{Motivation to Study the Analogue Behaviour of Transonic Black Hole Accretion}
\noindent
Since the publication of the seminal paper by Bondi in 1952 (Bondi 1952),
%\cite{bondi},
the transonic
behaviour of accreting fluid onto compact astrophysical objects has
been extensively studied in the astrophysics community, 
and the
pioneering work by Unruh in 1981 (Unruh 1981),
initiated a substantial number of works
in the theory of {\AHR}  with diverse fields of application stated in section 4 - 5.
It is surprising that no attempt was made to bridge
these two categories of research, astrophysical black hole accretion
and the theory of {\AHR},  by providing a self-consistent study of {\AHR}  for real
astrophysical fluid flows, i.e., by establishing the fact that
accreting black holes can be considered as a natural
example of analogue system. Since both the theory of transonic
astrophysical accretion and the theory of
{\AHR}  stem from almost
exactly the same physics, the propagation of a transonic fluid with
acoustic disturbances embedded into it, it is important to study
{\AHR} for transonic accretion onto astrophysical black
holes and to compute $T_{{AH}}$ for such accretion. 

In the following sections, we will describe the details of 
the transonic accretion and will show how the accreting black 
hole system can be considered as a classical analogue system.
We will first discuss general relativistic accretion of spherically
symmetric (mono-transonic Bondi (1952) type accretion) and axisymmetric (multi-transonic disc 
accretion) flow. We will then introduce a number of post-Newtonian 
pseudo-Schwarzschild black hole potential, and will discuss black hole accretion 
under the influence of such modified potentials.
\section{General Relativistic Spherical Accretion as an Analogue Gravity Model}
\noindent
In this section, we will demonstrate how one can construct and solve the 
equations governing the general relativistic, spherically symmetric, steady 
state accretion flow onto a Schwarzschild black hole. This 
section is largely based on Das 2004a.

Accretion flow described in this section 
is $\theta$ and $\phi$ symmetric and possesses only radial inflow velocity. 
In this section, we use the gravitational radius $r_g$ as
$r_g={2G{M_{BH}}}/{c^2}$.
The radial distances and velocities are scaled in units of $r_g$ and $c$
respectively and all other derived quantities are scaled accordingly;
$G=c=M_{BH}=1$ is used.
Accretion is governed by the radial part
of the general relativistic 
time independent Euler and continuity equations in Schwarzschild
metric. We will consider the stationary solutions.
We assume the dynamical in-fall time scale to be short compared with any
dissipation time scale during the accretion process.
\subsection{The Governing Equations}
\noindent
To describe the fluid,
 we use a 
 polytropic equation of state (this is common in 
the theory of relativistic black hole accretion) of the form
\begin{equation}
p=K{\rho}^\gamma ,
\label{eq92}
\end{equation}
where the polytropic index $\gamma$ (equal to the ratio of the two specific
heats $c_p$ and $c_v$) of the accreting material is assumed to be constant throughout the fluid.
A more realistic model of the flow 
would perhaps require  a variable polytropic index having a 
functional dependence on the radial
distance, i.e. $\gamma{\equiv}\gamma(r)$. However, we  have performed the
calculations for a sufficiently large range of $\gamma$ and we believe
that all astrophysically relevant
polytropic indices are covered in our analysis.

The constant $K$ in (\ref{eq92}) may be 
related to the specific entropy of the fluid,
 provided there is no entropy 
generation during the flow. 
If in addition to (\ref{eq92}) the
Clapeyron equation for an ideal gas 
holds
\begin{equation}
p=\frac{\kappa_B}{{\mu}m_p}{\rho}T\, ,
\label{eq93}
\end{equation}
 where $T$ is the locally measured temperature, $\mu$  the mean molecular weight,
$m_H{\sim}m_p$  the mass of the hydrogen atom, then the specific entropy, i.e. the entropy 
per particle, is given by (Landau \& Lifshitz 1959):
\begin{equation}
\sigma=\frac{1}{\gamma -1}\log K+
\frac{\gamma}{\gamma-1}+{\rm constant} ,
\label{eq94}
\end{equation}
where the constant depends on the chemical composition of the 
accreting material. 
Equation (\ref{eq94}) confirms that $K$ in (\ref{eq92})
is  a measure of the specific entropy of the accreting matter.

The specific enthalpy of the accreting matter can now be defined as
\begin{equation}
h=\frac{\left(p+\epsilon\right)}{\rho}\, ,
\label{eq95}
\end{equation}
where the energy density $\epsilon$ includes the rest-mass density and the internal 
energy and may be written as
\begin{equation}
\epsilon=\rho+\frac{p}{\gamma-1}\, .
\label{eq96}
\end{equation}
The adiabatic speed of sound is defined by
\begin{equation}
c_s^2=\frac{{\partial}p}{{\partial}{\epsilon}}{\Bigg{\vert}}_{\rm 
constant~entropy}\, .
\label{eq97}
\end{equation}
From (\ref{eq96}) we obtain
\begin{equation}
\frac{\partial{\rho}}{\partial{\epsilon}}
=\left(
\frac{\gamma-1-c_s^2}{\gamma-1}\right) .
\label{eq98}
\end{equation}
Combination of (\ref{eq97}) and (\ref{eq92}) gives
\begin{equation}
c_s^2=K{\rho}^{\gamma-1}{\gamma}\frac{\partial{\rho}}{\partial{\epsilon}}\, ,
\label{eq99}
\end{equation}
Using the above relations, one obtains the expression for the specific enthalpy 
\begin{equation}
h=\frac{\gamma-1}{\gamma-1-c_s^2}\, .
\label{eq100}
\end{equation}
The rest-mass density $\rho$, the pressure $p$, the temperature $T$
of the flow and the energy density $\epsilon$ 
may be expressed in terms of the speed of sound  $c_s$ as
\begin{equation}
\rho=K^{-\frac{1}{\gamma-1}}
\left(\frac{\gamma-1}{\gamma}\right)^{\frac{1}{\gamma-1}}
\left(\frac{c_s^2}{\gamma-1-c_s^2}\right)^{\frac{1}{\gamma-1}},
\label{eq101}
\end{equation}
\begin{equation}
p=K^{-\frac{1}{\gamma-1}}
\left(\frac{\gamma-1}{\gamma}\right)^{\frac{\gamma}{\gamma-1}}
\left(\frac{c_s^2}{\gamma-1-c_s^2}\right)^{\frac{\gamma}{\gamma-1}},
\label{eq102}
\end{equation}
\begin{equation}
T=\frac{\kappa_B}{{\mu}m_p}
\left(\frac{\gamma-1}{\gamma}\right)
\left(\frac{c_s^2}{\gamma-1-c_s^2}\right),
\label{eq103}
\end{equation}
\begin{equation}
\epsilon=
K^{-\frac{1}{\gamma-1}}
\left(\frac{\gamma-1}{\gamma}\right)^{\frac{1}{\gamma-1}}
\left(\frac{c_s^2}{\gamma-1-c_s^2}\right)^{\frac{1}{\gamma-1}}
\left[
1+\frac{1}{\gamma}
\left(
\frac{c_s^2}{\gamma-1-c_s^2}
\right)
\right].
\label{eq104}
\end{equation}

The conserved specific flow energy ${\cal E}$ (the relativistic 
analogue of Bernoulli's constant) along each stream line reads ${\cal E}=hu_t$, 
(Anderson 1989)
%(\cite{anderson}) 
where
$h$ and $u_\mu$ are the specific enthalpy and the four velocity, which can be 
re-cast in terms of the radial three velocity $u$ and the polytropic sound speed 
$c_s$ to obtain:
\begin{equation}
{\cal E}=\left[\frac{\gamma-1}{\gamma-\left(1+c^2_s\right)}\right]
\sqrt{\frac{1-1/r}{1-u^2}}
\label{eq71}
\end{equation}
One concentrates on positive Bernoulli constant solutions.
The mass accretion rate ${\dot M}$ may be obtained by integrating the continuity
equation:
\begin{equation}
{\dot M}=4{\pi}{\rho}ur^2\sqrt{\frac{r-1}{r\left(1-u^2\right)}}
\label{eq72}
\end{equation}
where $\rho$ is the proper mass density.  

We define the `entropy accretion rate'
${\dot {\Xi}}$
as a quasi-constant
multiple of the mass accretion rate in the following way:
\begin{equation}
{\dot {\Xi}}=K^{\displaystyle{\frac{1}{1-\gamma}}}{\dot M}=4{\pi}{\rho}ur^2\sqrt{\frac{r-1}{r\left(1-u^2\right)}}
\left[\frac{c^2_s\left(\gamma-1\right)}{\gamma-\left(1+c^2_s\right)}\right]
\label{eq73}
\end{equation}
Note that, in the absence of creation or annihilation of matter,
the mass accretion rate is a universal constant of motion,
whereas the entropy accretion
rate is not. As the expression for ${\dot {\Xi}}$ contains the quantity
$K\equiv p/\rho^\gamma$, which measures  the
specific entropy of the flow, the entropy rate ${\dot {\Xi}}$ remains constant
throughout the flow {\it only if} the entropy per particle
remains locally unchanged.
This latter condition may be violated if the accretion is
accompanied by a shock.
 Thus, ${\dot {\Xi}}$ is a
constant of motion for shock-free polytropic accretion and
becomes discontinuous (increases) at the shock location,
if a shock forms in the accretion.
One can solve the two conservation equations for ${\cal E}$ and
${\dot {\Xi}}$ to obtain the complete accretion profile.

\subsection{Transonicity}
Simultaneous solution of (\ref{eq71}-\ref{eq73})
 provides the dynamical three velocity gradient 
at any radial distance $r$:
\begin{equation}
\frac{du}{dr}=\frac{u\left(1-u^2\right)\left[c^2_s\left(4r-3\right)-1\right]}
{2r\left(r-1\right)\left(u^2-c^2_s\right)}=\frac{{\cal N}\left(r,u,c_s\right)}
{{\cal D}\left(r,u,c_s\right)}
\label{eq74}
\end{equation}
A real physical transonic flow must be smooth everywhere, except 
possibly at a shock. Hence, if the denominator ${{\cal D}\left(r,u,c_s\right)}$
of (\ref{eq74}) vanishes at a point, the numerator
${{\cal N}\left(r,u,c_s\right)}$ must also vanish at that point 
to ensure the physical continuity of the flow. Borrowing the terminology
from the dynamical systems theory (see, e.g., Jordan \& Smith 2005), one therefore arrives at the 
{\it critical point} conditions by making ${{\cal D}\left(r,u,c_s\right)}$
and ${{\cal N}\left(r,u,c_s\right)}$ of (\ref{eq74}) simultaneously equal 
to zero. We thus obtain the critical point conditions as:
\begin{equation}
u{\vc}=c_s{\vc}=\sqrt{\frac{1}{4r_c-3}},
\label{eq75}
\end{equation}
$r_c$ being the location of the critical point or the so called
`fixed point' of the differential equation (\ref{eq74}).

From (\ref{eq75}), one easily obtains that $M_c$, the Mach number at the 
critical point, is {\it exactly} equal to {\it unity}. This ensures that
the critical points {\it are} actually the sonic points, and thus, $r_c$ is
actually the location of the acoustic event horizon. In this section, hereafter, 
we will thus use $r_h$ in place of $r_c$. Note, however, that the equivalence 
of the critical point with the sonic point (and thus with the acoustic horizon)
is {\it not} a generic feature. Such an equivalence strongly depends on 
the flow geometry and the equation of state used. For spherically symmetric 
accretion (using any equation of state), or polytropic disc accretion where the expression 
for the
disc height is taken to be constant (Abraham, Bili\'c \& Das 2006), or 
isothermal disc accretion with variable disc height, 
such an equivalence holds good. For all other kind of disc accretion,
critical points and the sonic points are {\it not} equivalent, and the 
acoustic horizon forms at the sonic points and not at the critical point.
We will get back to this issue in greater detail in section 13.3.

Substitution of $u{\vh}$ and $c_s{\vh}$ into (\ref{eq71}) for $r=r_h$ provides:
\begin{equation}
r_h^3+r_h^2\Gamma_1+r_h\Gamma_2+\Gamma_3=0
\label{eq76}
\end{equation}
where
\begin{eqnarray}
\Gamma_1=\left[\frac{2{\cal E}^2\left(2-3\gamma\right)+9\left(\gamma-1\right)}
         {4\left(\gamma-1\right)\left({\cal E}^2-1\right)}\right], && \nonumber \\
\Gamma_2=\left[\frac{{\cal E}^2\left(3\gamma-2\right)^2-
          27\left(\gamma-1\right)^2}
          {32\left({\cal E}^2-1\right)\left(\gamma-1\right)^2}\right],~
\Gamma_3=\frac{27}{64\left({\cal E}^2-1\right)}.
\label{eq77}
\end{eqnarray}
Solution of (\ref{eq76}) provides the location of the {\AH} in terms of only two accretion parameters
$\{{\cal E},\gamma\}$, which is the two parameter input set to study the flow.

We now
set the appropriate limits on \egam  to model the realistic situations
encountered in astrophysics. As ${\cal E}$ is scaled in terms of
the rest mass energy and includes the rest mass energy,
${\cal E}<1$ corresponds to the negative energy accretion state where
radiative extraction of rest mass energy from the fluid is required. For such extraction
to be made possible, the accreting fluid has to
possess viscosity or other dissipative mechanisms, which may violate the Lorentzian invariance.
On the other hand, although almost any ${\cal E}>1$ is mathematically allowed, large
values of ${\cal E}$ represents flows starting from infinity
with extremely high thermal energy (see section 13.4 for 
further detail), and ${\cal E}>2$ accretion represents enormously
hot flow configurations at very large distance from the black hole,
which are not properly conceivable in realistic astrophysical situations.
Hence one sets $1{\lsim}{\cal E}{\lsim}2$. Now, $\gamma=1$ corresponds to isothermal accretion
where accreting fluid remains optically thin. This is the physical lower limit for
$\gamma$, and $\gamma<1$ is not realistic in accretion
astrophysics. On the other hand,
$\gamma>2$ is possible only for superdense matter
with substantially large magnetic
field (which requires the accreting material to be governed by general relativistic
magneto-hydrodynamic
equations, dealing with which
is beyond the scope of this article) and direction dependent anisotropic pressure. One thus
sets $1{\lsim}\gamma{\lsim}2$ as well, so \egam has the boundaries
$1{\lsim}\{{\cal E},\gamma\}{\lsim}2$. However, one should note that the most preferred
values of $\gamma$ for realistic black hole accretion ranges from $4/3$
to $5/3$
(Frank, King \& Raine 1992).

For any specific value of
$\{{\cal E},\gamma\}$,
(\ref{eq76}) can be solved {\it completely analytically} 
by employing the Cardano-Tartaglia-del Ferro technique. One defines:
\begin{eqnarray}
\Sigma_1=\frac{3\Gamma_2-\Gamma_1^2}{9},~
\Sigma_2=\frac{9\Gamma_1\Gamma_2-27\Gamma_3-2\Gamma_1^3}{54},~
\Psi=\Sigma_1^3+\Sigma_2^2,~
 \Theta={\rm cos}^{-1}\left(\frac{\Sigma_2}{\sqrt{-\Sigma_1^3}}\right) && \nonumber \\
% \Theta={\rm cos}^{-1}\left(\frac{\Sigma_2}{\sqrt{-\Sigma_1^3}}\right)
\Omega_1=\sqrt[3]{\Sigma_2+\sqrt{\Sigma_2^2+\Sigma_1^3}},~
\Omega_2=\sqrt[3]{\Sigma_2-\sqrt{\Sigma_2^2+\Sigma_1^3}},~
\Omega_{\pm}=\left(\Omega_1\pm\Omega_2\right)
\label{eq78}
\end{eqnarray}
so that the three roots for $r_h$ come out to be:
\begin{equation}
^1\!r_h=-\frac{\Gamma_1}{3}+\Omega_+, \quad \quad
^2\!r_h=-\frac{\Gamma_1}{3}-\frac{1}{2}\left(\Omega_+-i\sqrt{3}\Omega_-\right), \quad \quad
^3\!r_h=-\frac{\Gamma_1}{3}-\frac{1}{2}\left(\Omega_--i\sqrt{3}\Omega_-\right)
\label{eq79}
\end{equation}
However, note that not all $^i\!r_h\{i=1,2,3\}$ would be real for all \egam. It is
easy to show that if $\Psi>0$, only one root is real; if $\Psi=0$, all roots are
real and at least two of them are identical; and if $\Psi<0$, all roots are real 
and distinct. 
Selection of the real physical ($r_h$ has to be greater than unity) roots
requires a close look at the
solution for $r_h$ for 
the astrophysically relevant range 
of \egam.
One finds that for the preferred range of \egam,
one {\it always} obtains $\Psi<0$. Hence the roots are always real and three real
unequal roots can be computed as:
\begin{eqnarray}
^1\!{{r}}_h=2\sqrt{-\Sigma_1}{\rm cos}\left(\frac{\Theta}{3}\right)
                  -\frac{\Gamma_1}{3},~
^2\!{{r}}_h=2\sqrt{-\Sigma_1}{\rm cos}\left(\frac{\Theta+2\pi}{3}\right)
                  -\frac{\Gamma_1}{3}, && \nonumber \\
^3\!{{r}}_h=2\sqrt{-\Sigma_1}{\rm cos}\left(\frac{\Theta+4\pi}{3}\right)
                  -\frac{\Gamma_1}{3}
\label{eq80}
\end{eqnarray}
One finds that for {\it all} $1{\lsim}${\egam}${\lsim}2$, $^2\!{{r}}_h$ becomes negative. 
It is observed 
that $\{^1\!{{r}}_h,^3\!{{r}}_h\}>1$ for most values of the astrophysically 
tuned \egam.
However, it is also found that $^3\!{{r}}_h$ does not allow steady physical flows to pass
through it; either $u$, or $a_s$, or both, becomes superluminal before the flow reaches
the actual event horizon, or the Mach number profile shows intrinsic fluctuations for 
$r<r_h$. This information is obtained by numerically integrating the 
complete flow profile passing through $^3\!{{r}}_h$. Hence it turns out that one needs to
concentrate {\it only} on  $^1\!{{r}}_h$ for realistic astrophysical black hole accretion. 
Both large ${\cal E}$ and large $\gamma$ enhance the thermal energy of the flow 
so that the 
accreting fluid acquires the large radial velocity to overcome $a_s$ only in the 
close vicinity of the black hole . Hence $r_h$ anti-correlates with \egam. 

The critical properties and stability of such acoustic horizons has 
recently been studied using a dynamical systems approach, see Mandal,
Ray \& Das 2007 for further details.

To obtain
$(du/dr)$ and $(dc_s/dr)$ on the {\AH}, L' Hospital's rule is applied to 
(\ref{eq74}) to have:
\begin{equation}
\left(\frac{du}{dr}\right)_{r=r_h}=\Phi_{12}-\Phi_{123},~
\left(\frac{dc_s}{dr}\right)_{r=r_h}=\Phi_4\left(\frac{1}{\sqrt{4r_h-3}}+\frac{\Phi_{12}}{2}
                            -\frac{\Phi_{123}}{2}\right)
\label{eq81}
\end{equation}
where
\begin{eqnarray}
\Phi_{12}=-\Phi_2/2\Phi_1,~
\Phi_{123}=\sqrt{\Phi_2^2-4\Phi_1\Phi_3}/2\Phi_1,~
\Phi_1=\frac{6r_h\left(r_h-1\right)}{\sqrt{4r_h-3}}, ~ && \nonumber \\ 
\Phi_2=\frac{2}{4r_h-3}\left[4r_h\left(\gamma-1\right)-
        \left(3\gamma-2\right)\right],~ && \nonumber \\ 
\Phi_3=\frac{8\left(r_h-1\right)}{\left(4r_h-3\right)^{\frac{5}{2}}}
       \left[r_h^2\left(\gamma-1\right)^2-r_h\left(10\gamma^2-19\gamma+9\right)
       +\left(6\gamma^2-11\gamma+3\right)\right],~ && \nonumber \\
\Phi_4=\frac{2\left(2r_h-1\right)-\gamma\left(4r_h-3\right)}
       {4\left(r_h-1\right)}
\label{eq82}
\end{eqnarray}
\subsection{Analogue Temperature}
\noindent
For spherically symmetric general relativistic flow onto Schwarzschild black holes,
one can evaluate the exact value of the Killing fields and
Killing vectors to calculate the surface gravity for that geometry. The
analogue Hawking temperature for such geometry comes out to be (Das 2004a)
%(\cite{das2004a}):
\begin{equation}
T_{{AH}}=\frac{\hbar{c^3}}{4{\pi}{\kappa_B}GM_{BH}}
    \left[\frac{r_h^{1/2}\left(r_h-0.75\right)}
    {\left(r_h-1\right)^{3/2}}\right]
    \left|\frac{d}{dr}\left(c_s-u\right)\right|_{r=r_h},
\label{eq83}
\end{equation}
where the values of $r_h, (du/dr)_h$ and $(dc_s/dr)_h$ are obtained using the system of
units and scaling used in this article.

It is evident from (\ref{eq83}) that the {\it exact} value of $T_{AH}$ can
be {\it analytically} calculated from the results obtained in the
previous section. While (\ref{eq79})
provides the location of the acoustic horizon ($r_h$), the value of
$\left|\frac{d}{dr}\left(c-u\right)\right|_{r=r_h}$ is obtained 
from (\ref{eq81}-\ref{eq82})
as a function of ${\cal E}$ and $\gamma$, both of which are real, physical,
measurable quantities.
Note again, that, since $r_h$ and other quantities
appearing in (\ref{eq83}) are {\it analytically} calculated as a function of
\egam, (\ref{eq83}) provides an {\it exact analytical value} of the
general relativistic analogue Hawking temperature for {\it all possible solutions} of an
spherically accreting
astrophysical black hole system, something which has never been done in the literature
before.
If $\sqrt{4r_h-3}(1/2-1/\Phi_4)(\Phi_{12}-\Phi_{123})>1$,
one {\it always} obtains $(dc_s/dr<du/dr)_h$ from (\ref{eq81}), 
which indicates the presence of the
{\it acoustic white holes} at $r_h$. This inequality holds good for
certain astrophysically relevant range of \egam, thus 
acoustic {\it white hole solutions} 
are obtained for general relativistic, spherically symmetric
black hole accretion, see Das 2004a for further detail.
                                                                                                       
For a particular value of \egam, one can define the quantity $\tau$ to be the ratio of
$T_{{AH}}$ and $T_H$ as:
\begin{equation}
\tau=\frac{T_{{AH}}}{T_H}.
\label{eq84}
\end{equation}
It turns out that $\tau$ is {\it independent} of the mass
of the black hole.
Thus, by computing the value of $\tau$,
we can
compare the properties of the acoustic versus event horizon of
an
accreting black hole of {\it any} mass, starting from the 
primordial black hole to the super massive black holes at the
dynamical centre of the galaxies.

For general relativistic spherical accretion, one finds 
that for certain range of 
\egam, $T_{AH}$ {\it exceeds} (i.e., $\tau>1$)
the value of 
$T_H$, hence the analogue Hawking temperature can be 
{\it larger} than the actual Hawking temperature,
see Das 2004a for further details.
\section{Multi-transonic, Relativistic Accretion Disc as Analogue 
Gravity Model}
\subsection{The Stress Energy Tensor and Flow Dynamics}
\noindent
To provide a generalized description of axisymmetric fluid flow in strong
gravity, one needs to solve the equations of motion for the
fluid and the Einstein equations. The problem may be made
tractable by assuming the accretion to be non-self
gravitating so that the fluid dynamics may be dealt in a
metric without back-reactions.
To describe the flow, we use the Boyer-Lindquist
co-ordinate (Boyer \& Lindquist 1967),
% with signature ($-+++$), 
and an
azimuthally Lorentz boosted orthonormal tetrad basis co-rotating
with the accreting fluid. We define $\lambda$ to be the specific
angular momentum of the flow. Since we are not interested in non-axisymmetric
disc structure, we neglect any gravo-magneto-viscous
non-alignment between $\lambda$ and black hole spin angular
momentum. We consider the flow to be non-self-gravitating to exclude
any back reaction on the metric.
For this section, the gravitational radius $r_g$ is taken to 
be $GM_{BH}/c^2$.

The most general form of the energy momentum
tensor for the compressible hydromagnetic astrophysical
fluid (with a frozen in magnetic field) vulnerable to the shear,
bulk viscosity and generalized energy exchange, may be
expressed as (Novikov \& Thorne 1973):
\begin{equation}
{\Im}^{{\mu}{\nu}}={\Im}^{{\mu}{\nu}}_{M}+{\Im}^{{\mu}{\nu}}_{\bf B}
\label{eq85}
\end{equation}
where ${\Im}^{{\mu}{\nu}}_M$ and ${\Im}^{{\mu}{\nu}}_{\bf B}$
are the fluid (matter) part and the Maxwellian
(electromagnetic) part of the energy momentum
tensor.
${\Im}^{{\mu}{\nu}}_{M}$ and ${\Im}^{{\mu}{\nu}}_{\bf B}$
may be expressed as:
\begin{equation}
{\Im}^{{\mu}{\nu}}_M={\rho}v^{\mu}v^{\nu}+\left(p-\varsigma{\theta}\right)h^{\mu\nu}
-2\eta{\sigma}^{\mu\nu}+{\rm q}^{\mu}v^\nu+v^\mu{{\rm q}^\nu},~
{\Im}^{{\mu}{\nu}}_{\bf B}=\frac{1}{8\pi}\left({\rm B}^2v^{\mu}v^\nu+{\rm B}^2h^{\mu\nu}
-2{\rm B}^\mu{\rm B}^\nu\right)
\label{eq86}
\end{equation}
In the above expression, ${\rho}v^{\mu}v^{\nu}$ is the total mass energy density excluding the 
frozen-in magnetic field mass energy density as measured in the local rest frame of the 
baryons (local orthonormal frame, hereafter LRF, 
 in which there is no net baryon flux in any direction).
$ph^{\mu\nu}$ is the isotropic pressure for incompressible gas (had it been the case that 
$\theta$ would be zero). $\varsigma$ and $\eta$ are the co-efficient of bulk viscosity 
and of dynamic viscosity, respectively. Hence $-\varsigma{\theta}h^{\mu\nu}$ and 
$-2{\eta}{\sigma^{\mu\nu}}$ are the isotropic viscous stress and the viscous shear 
stress, respectively. ${\rm q}^{\mu}v^\nu+v^\mu{{\rm q}^\nu}$ is the energy and 
momentum flux, respectively, in LRF of the
baryons. In the expression for ${\Im}^{{\mu}{\nu}}_{\bf B}$,
${\rm B}^2/8\pi$ in the first term represents the energy density, in the second
term represents the magnetic pressure orthogonal to the magnetic field lines, 
and in third term magnetic tension along the field lines (all terms expressed in LRF),
respectively. 

Here, the electromagnetic field is described by the field tensor
${\cal F}^{{\mu}{\nu}}$ and it's dual
${\cal F}^{{\ast}{\mu}{\nu}}$ (obtained from
${\cal F}^{{\mu}{\nu}}$ using Levi-Civita `flipping' tensor
${\epsilon}^{{\mu}{\nu}{\alpha}{\beta}}$)
satisfying the Maxwell equations through the vanishing of the
four-divergence of ${\cal F}^{{\ast}{\mu}{\nu}}$.
A complete description of flow behaviour could be obtained
by taking the co-variant derivative of ${\Im}^{{\mu}{\nu}}$
and ${\rho}v^{\mu}$ to obtain the energy momentum
conservation equations and the conservation of baryonic mass.

However, at this stage, the complete solution remains
analytically untenable unless we are forced
to adopt a number
of simplified approximations.
We would like to study the {\it inviscid}
accretion of {\it hydrodynamic} fluid.
Hence ${\Im}^{{\mu}{\nu}}$ 
may be described by
the standard form of the energy momentum (stress-energy)
tensor of a perfect perfect fluid:
\begin{equation}
{\Im}^{{\mu}{\nu}}=\left(\epsilon+p\right)v_{\mu}v_{\nu}+pg_{{\mu}{\nu}},
~
{\rm or,}~{\bf T}=\left(\epsilon+p\right){\bf v}{\otimes}{\bf v}+p{\bf g}
\label{eq87}
\end{equation}
Our calculation will thus be 
focused on the stationary
axisymmetric solution of the energy momentum
and baryon number conservation equations
\begin{equation}
{{\Im}^{{\mu}{\nu}}}_{;\nu}
=0;
\;\;\;\;\;
\left({\rho}{v^\mu}\right)_{;\mu}=0,
\label{eq88}
\end{equation}
%%\eqno{(8)}
Specifying the metric to be stationary and axially symmetric,
 the two
generators
 $\xi^{\mu}\equiv (\partial/\partial t)^{\mu}$ and
 $\phi^{\mu}\equiv (\partial/\partial \phi)^{\mu}$ of the temporal and
axial isometry, respectively, are
Killing vectors.

We consider the flow to be
`advective', i.e. to possess considerable radial three-velocity.
The above-mentioned advective velocity, which we hereafter denote by $u$
and  consider it to be confined on the equatorial plane, is essentially the 
three-velocity component perpendicular to the set of hypersurfaces
$\{\Sigma_v\}$ defined by
$v^2={\rm const}$, where $v$ is the magnitude of the 3-velocity.
Each $\Sigma_v$ is timelike since 
its normal $\eta_{\mu}\propto \partial_{\mu} v^2$ 
is spacelike and may be normalized as
$\eta^{\mu}\eta_{\mu}=1$. 

We then define the specific angular momentum $\lambda$ and the angular
velocity $\Omega$ as
\begin{equation}
\lambda=-\frac{v_\phi}{v_t}; \;\;\;\;\;
\Omega=\frac{v^\phi}{v^t}
=-\frac{g_{t\phi}+\lambda{g}_{tt}}{{g_{\phi{\phi}}+\lambda{g}_{t{\phi}}}}\, ,
\label{eq89}
\end{equation}

The metric on the equatorial plane is given by (Novikov \& Thorne 1973)
\begin{equation}
ds^2=g_{{\mu}{\nu}}dx^{\mu}dx^{\nu}=-\frac{r^2{\Delta}}{A}dt^2
+\frac{A}{r^2}\left(d\phi-\omega{dt}\right)^2
+\frac{r^2}{\Delta}dr^2+dz^2 ,
\label{eq90}
\end{equation}
where $\Delta=r^2-2r+a^2, ~A=r^4+r^2a^2+2ra^2$,
and $\omega=2ar/A$, $a$ being the Kerr parameter related to the black-hole spin. 
The normalization condition $v^\mu{v}_\mu=-1$, together with  
the expressions for  
$\lambda$ and $\Omega$  in (\ref{eq89}), provides the relationship between the
advective velocity $u$ and the temporal component of the four velocity
\begin{equation}
v_t=
\left[\frac{Ar^2\Delta}
{\left(1-u^2\right)\left\{A^2-4\lambda arA
+\lambda^2r^2\left(4a^2-r^2\Delta\right)\right\}}\right]^{1/2} .
\label{eq91}
\end{equation}
%\subsection{Thermodynamics}
\noindent
In order to solve (\ref{eq88}), we need to specify a realistic equation of
state. In this work, we concentrate on polytropic accretion. However, polytropic 
accretion is not the only choice to describe the general relativistic axisymmetric
black-hole accretion. Equations of state other than the adiabatic one,  such as 
the isothermal equation (Yang \& Kafatos 1995)
or the two-temperature plasma (Manmoto 2000),
have also been used to study the black-hole accretion flow.

Like spherical accretion, here also we assume the dynamical in-fall time scale to be short compared with any 
dissipation time scale during the accretion process.
% of the form
%\begin{equation}
%p=K{\rho}^\gamma ,
%\label{eq92}
%\end{equation}
%where the polytropic index $\gamma$ equal to the ratio of the two specific
%heats $c_p$ and $c_v$ of the accreting material is assumed to be constant throughout the fluid.
%A more realistic model of the flow 
%would perhaps require  a variable polytropic index having a 
%functional dependence on the radial
%distance, i.e. $\gamma=\gamma(r)$. However, 
We  have performed the
calculations for a sufficiently large range of $\gamma$ and we believe
that all astrophysically relevant
polytropic indices are covered in our work.
%\subsection{Disc Geometry and Conservation Equations \label{sec:tbhadgr_dgce}}
\subsection{Disc Geometry and the Conservation Equations}
\noindent
We assume that
the disc has a radius-dependent local
thickness $H(r)$, and its central plane coincides with
the equatorial plane of the black hole.
It is a standard practice
in accretion disc theory
(Matsumoto et. al. 1984; Paczy'nski 1987; Abramowicz,
Czerny, Lasota \& Szuszkiewicz 1988;
Chen \& Taam 1993;
Kafatos \& Yang 1994;
Artemova, Bj\"{o}rnsson \& Novikov 1996;
Narayan, Kato \& Honma 1997;
Wiita 1999;
Hawley \& Krolik 2001;
Armitage, Reynolds \& Chiang 2001)
to
use the vertically integrated
model in
describing the black-hole accretion discs where the equations of motion
apply to the equatorial plane of the black hole, assuming the flow to
be in hydrostatic equilibrium in the transverse direction. 
The assumption of hydrostatic
equilibrium is justified for a thin flow because for such flows, the infall
time scale is expected to exceed the local sound crossing time
scale in the direction transverse to the flow.
We follow the same 
procedure here. 
The thermodynamic 
flow variables are averaged over the disc height,
i.e.,
a thermodynamic 
quantity $y$ used in our model is vertically integrated over the disc height and averaged as
$\bar{y}=\int^H(r)_0 (ydh)/\int^H(r)_0 H(r)$.

%\begin{figure}
%\vbox{
%\vskip -1.0cm
%\centerline{
%\psfig{file=figure2.eps,height=11cm,width=14.2cm,angle=0.0}}
%{\bf Figure 2:} Height averaged thermodynamic quantities for disc accretion}
%\end{figure}
%\vskip 0.25truecm

\begin{center}
\begin{figure}[h]
\includegraphics[scale=0.4,angle=0.0]{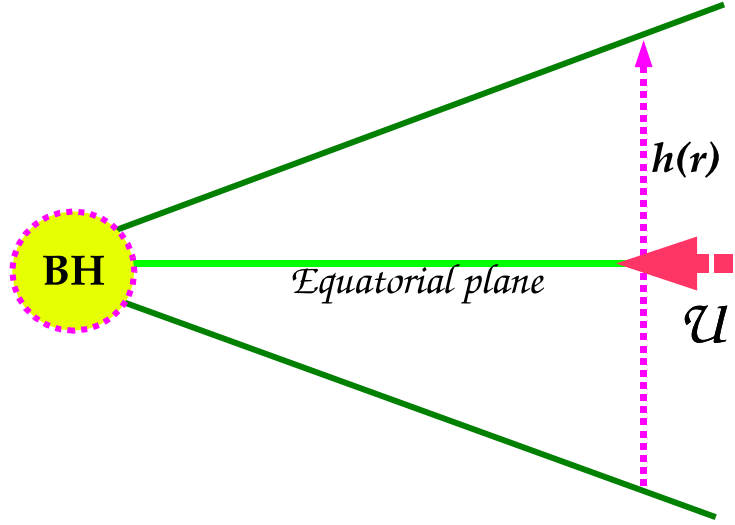}
\caption[]{Height averaged thermodynamic quantities for disc accretion.}
\label{fig2}
\end{figure}
\end{center}

In figure 2, we schematically represent the above mentioned modelling.
The yellow circular patch with BH written inside represents the black hole 
and the pink dashed boundary mimics the event horizon. The wedge shaped dark green 
lines represents the envelop of the accretion disc. The light green line centrally 
flanked by the two dark green disk boundaries, is the equatorial plane, on which all 
of the dynamical quantities (e.g., the advective velocity $u$) are assumed to be 
confined. Any thermodynamic quantity (e.g., the flow density) is 
averaged over the local disc height ${\bf h}(r)$ as shown in the figure.

We follow  Abramowicz, Lanza \& Percival (1997)
to derive an expression for the disc height $H(r)$ 
in our flow geometry since the relevant equations in 
Abramowicz, Lanza \& Percival (1997)
are non-singular on the horizon and can accommodate both the axial and  
a quasi-spherical flow geometry. In the Newtonian framework, the disc height
 in vertical 
equilibrium is obtained from the $z$ component of the non-relativistic Euler 
equation where all the terms involving velocities and the
higher powers of $\left({z}/{r}\right)$ are neglected. 
In the case of a general relativistic disc, the vertical pressure
gradient in the comoving frame is compensated by the tidal gravitational
field. We then obtain the disc height 
\begin{equation}
H(r)=\sqrt{\frac{2}{\gamma + 1}} r^{2} \left[ \frac{(\gamma - 1)c^{2}_{c}}
{\{\gamma - (1+c^{2}_{s})\} \{ \lambda^{2}v_t^2-a^{2}(v_{t}-1) \}}\right] ^{\frac{1}{2}} ,
\label{eq105}
\end{equation}
which, by making use of
(\ref{eq91}), 
may be be expressed in terms of 
the advective velocity $u$. 

The temporal  component of the energy momentum tensor conservation equation 
leads to the
constancy along each streamline of the flow specific energy 
${\cal E}$ (${\cal E}=hv_t$), and hence
% (relativistic analogue of Bernoulli's constant) defined as 
%(Anderson 1989)
%\begin{equation}
%{\cal E}=hv_t. 
%\label{eq106}
%\end{equation}
from (\ref{eq91}) and (\ref{eq100}) it follows that:
\begin{equation}
{\cal E} =
\left[ \frac{(\gamma -1)}{\gamma -(1+c^{2}_{s})} \right]
\sqrt{\left(\frac{1}{1-u^{2}}\right)
\left[ \frac{Ar^{2}\Delta}{A^{2}-4\lambda arA +
\lambda^{2}r^{2}(4a^{2}-r^{2}\Delta)} \right] } \, .
\label{eq107}
\end{equation}
The rest-mass accretion rate ${\dot M}$ is obtained by integrating the relativistic 
continuity equation (\ref{eq88}). One finds
\begin{equation}
{\dot M}=4{\pi}{\Delta}^{\frac{1}{2}}H{\rho}\frac{u}{\sqrt{1-u^2}} \, ,
\label{eq108}
\end{equation}
Here, we adopt the sign convention that a positive $u$ corresponds to
accretion.
The entropy accretion rate ${\dot \Xi}$ 
can be expressed as:
\begin{equation}
{\dot \Xi}
 = \left( \frac{1}{\gamma} \right)^{\left( \frac{1}{\gamma-1} \right)}
4\pi \Delta^{\frac{1}{2}} c_{s}^{\left( \frac{2}{\gamma - 1}\right) } \frac{u}{\sqrt{1-u^2}}\left[\frac{(\gamma -1)}{\gamma -(1+c^{2}_{s})}
\right] ^{\left( \frac{1}{\gamma -1} \right) } H(r)
\label{eq109}
\end{equation}
One can solve the conservation equations for ${\cal E}, {\dot M}$ and
${\dot \Xi}$ to obtain the complete accretion profile. 
\subsection{Transonicity}
\noindent
The gradient of the acoustic velocity can be computed by 
differentiating (\ref{eq109}) and can be obtained as:
\begin{equation}
\frac{dc_s}{dr}=
\frac{c_s\left(\gamma-1-c_s^2\right)}{1+\gamma}
\left[
\frac{\chi{\psi_a}}{4} -\frac{2}{r} 
-\frac{1}{2u}\left(\frac{2+u{\psi_a}}{1-u^2}\right)\frac{du}{dr} \right]
\label{eq110a}
\end{equation}
The dynamical velocity gradient can then be calculated by differentiating (\ref{eq108}) 
with the help of (\ref{eq110a}) as:
\begin{equation}
\frac{du}{dr}=
\frac{\displaystyle
\frac{2c_{s}^2}{\left(\gamma+1\right)}
  \left[ \frac{r-1}{\Delta} + \frac{2}{r} -
         \frac{v_{t}\sigma \chi}{4\psi}
  \right] -
  \frac{\chi}{2}}
{ \displaystyle{\frac{u}{\left(1-u^2\right)} -
  \frac{2c_{s}^2}{ \left(\gamma+1\right) \left(1-u^2\right) u }
   \left[ 1-\frac{u^2v_{t}\sigma}{2\psi} \right] }},
%= \frac{{\cal N}}{{\cal D}}
\label{eq110}
\end{equation}
\noindent where
\begin{eqnarray}
\psi=\lambda^2{v_t^2}-a^2\left(v_t-1\right),~
\psi_a=\left(1-\frac{a^2}{\psi}\right),~
\sigma = 2\lambda^2v_{t}-a^2,
~
& & \nonumber \\
\chi =
\frac{1}{\Delta} \frac{d\Delta}{dr} +
\frac{\lambda}{\left(1-\Omega \lambda\right)} \frac{d\Omega}{dr} -
\frac{\displaystyle{\left( \frac{dg_{\phi \phi}}{dr} + \lambda \frac{dg_{t\phi}}{dr} \right)}}
     {\left( g_{\phi \phi} + \lambda g_{t\phi} \right)}.
\label{eq111}
\end{eqnarray}
\noindent
The critical point conditions can be obtained as:
\begin{equation}
{c_{s}}_{\vc}={\left[\frac{u^2\left(\gamma+1\right)\psi}
                                  {2\psi-u^2v_t\sigma}
                       \right]^{1/2}_{\vc}  },
~~u{\vc}= {\left[\frac{\chi\Delta r} {2r\left(r-1\right)+ 4\Delta} \right]
^{1/2}_{\rm r=r_c}  },
\label{eq112}
\end{equation}
\noindent
For any value of
\eker,
substitution of the values of $u{\vc}$ and $c_{s}{\vert}_{\rm r=r_c}$  in terms of $r_c$
in the expression
for ${\cal E}$ (\ref{eq107}),
provides
a polynomial in $r_c$, the solution of which determines
the location of the critical point(s) $r_c$.

It is obvious from (\ref{eq112}) that, unlike relativistic spherical accretion,
$u_{\vc}{\ne}{c_s}_{\rm r=r_c}$, and hence the Mach number at the critical point is {\it not} 
equal to unity in general. This phenomena can more explicitly be 
demonstrated for $a=0$, i.e., for relativistic disc accretion in the Schwarzschild metric.

For Schwarzschild black hole,
one can calculate the Mach number of the flow at the critical point as
\footnote{The same expression can be obtained by putting $a=0$ in (\ref{eq112}).}
(Das, Bili\'c \& Dasgupta 2006)
\begin{equation}
M_c=
%{\frac{2}{\gamma+1}}
\sqrt{
\left({\frac{2}{\gamma+1}}\right)
\frac
{{f_{1}}(r_c,\lambda)}
{{{f_{1}}(r_c,\lambda)}+{{f_{2}}(r_c,\lambda)}}
}\, .
\label{eq115}
\end{equation}
where 
\begin{equation}
{f_{1}}(r_c,\lambda) = \frac{3r_c^{3}-2\lambda^{2}r_c+
3{\lambda^2}}{r_c^{4}-\lambda^{2}r_c(r_c-2)},~
{f_{2} } (r_c,\lambda) = \frac{2r_c-3}{r_c(r_c-2)} -
\frac{2r_c^{3}-\lambda^{2}r_c+\lambda^{2}}{r_c^{4}-\lambda^{2}r_c(r_c-2)}
\label{eq115a}
\end{equation}
Clearly, $M_c$ is generally not equal to unity, and for $\gamma\geq 1$, is always less 
than one. 

Hence  we distinguish a sonic point from a critical point.
In the literature on transonic black-hole accretion discs, the concepts of critical 
and sonic points are often made synonymous by defining an `effective' sound speed 
leading to the `effective'  Mach number (for further details, see, eg.
Matsumoto et. al. 1984, Chakrabarti 1989).
Such  definitions were proposed as effects of a 
specific disc geometry. We, however, prefer to maintain the usual definition of the Mach 
number for two reasons.
 
First, in the existing literature on transonic disc accretion, 
the Mach number at the critical point  turns out to be a function of 
$\gamma$ only, and hence $M_c$ remains  constant  if $\gamma$ is constant.
 For example, 
using the Paczy\'nski and Wiita (1980) pseudo-Schwarzschild potential to 
describe the adiabatic accretion phenomena leads to 
(see section 16.1.1 for the derivation and for further details)
\begin{equation}
M_c=\sqrt{\frac{2}{\gamma+1}}\, .
\label{eq116}
\end{equation}
The above expression does not depend on the location of the 
critical point and depends only on the value of the 
adiabatic index chosen to describe the flow. Note that 
for isothermal accretion $\gamma=1$ and hence the sonic points and 
the critical points {\it are} equivalent (since $M_c=1$), 
see (\ref{kk18}) in section 16.1.2 for further details.

However, the quantity $M_c$ in Eq. (\ref{eq115}) 
as well as in (\ref{eq112})
is clearly a function of $r_c$, and hence, generally, it takes  different 
values for different $r_c$ for transonic accretion.
The difference between the 
radii of the critical 
point and the sonic point may be quite significant.
One defines the
radial difference of the critical and the sonic point 
(where the Mach number is exactly equal to unity) as
\begin{equation}
{\Delta}r_c^s=|r_s-r_c|.
\label{eq117}
\end{equation}
The quantity ${\Delta}r_c^s$ may be  
 a complicated  function of \eker, the  form of which can not 
be expressed analytically. 
The radius $r_s$ in Eq. (\ref{eq117}) is the radius of the
sonic point  corresponding to the same \eker for which the
radius of the critical point $r_c$ is evaluated.
 Note, however, that since $r_s$ is calculated by integrating the 
flow from $r_c$, ${\Delta}r_c^s$ is defined only for saddle-type
critical points (see subsequent paragraphs for 
further detail). This is because, 
 a physically acceptable transonic solution
can be constructed only through a saddle-type critical point. 
 One can then show that ${\Delta}r_c^s$ can be as large as $10^2$ $r_g$ or even
  more (for further details, see Das, Bili\'c \& Dasgupta 2006).

The second and perhaps the more important reason for keeping $r_c$ and $r_s$ 
distinct
is the following. 
In addition to studying the dynamics of general relativistic transonic 
black-hole accretion, we are also interested in studying the 
analogue Hawking effects for such accretion flow.
We need to identify the 
 location of the acoustic horizon 
 as a radial distance at which  the Mach equals to one, hence, a {\it sonic
  point}, and not a {\em  critical point} 
will be of our particular interest.
 To this end, we first calculate the critical point $r_c$
  for a particular \eker 
following the procedure discussed above, and then we compute the location 
of the sonic point (the radial distance where the 
Mach number exactly equals to unity)
by integrating the flow equations starting from the critical points.
The dynamical and the acoustic velocity, as well as their
space derivatives, at the {\it sonic} point, are then
evaluated.
The details of this procedure for the
Schwarzschild metric are provided in Das, Bili\'c \& Dasgupta 2006.

Furthermore, the definition of the acoustic metric in terms of 
the sound speed does not seem to be mathematically consistent with the idea of
an `effective' sound speed, irrespective of whether one deals with
 the Newtonian,  post Newtonian, or a relativistic description
of the accretion disc. Hence, we do not adopt
the idea of identifying critical a point with a sonic point.
However,  for saddle-type
critical points, $r_c$ and $r_s$ should always have one-to-one correspondence, 
in the sense that
every  critical point that  allows a steady solution to pass through it 
is accompanied by a sonic point, generally at a 
different radial distance
$r$.

It is worth emphasizing that the distinction between critical and  
sonic points is a direct manifestation of the non-trivial 
functional dependence of the disc thickness on
the fluid velocity, the sound speed
and the radial distance.
 In the simplest idealized case when
the disc thickness is assumed to be constant,
one would expect no distinction
between critical and sonic points. 
In this case, as
has  been demonstrated for a thin disc
accretion onto the Kerr black hole  (Abraham, Bili\'c \& Das 2006),
the quantity $\Delta r_c^s$ vanishes identically for any astrophysically 
relevant value of \eker.
Hereafter, we will use $r_h$ to denote the sonic point $r_s$,
since a sonic point is actually the location of the acoustic horizon. 
%\begin{figure}
%\vbox{
%\vskip -0.0cm
%\centerline{
%\psfig{file=figure3.ps,height=9cm,width=12.2cm,angle=270.0}}
%\vskip -0.0cm
%{\bf Figure 3:} Parameter space for general relativistic 
%multi-transonic accretion and wind in Kerr geometry, see text for detail. This 
%figure is reproduced from Goswami, Khan, Ray \& Das 2007.} 
%\end{figure}

\begin{center}
\begin{figure}[h]
\includegraphics[scale=0.5,angle=270.0]{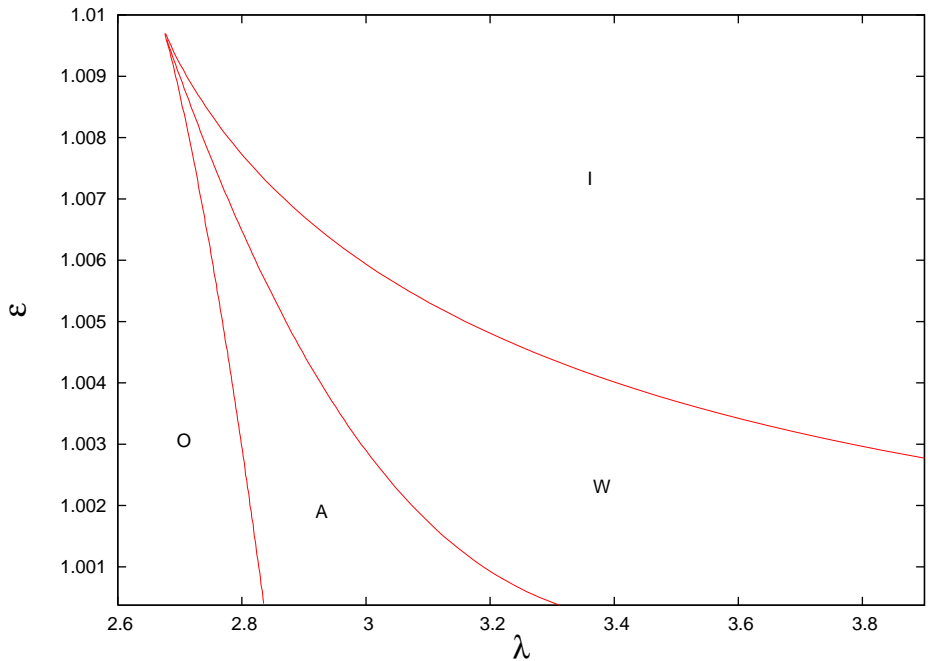}
\caption[]{Parameter space for general relativistic
multi-transonic accretion and wind in Kerr geometry, see text for detail. This
figure is reproduced from Goswami, Khan, Ray \& Das 2007.}
\label{fig3}
\end{figure}
\end{center}
%\vskip 0.25truecm
\subsection{Multi-transonic Behaviour}
\noindent
Unlike spherical accretion, one finds {\it three} (at most)
critical points for relativistic disc accretion for some 
values of \eker. 
In figure 3, we classify the 
$\left[{\cal E},\lambda\right]$ parameter space, for a 
fixed value of adiabatic index ($\gamma=4/3$) and 
the Kerr parameter ($a=0.3$), to show the formation of 
various kind of critical points. The regions marked by 
O and I correspond to the 
formation of a single critical point, and hence the {\it mono-transonic}
disc accretion is produced for such region. 
In  the region marked by
{\bf I}, the critical points are called `inner type' critical points since
these points are  quite close to the event horizon,  approximately
in the range
$2<r_c^{\rm in}{\le}10$.
In the region marked by {\bf O}, the
critical points are called `outer type' critical points, because these points are
located considerably far away from the black hole.
Depending on the value of
\eker, an outer critical point may be as far as $10^4r_g$,
or more. 

The outer type critical points for the mono-transonic region are formed, 
as is obvious from the figure, for weakly rotating flow. For low angular 
momentum, accretion flow contains less amount of rotational energy, thus 
most of the kinetic energy in utilized to increase the radial 
dynamical velocity $u$ at a faster rate, leading to a higher value 
of $d^2u/dr^2$. Under such circumstances, the dynamical velocity $u$ becomes
large enough to overcome the acoustic velocity $c_s$ at a larger radial distance
from the event horizon, leading to the generation of supersonic flow at a large
value of $r$, which results the formation of the sonic point 
(and hence the corresponding critical point) far away from the black hole event horizon.
On the contrary, the inner type critical points are formed,
as is observed from the figure, for strongly rotating flow in general. Owing to 
the fact that such flow would posses a large amount of rotational energy, only a small 
fraction of the total specific energy of the flow will be spent to increase the 
radial dynamical velocity $u$. Hence for such flow, $u$ can overcome $c_s$ only at a 
very small distance (very close to the event horizon) where the intensity of the 
gravitational field becomes enormously large, producing a very high value of the 
linear kinetic energy of the flow (high $u$), over shedding the contribution to the 
total specific energy from all other sources. However, from the figure it is 
also observed that the inner type sonic points are formed also for moderately 
low values of the angular momentum as well (especially in the region close to 
the vertex of the wedge shaped zone marked by W). For such regions, the total conserved specific 
energy is quite high. In the asymptotic limit, the expression for the total specific 
energy is governed by the Newtonian mechanics and one can have:
\begin{equation}
{\cal E}=
\left(\frac{u^2}{2}\right)_{\rm linear}
+
\left(\frac{c_s^2}{\gamma-1}\right)_{\rm thermal}
+
\left(\frac{\lambda^2}{2r^2}\right)_{\rm rotational}
+
\left(\Phi\right)_{\rm gravitational}
\label{eqN1}
\end{equation}
where $\Phi$ is the gravitational potential energy in the 
asymptotic limit, see section 16.1.1 for further detail. 
From (\ref{eqN1}) it is obvious that at a considerably large 
distance from the black hole, the contribution to the total energy of 
the flow comes mainly (rather entirely) from the thermal energy. 
A high value of ${\cal E}$ (flow energy in excess to its rest mass energy)
corresponds to a `hot' flow starting from infinity. Hence the acoustic velocity
corresponding to the `hot' flow obeying such outer boundary condition would 
be quite large. For such accretion, flow has to travel a large distance 
subsonically and can acquire a supersonic dynamical velocity 
$u$ only at a very close proximity to the event horizon, where the gravitational 
pull would be enormously strong.

The $\left[{\cal E},\lambda\right]$ corresponding to the  
wedge shaped regions marked by A and W produces {\it three} critical points, among which the
largest and the smallest values correspond to the X type (saddle type), the outer
$r_c^{out}$ and the inner $r_c^{in}$, critical points respectively. The 
O type (centre type) middle critical point, $r_c^{mid}$, which is unphysical 
in the sense that no steady transonic solution passes through it, lies 
in between $r_c^{in}$ and $r_c^{out}$. 
The following discussion provides the methodology for finding out the nature 
(whether saddle/centre type) in brief, see Goswami, Khan, Ray \& Das 2007 
for further detail.

Eq. (\ref{eq110}) could be recast as 
\begin{equation}
\frac{du^{2}}{dr}={\frac{\frac{2}{\gamma
+1}c_{s}^{2}\left[\frac{g^{\prime}_{1}}{g_{1}}-\frac{1}{g_{2}}\frac{\partial
g_{2}}{\partial
r}\right]-\frac{f^{\prime}}{f}}
{\frac{1}{1-u^{2}}\left(1-\frac{2}{\gamma
+1}\frac{c_{s}^{2}}{u^{2}}\right)+\frac{2}{\gamma
+1}\frac{c_{s}^{2}}{g_{2}}\left(\frac{\partial g_{2}}{\partial
u^{2}}\right)}}
\label{mod1}
\end{equation}
\begin{equation}
\frac{du^{2}}{d{\bar{\tau}}}=\frac{2}{\gamma
+1}c_{s}^{2}\left[\frac{g^{\prime}_{1}}{g_{1}}-\frac{1}{g_{2}}\frac{\partial
g_{2}}{\partial
r}\right]-\frac{f^{\prime}}{f}
\label{mod2}
\end{equation}
with the primes representing full derivatives with respect to $r$,
and $\bar{\tau}$ is an arbitrary mathematical parameter.
Here, 
\begin{eqnarray}
f(r)=\frac{Ar^{2}\Delta}{A^{2} - 4\lambda arA +
\lambda^{2}r^{2}(4a^{2}-r^{2}\Delta)},~ && \nonumber \\
g_{1}i(r)=\Delta r^{4},~g_{2}(r,u)=\frac{\lambda^{2}f}{1-u^{2}}
-\frac{a^{2}f^{\frac{1}{2}}}{\sqrt{1-u^{2}}} + a^{2}
\label{mod3}
\end{eqnarray}

The critical conditions are obtained with the simultaneous vanishing
of the right hand side, and the coefficient of ${d(u^2)/dr}$ in the left 
hand side in (\ref{mod1}). This will provide
\begin{equation}
\left|
\frac{2c_s^2}{\gamma+1}
\left[\frac{g_1^{\prime}}{g_1}-\frac{1}{g_2}\left(\frac{\partial{g_2}}{\partial{r}}\right)\right]
-\frac{f^{\prime}}{f}
\right|_{\rm r=r_c} 
=
\left|
\frac{1}{1-u^2}\left(1-\frac{2}{\gamma+1}\frac{c_s^2}{u^2}\right)
+\frac{2}{\gamma+1}\frac{c_s^2}{g_2}\left(\frac{\partial{g_2}}{\partial{u^2}}\right)
\right|_{\rm r=r_c}=0
\label{mod4}
\end{equation}
as the two critical point conditions.
Some simple algebraic manipulations will show that
\begin{equation}
u_c^2=\frac{f^{\prime}g_1}{f{g_1^{\prime}}}
\label{mod5}
\end{equation}
following which $c_{s}^2|_{\rm r=r_c}$ can be rendered as a function of $r_c$ only, 
and further, by use of (\ref{eq107})., $r_c$, $c_{sc}^2$ and $u_c^2$ can 
all be fixed in terms of the constants of motion like $E$, $\gamma$, 
$\lambda$ and $a$. Having fixed the critical points it should now be 
necessary to study their nature in their phase portrait of $u^2$ 
versus $r$. To that end one applies a perturbation about the fixed point 
values, going as,
\begin{equation}
u^2=u^2|_{\rm r=r_c}+\delta{u^2},~c_s^2=c_s^2|_{\rm r=r_c}+\delta{c_s^2},~
r=r_c+\delta{r}
\label{mod6}
\end{equation}
in the parameterized set of autonomous first-order differential equations,
\begin{equation}
\frac{d({u^2})}{d{\bar{\tau}}}
=\frac{2}{\gamma+1}c_s^2
\left[\frac{g_1^{\prime}}{g_1}
-\frac{1}{g_2}\left(\frac{\partial{g_2}}{\partial{r}}\right)\right]
\frac{f^{\prime}}{f}
\label{mod7}
\end{equation}
and
\begin{equation}
\frac{dr}{d{\bar{\tau}}}=
\frac{1}{1-u^2}\left(1-\frac{2}{\gamma+1}\frac{c_s^2}{u^2}\right)
+\frac{1}{\gamma+1}\frac{c_s^2}{g_2}\left(\frac{\partial{g_2}}{\partial{u^2}}\right)
\label{mod8}
\end{equation}
with ${\bar{\tau}}$  being an arbitrary parameter. In the two equations above 
$\delta c_s^2$ can be closed in terms of $\delta u^2$ and $\delta r$ 
with the help of (\ref{eq110a}). Having done so, one could then make use of 
solutions of the form, $\delta r \sim \exp ({\bar{\Omega}} \tau)$ and 
$\delta u^2 \sim \exp ({\bar{\Omega}} \tau)$,  
from which, ${\bar{\Omega}}$ would give the eigenvalues --- growth rates of 
$\delta u^2$ and $\delta r$ in ${\bar{\tau}}$ space --- of the stability matrix 
implied by (\ref{mod7}-\ref{mod8}). Detailed calculations will show the eigenvalues 
to be 
\begin{equation}
{\bar{\Omega}}^{2}=\left|{\bar{\beta}}^{4}c_{s}^{4}\chi_1^{2}+\xi_{1}\xi_{2}\right|_{\rm r=r_c}
\label{mod9}
\end{equation}
where ${\bar{\beta}}^2=\frac{2}{\gamma+1}$ and $\chi_1,\xi_1$ and $\xi_2$ can be 
expressed as polynomials of $r_c$ (see Goswami, Khan, Ray \& Das 2007 for
the explicit form of the polynomial), hence ${\bar{\Omega}}^2$ can be evaluated
for any \eker once the value of the corresponding critical point $r_c$ is known.
The structure of (\ref{mod9}) immediately shows that the only admissible 
critical points in the conserved Kerr system will be either saddle points 
or centre type points.
For a saddle point, ${\bar{\Omega}}^2 > 0$, while for a centre-type point,
${\bar{\Omega}}^2 < 0$.

For multi-transonic flow characterized by a specific set of \eker, one can 
obtain the value of ${\bar{\Omega}}^2$ to be positive for $r_c^{in}$ and $r_c^{out}$,
showing that those critical points are of saddle type in nature. ${\bar{\Omega}}^2$
comes out to be negative for $r_c^{mid}$, confirming that the middle sonic 
point is of centre type and hence no transonic solution passes through it. One
can also confirm that {\it all} mono-transonic 
flow (flow with a single critical point characterized by $\left[{\cal E},\lambda\right]$
used from the green tinted region, either {\bf I} or {\bf O}) 
corresponds to saddle type critical point.

However, there is a distinct difference between the multi-transonic flow characterized by 
$\left[{\cal E},\lambda\right]$ taken from the region marked by {\bf A}, and the 
region marked by {\bf W}. For region marked by {\bf A}, the 
entropy accretion rate ${\dot {\Xi}}$ for flows passing through the 
inner critical point is {\it greater} than that of the outer critical point
\begin{equation}
{\dot {\Xi}}\left(r_c^{in}\right)>{\dot {\Xi}}\left(r_c^{out}\right) 
\label{eqN2}
\end{equation}
while for the region marked by {\bf W}, the following relation holds
\begin{equation}
{\dot {\Xi}}\left(r_c^{in}\right)<{\dot {\Xi}}\left(r_c^{out}\right)
\label{eqN3}
\end{equation}
The above two relations show that  $\left[{\cal E},\lambda\right]$ region 
marked by ${\cal A}$ represents multi-transonic accretion,
while $\left[{\cal E},\lambda\right]{\in}\left[{\cal E},\lambda\right]_{\bf W}$
corresponds to the mono-transonic accretion but multi-transonic wind.
More details about such classification will be discussed in the following paragraphs.

There are other regions for $\left[{\cal E},\lambda\right]$ space for which either 
no critical points are formed, or two critical points are formed. These regions 
are not shown in the figure. However, none of these regions is of our interest.
If no critical point is found, it is obvious that 
transonic accretion does not form for those set of $\left[{\cal E},\lambda\right]$.
For two critical point region, one of the critical points are always of `O' type, since 
according to the standard dynamical systems theory two successive critical points can 
not be of same type (both saddle, or both centre). Hence the solution which passes through the 
saddle type critical point would encompass the centre type critical point by forming a 
loop (see, e.g., Das, Bili\'c \& Dasgupta 2006 for such loop formation in Schwarzschild
metric) like structure and hence such solution would not be physically acceptable since
that solution will form a {\it closed} loop and will not connect infinity to the 
event horizon.
\subsection{Multi-transonic Flow Topology and Shock Formation}
\noindent
To obtain the dynamical velocity gradient at the 
critical point, we apply l'Hospital's rule on (\ref{eq110}). 
After some algebraic manipulations,
the following quadratic equation
is formed,
which can be solved
to obtain $(du/dr)_{\vc}$ (see Barai, Das \& Wiita 2004 for further details):
\begin{equation}
\alpha \left(\frac{du}{dr}\right)_{\vc}^2 + \beta \left(\frac{du}{dr}\right)_{\vc} + \zeta = 0,
\label{eq113}
\end{equation}
\noindent
where the coefficients are:
\begin{eqnarray}
\alpha=\frac{\left(1+u^2\right)}{\left(1-u^2\right)^2} - \frac{2\delta_1\delta_5}{\gamma+1}, 
 \quad \quad \beta=\frac{2\delta_1\delta_6}{\gamma+1} + \tau_6,
 \quad \quad \zeta=-\tau_5;
& & \nonumber \\
%\delta_1 = \frac{u}{1-u^2}, 
\delta_1=\frac{c_s^2\left(1-\delta_2\right)}{u\left(1-u^2\right)}, \quad \quad
\delta_2 = \frac{u^2 v_t \sigma}{2\psi}, \quad \quad
\delta_3 = \frac{1}{v_t} + \frac{2\lambda^2}{\sigma} - \frac{\sigma}{\psi} ,
\quad \quad \delta_4 = \delta_2\left[\frac{2}{u}+\frac{u v_t \delta_3}{1-u^2}\right],
& & \nonumber \\
~
\delta_5 = \frac{3u^2-1}{u\left(1-u^2\right)} - \frac{\delta_4}{1-\delta_2} -
           \frac{u\left(\gamma-1-c_s^2\right)}{a_s^2\left(1-u^2\right)},
\quad \quad \delta_6 = \frac{\left(\gamma-1-c_s^2\right)\chi}{2c_s^2} +
           \frac{\delta_2\delta_3 \chi v_t}{2\left(1-\delta_2\right)},
& & \nonumber \\
\tau_1=\frac{r-1}{\Delta} + \frac{2}{r} - \frac{\sigma v_t\chi} {4\psi},
\quad \quad
\tau_2=\frac{\left(4\lambda^2v_t-a^2\right)\psi - v_t\sigma^2} {\sigma \psi},
& & \nonumber \\
\tau_3=\frac{\sigma \tau_2 \chi} {4\psi},
\quad \quad
\tau_4 = \frac{1}{\Delta} 
       - \frac{2\left(r-1\right)^2}{\Delta^2}
       -\frac{2}{r^2} - \frac{v_t\sigma}{4\psi}\frac{d\chi}{dr},
& & \nonumber \\
\tau_5=\frac{2}{\gamma+1}\left[c_s^2\tau_4 -
     \left\{\left(\gamma-1-c_s^2\right)\tau_1+v_tc_s^2\tau_3\right\}\frac{\chi}{2}\right]
   - \frac{1}{2}\frac{d\chi}{dr},
& & \nonumber \\
\tau_6=\frac{2 v_t u}{\left(\gamma+1\right)\left(1-u^2\right)}
       \left[\frac{\tau_1}{v_t}\left(\gamma-1-c_s^2\right) + c_s^2\tau_3\right].
\label{eq114}
\end{eqnarray}
Note that all the above quantities are evaluated at the critical point.
%\begin{figure}
%\vbox{
%\vskip -1.0cm
%\centerline{
%\psfig{file=figure4.ps,height=11cm,width=16.2cm,angle=270.0}}
%\vskip 0.0cm
%{\bf Figure 4:} Solution topology for multi-transonic accretion 
%in Kerr geometry for a specific set of \eker as shown in the figure.
%See text for detail.}
%\end{figure}
%
\begin{center}
\begin{figure}[h]
\includegraphics[scale=0.6,angle=270.0]{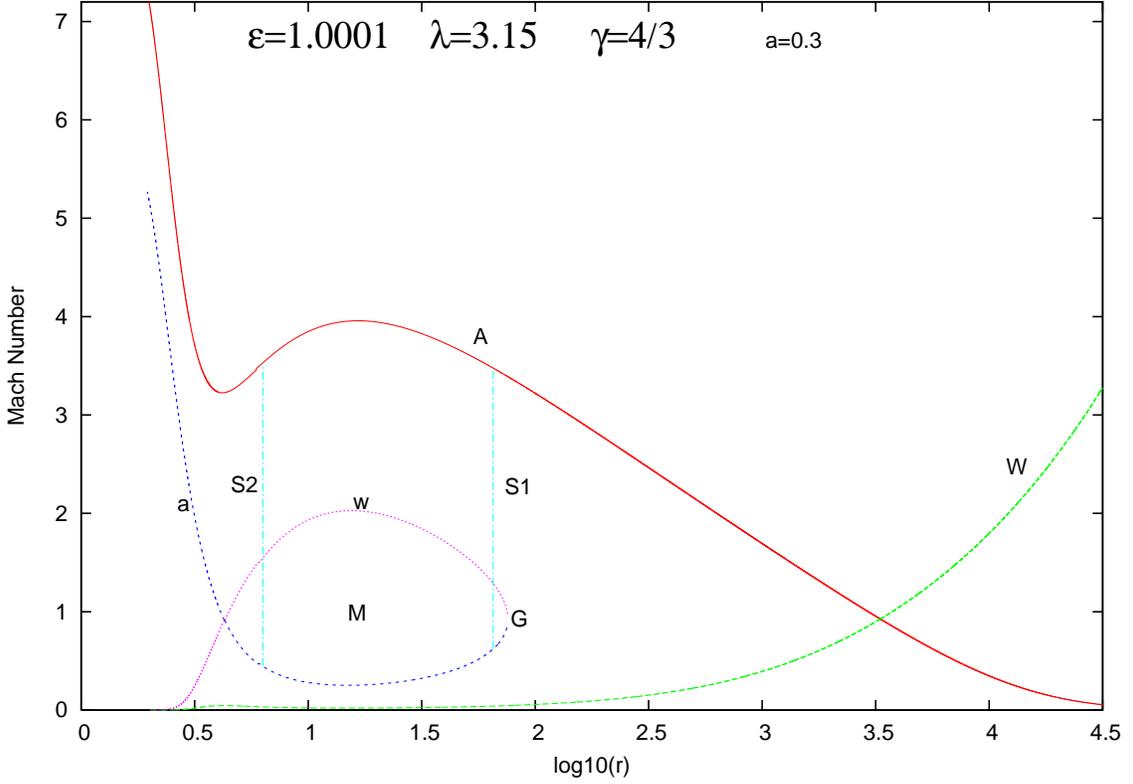}
\caption[]{Solution topology for multi-transonic accretion
in Kerr geometry for a specific set of \eker as shown in the figure.
See text for detail.}
\label{fig4}
\end{figure}
\end{center}

Hence we compute the critical advective velocity gradient as
\begin{equation}
\left(\frac{du}{dr}\right)_{\rm r=r_c}
=-\frac{\beta}{2\alpha}
{\pm}
\sqrt{\beta^2-4\alpha{\zeta}}
\label{eq113a}
\end{equation}
where the `+' sign corresponds to the accretion solution and
the `-' sign corresponds to the wind solution, see the
following discussion for further details.
Similarly, the space gradient of the acoustic velocity
$dc_s/dr$ and its value at the critical point can also 
be calculated.

The flow topology characterized by $\left[{\cal E},\lambda\right]$ corresponding 
to the {\bf I} or {\bf O} region (green tinted) is easy to obtain since the flow passes 
through only one saddle type critical point. Some of such representative topologies 
will be found in Das, Bili\'c \& Dasgupta 2006 for the Schwarzschild metric. For Kerr
metric, the flow profile would exactly be the same, only the numerical values for the
critical/sonic  point would be different, leading to the different values of $u$, $c_s$ and
other corresponding accretion parameters at the same radial distance $r$.
In this section we concentrate on multi-transonic flow topology, i.e., flow topology 
for $\left[{\cal E},\lambda\right]{\in}\left[{\cal E},\lambda\right]_{\bf A}$
or $\left[{\cal E},\lambda\right]{\in}\left[{\cal E},\lambda\right]_{\bf W}$.
Figure 4 represents one such topology. While the local radial flow Mach number has
been plotted along the $Y$ axis, the distance from the event horizon (scaled in the
unit of $GM_{BH}/c^2$) in logarithmic unit has been plotted along the $X$ axis.

The solid red line marked by A corresponds to the transonic accretion passing through 
the outer critical point $r_c^{out}$. Using a specific set of \eker as shown in the figure,
one first solve the equation for ${\cal E}$ at the critical point (using the critical 
point conditions as expressed in (\ref{eq112}) to find out the corresponding three 
critical points, saddle type $r_c^{in}$ (4.279 $r_g$), centre type $r_c^{mid}$ (14.97 $r_g$) and saddle type
$r_c^{out}$ 
(3315.01 $r_g$). The critical value of the advective velocity gradient at $r_c^{out}$ is 
then calculated using (\ref{eq113}-\ref{eq113a}). Such $u_{\vc},{c_s}_{\vc}$ and 
$du/dr_{\vc}$ serve as the initial value condition for performing the numerical
integration of the advective velocity gradient (\ref{eq110}) using the fourth-order 
Runge-Kutta method. Such integration provides the outer {\it sonic} point $r_s^{out}$
($r_s^{out}<r_c^{out}$), the
local advective velocity, the polytropic sound speed,
the Mach number, the fluid density, the disc height, the bulk temperature of the
flow, and any other relevant dynamical and thermodynamic quantity
characterizing the flow. 

The dotted green line marked by W is obtained for the 
value of $du/dr_{\vc}$ corresponds to the `-' sign in (\ref{eq113a}). Such a solution is called 
the corresponding `wind' solution. The accretion and the wind solution intersects at the
critical point (here, at $r_c^{out}$). This wind branch is just a
mathematical counterpart of the accretion solution (velocity reversal
symmetry of accretion),
owing to the presence of the quadratic term
of the dynamical velocity in the equation governing the
energy momentum conservation.
The term `wind solution' has
a historical origin.
The solar wind solution first introduced
by Parker  (1965)
has the same  topology profile as that of the
wind solution obtained in classical Newtonian Bondi accretion (Bondi 1952). Hence the
name `wind solution' has been adopted in a more general sense.
The wind solution thus represents a hypothetical process,
in which, instead of starting from infinity
and heading towards the black hole, the flow
 generated near the black-hole event horizon would fly away from the
black hole towards infinity.

The dashed blue line marked by `a' and the dotted magenta line marked by `w' are the respective
accretion and the wind solutions passing through the inner critical point $r_c^{in}$
(the intersection of the accretion and the wind branch is the location of $r_c^{in}$). 
Such accretion and wind profile are obtained following exactly the same procedure as has been
used to draw the accretion and wind topologies (red and green lines) passing through the
outer critical point. Note, however, that the accretion solution through $r_c^{in}$ 
{\it folds back onto the wind solution} and the accretion-wind closed loop 
encompasses the middle sonic point, location of which is represented by $M$ in the figure.
One should note that an `acceptable'  physical
transonic solution must be globally consistent, i.e. it must connect
 the radial infinity
$r{\rightarrow}\infty$ with the black-hole event horizon $r=2r_g$.
\footnote{This  acceptability    constraint
further demands that  the critical point corresponding to the flow
should be
of a saddle or a nodal type.
This condition is necessary although not sufficient.}.
Hence, for multi-transonic accretion, there is no {\it individual} existence
of physically acceptable accretion/wind solution passing through the inner 
critical (sonic) point, although such solution can be `clubbed' with the 
accretion solution passing through $r_c^{out}$ through shock formation, see 
the following discussions for further details.

The set $\left[{\cal E},\lambda\right]_{\bf A}$ 
(or more generally $\left[{\cal E},\lambda,\gamma,a\right]_{\bf A}$)
thus
produces doubly degenerate accretion/wind solutions.
Such two
fold degeneracy may be removed by the entropy considerations since
the entropy accretion rates ${\dot \Xi}$($r_c^{\rm in}$) and
${\dot \Xi}$($r_c^{\rm out}$) are generally not equal.
For any $\left[{\cal E},\lambda,\gamma,a\right]{\in}
\left[{\cal E},\lambda,\gamma,a\right]_{\bf A}$ 
we find that the entropy accretion rate  ${\dot \Xi}$ evaluated for the
complete accretion solution passing through the outer critical point
is {\it less} than that of the rate  evaluated for the incomplete accretion/wind solution
passing through the inner critical point.
Since the quantity ${\dot \Xi}$
is a measure of the specific entropy density of the flow,
the solution passing through $r_c^{\rm out}$ will naturally tend
to make
a transition to its higher entropy counterpart,
i.e. the incomplete accretion solution
passing through $r_c^{\rm in}$.
Hence, if there existed a mechanism for
the accretion solution passing through the outer critical point
(solid red line marked with A) to increase
its entropy accretion rate by an amount
\begin{equation}
{\Delta}{\dot \Xi}=
{\dot \Xi}(r_c^{\rm in})-{\dot \Xi}(r_c^{\rm out}),
\label{eq48}
\end{equation}
there would be a transition to the
incomplete accretion solution 
(dashed blue line marked with `a') passing through the 
inner critical point.
Such a transition would take place at
a radial distance somewhere  between the radius of the inner  sonic point
 and the
radius
of the accretion/wind turning point (75.7 $r_g$) marked by G in the
figure.
In this way one would obtain a combined accretion solution connecting
$r{\rightarrow}{\infty}$ with $r=2$ (the event horizon)
which includes a part of the accretion
solution passing through the inner critical, and hence the inner sonic point.
One finds that for some specific values of 
$\left[{\cal E},\lambda,\gamma,a\right]_{\bf A}$,
a standing Rankine-Hugoniot shock may accomplish this task.
A supersonic accretion through the outer {\it sonic} point $r_s^{\rm out}$
(which in obtained by integrating the flow starting from the outer 
critical point $r_c^{out}$)
can generate
entropy through such a shock formation and can join the flow passing through
the inner {\it sonic} point $r_s^{\rm in}$ 
(which in obtained by integrating the flow starting from the outer
critical point $r_c^{in}$). Below we will carry on a detail discussion on such 
shock formation.

In this article, the basic equations governing the flow
are the energy and baryon number
conservation equations which contain no dissipative
terms and the flow is assumed to be inviscid.
Hence, the shock
which may be produced in this way can only be of Rankine-Hugoniot type
which conserves energy. The shock thickness must be very small
in this case, otherwise non-dissipative
flows may radiate energy through the upper and the lower boundaries because
of the presence of strong temperature gradient in between the inner and
outer boundaries of the shock thickness.
In the  presence of a shock
the flow may have the following profile.
A subsonic flow starting from infinity first becomes supersonic after crossing
the outer sonic point and somewhere in between the outer sonic point and the inner
sonic point
the shock transition takes place and forces the solution
to jump onto the corresponding subsonic branch. The hot and dense post-shock
subsonic flow produced in this way becomes supersonic again after crossing
the inner sonic point and ultimately dives supersonically into the
black hole.
A flow heading towards a neutron star can have the liberty of undergoing
another shock transition
after it crosses the inner sonic point
\footnote{
Or, alternatively, a shocked flow heading towards a neutron star
need not have to encounter the inner sonic point at all.}, because the hard surface boundary
condition of a neutron star by no means prevents the flow
from hitting the star surface subsonically.

For the complete general relativistic accretion flow discussed in this article,
the energy momentum tensor ${\Im}^{{\mu}{\nu}}$, the four-velocity $v_\mu$,
and the speed of sound $c_s$ may have discontinuities at a
hypersurface $\Sigma$ with its normal $\eta_\mu$.
 Using the energy momentum conservation and the
continuity equation, one has
\begin{equation}
\left[\left[{\rho}v^{\mu}\right]\right] {\eta}_{\mu}=0,
\left[\left[{\Im}^{\mu\nu}\right]\right]{\eta}_{\nu}=0.
\label{eqa51}
\end{equation}
For a perfect fluid, one can thus formulate the relativistic
 Rankine-Hugoniot conditions as
\begin{equation}
\left[\left[{\rho}u\Gamma_{u}\right]\right]=0,
\label{eqa52}
\end{equation}
\begin{equation}
\left[\left[{\Im}_{t\mu}{\eta}^{\mu}\right]\right]=
\left[\left[(p+\epsilon)v_t u\Gamma_{u} \right]\right]=0,
\label{eqa53}
\end{equation}
\begin{equation}
\left[\left[{\Im}_{\mu\nu}{\eta}^{\mu}{\eta}^{\nu}\right]\right]=
\left[\left[(p+\epsilon)u^2\Gamma_{u}^2+p \right]\right]=0,
\label{eqa54}
\end{equation}
where $\Gamma_u=1/\sqrt{1-u^2}$ is the Lorentz factor.
The first two conditions (\ref{eqa52})
and (\ref{eqa53})
are trivially satisfied owing to the constancy of the
specific energy and mass accretion rate.
The constancy of mass accretion yields
\begin{equation}
\left[\left[
K^{-\frac{1}{\gamma-1}}
\left(\frac{\gamma-1}{\gamma}\right)^{\frac{1}{\gamma-1}}
\left(\frac{c_s^2}{\gamma-1-c_s^2}\right)^{\frac{1}{\gamma-1}}
\frac{u}{\sqrt{1-u^2}}
H(r)\right]\right]=0.
\label{eqa55}
\end{equation}
The third Rankine-Hugoniot condition
(\ref{eqa54})
may now be  written as
\begin{equation}
\left[\left[
K^{-\frac{1}{\gamma-1}}
\left(\frac{\gamma-1}{\gamma}\right)^{\frac{\gamma}{\gamma-1}}
\left(\frac{c_s^2}{\gamma-1-c_s^2}\right)^{\frac{\gamma}{\gamma-1}}
\left\{\frac{u^2\left(\gamma-c_s^2\right)+c_s^2}{c_s^2\left(1-u^2\right)}\right\}
\right]\right]=0.
\label{eqa56}
\end{equation}
Simultaneous solution of Eqs. (\ref{eqa55}) and (\ref{eqa56}) yields the `shock invariant'
quantity
\begin{equation}
{\cal S}_h=
c_s^{\frac{2\gamma+3}{\gamma-1}}
\left(\gamma-1-c_s^2\right)^{\frac{3\gamma+1}{2\left(1-\gamma\right)}}
u\left(1-u^2\right)^{-\frac{1}{2}}
\left[\lambda^2v_t^2-a^2\left(v_t-1\right)\right]^{-\frac{1}{2}}
\left[
\frac{u^2\left(\gamma-c_s^2\right)+c_s^2}{c_s^2\left(1-u^2\right)}
\right]
\label{eqa57}
\end{equation}
which changes continuously across the shock surface.
We also define
 the {\em shock strength} ${\cal S}_i$ and the
{\em entropy enhancement} $\Theta$  as the ratio of the pre-shock
to post-shock Mach numbers (${\cal S}_i=M_{-}/M_{+}$),
 and as the ratio of the post-shock to pre-shock
entropy accretion rates ($\Theta={\dot {\Xi}}_{+}/{\dot {\Xi}}_{-}$) of the
flow, respectively.
Hence, $\Theta={\dot {\Xi}}{(r_c^{in})}/{\dot {\Xi}}{r_c^{out}}$
for accretion and $\Theta={\dot {\Xi}}{(r_c^{out})}/{\dot {\Xi}}{r_c^{in}}$
for wind, respectively.

The shock location in
multi-transonic accretion
is found in the following way.
Consider the multi-transonic
flow topology as depicted in the Fig. 4.
Integrating along the solution passing through the outer 
critical point, we calculate the shock invariant
${\cal S}_h$ in addition to
$u$, $c_s$ and $M$. We also calculate  ${\cal S}_h$
while integrating along the solution passing through the inner critical 
point, starting from the inner {\it sonic}
point up to the point of inflexion G.
 We then determine
the radial distance $r_{sh}$, where the numerical values of ${\cal S}_h$,
obtained by integrating the two different sectors described above, are
equal. Generally,
 for
any value of \eker 
allowing shock
formation,  one finds {\it two} shock locations
marked by S1 (the `outer'
shock, formed at 65.31$r_g$ -- between the outer and the middle
sonic points) and
S2 (the `inner' shock, formed at 6.31 $r_g$ -- between the
inner and the middle sonic points) in the figure.
According to a standard
local stability analysis (Yang \& Kafatos 1995),
for a multi-transonic accretion, one can show that
only the shock formed  between
the middle
and the outer sonic point is stable.
The shock strength is {\it different} for the inner and 
for the outer shock. For the stable (outer) shock, the 
shock strength for the case shown in the figure is 5.586,
hence it is a strong shock.
Therefore, in the multi-transonic accretion
with the topology shown in Fig. 4.,
the shock at  S1 is stable and that
at  S2 is unstable.
Hereafter, whenever we mention the shock
location, we  refer
to the stable shock location only.
%\begin{figure}
%\vbox{
%\vskip -6.0cm
%\centerline{
%\psfig{file=figure5.eps,height=20cm,width=17.2cm,angle=0.0}}
%\vskip -5.0cm
%{\bf Figure 5:} Pre- and post-shock disc geometry with thermally 
%driven optically thick halo. See text for further detail.}
%\end{figure}
%
\vskip -4.0cm
\begin{center}
\vskip -0.0cm
\begin{figure}[h]
\includegraphics[scale=0.6,angle=0.0]{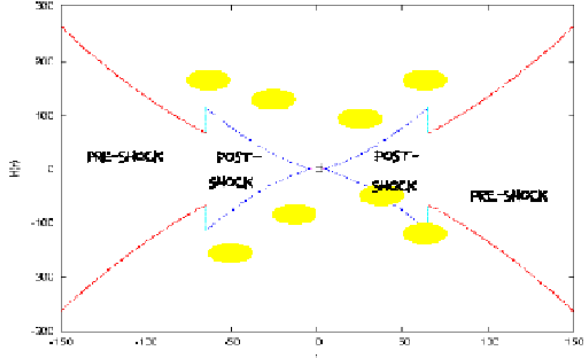}
\vskip -4.0cm
\caption[]{Pre- and post-shock disc geometry with thermally
driven optically thick halo. See text for further detail.}
\label{fig5}
\end{figure}
\end{center}

\subsection{Disc Geometry and Shock Generated Outflow}
\noindent
As a consequence of the shock formation in an accretion flow,
the post-shock flow temperature will also increase abruptly.
The post-
to pre-shock temperature ratio $T_+/T_-$ is, in general,
a sensitive function of \eker.
In Fig. 5., we present the disc structure obtained by
solving (\ref{eq105}) for the
combined shocked accretion flow.
The point {\bf B} represents the black-hole event horizon. The pre- and post-shock
regions of the disc  are clearly distinguished in the figure
and show
that the post-shock disc puffs up significantly. The pre-shock supersonic 
disc is shown by the red lines. The post-shock subsonic part of the 
disc is shown by dotted blue lines and the post-shock supersonic 
part (very close to the event horizon since $r_c^{in}=4.279 r_g$)
is shown by dotted magenta lines (not well resolved in the figure though).
The bulk flow temperature
will be increased in the post-shock region.
Such an  increased disc temperature
may lead to a disc evaporation resulting
in the formation of an optically thick halo,
which are schematically shown using yellow 
coloured elliptic structures.
Besides,
 a strong temperature enhancement  may lead to the
formation of thermally driven outflows.
  The generation of centrifugally
driven and thermally driven outflows from black-hole accretion discs
 has been discussed
 in the
post-Newtonian framework (Das \& Chakrabarti 1999; Das, Rao \& Vadawale 2003)
The post-Newtonian approach
may  be extended to general relativity using
the formalism
presented here.

Owing to the
 very high
 radial component of the
infall velocity of accreting material close to the black hole,
the viscous time scale is much larger than the infall time scale.
Hence, in the vicinity of the black hole,
a rotating inflow
entering the black hole will  have  an almost constant
specific angular momentum for any moderate viscous stress.
This angular momentum
yields a very strong centrifugal force  which
increases much faster than the gravitational force.
These two forces become comparable 
 at
some specific radial distance.
At that point
 the matter starts
piling up and produces a boundary layer supported by the centrifugal pressure,
which may break the inflow to produce the shock.
This actually happens  not quite at
the point where the gravitational and centrifugal forces become equal but
slightly farther out
 owing to the thermal pressure.
Still closer to the black hole, gravity inevitably wins
and matter enters the horizon supersonically after passing
through a sonic point.
The formation of such a layer
may be attributed to the shock formation in accreting fluid.
The  post-shock flow becomes hotter and denser,
and for all practical purposes,
behaves as the stellar atmosphere as far as the formation of
outflows is concerned.
A part of the hot and dense shock-compressed in-flowing material
is then `squirted' as an outflow from the post-shock region.
Subsonic outflows originating
from the puffed up
hotter post-shock accretion disc (as shown in the figure)
pass through the outflow sonic points and reach large distances
as in a wind solution.

The generation of such shock-driven outflows
is a reasonable assumption. A calculation describing the change
of linear momentum of the accreting material in the direction perpendicular to the
plane of the disc is beyond the scope of the disc model described in
this article because the explicit variation of dynamical variables along the Z axis
(axis perpendicular to the equatorial plane of the
disc)
cannot be treated
analytically.
The enormous post-shock thermal pressure
is capable of providing a substantial amount of `hard push' to the accreting
material against the gravitational attraction of the black hole. This `thermal
kick' plays an important role in re-distributing the linear momentum of the
inflow and generates a non-zero component along the Z direction.
In other words,
the thermal pressure at the post-shock region,
being anisotropic in nature, may deflect a part of the inflow
perpendicular to the equatorial plane of the disc.
Recent work shows that (Moscibrodzka, Das \& Czerny 2006)
such shock-outflow model can
be applied to successfully investigate the origin and dynamics of
the strong X-ray flares emanating out from our galactic centre.

\subsection{Multi-transonic Wind}
\noindent
The blue coloured wedge shaped region marked by 
{\bf W} represents the \eker zone 
for which
three critical points, the inner, the middle and the outer are also found.
However, in contrast to $\left[{\cal E},\lambda,\gamma,a\right]{\in}
\left[{\cal E},\lambda,\gamma,a\right]_{\bf A}$,
the set $\left[{\cal E},\lambda,\gamma,a\right]{\in}
\left[{\cal E},\lambda,\gamma,a\right]_{\bf W}$
yields  solutions
for which
${\dot \Xi}(r_c^{\rm in})$ is {\it less} than
${\dot \Xi}(r_c^{\rm out})$.
Besides,
the topological flow profile of
these solutions
is   different.
Here the closed loop-like structure is formed through the {\it outer} critical point.
One such solution topology is presented 
in Fig. 6 for a specific set of 
\eker as shown in the figure. The same colour-scheme which has been 
used to denote various accretion and wind branches (through various critical 
points) of multi-transonic accretion (Fig. 4.), has been used here as
well to manifest how the loop formation switches from flow through 
$r_c^{in}$ (multi-transonic
accretion, Fig. 4.) to flow through $r_c^{out}$ (multi-transonic wind, Fig. 6.).
This  topology
is interpreted in the following way.
The flow (blue dashed line marked by `a') passing through the inner critical point
(3.456 $r_g$) is the complete mono-transonic
accretion flow, and the dotted magenta line marked by `w'
is its corresponding wind solution. The solutions 
passing through
the outer critical point (3307.318 $r_g$),
represents the incomplete accretion (solid red line marked by `A')/wind (dashed
green line marked by `W') solution.
 However, as
${\dot \Xi}(r_c^{in})$ turns out to be less than
${\dot \Xi}(r_c^{out})$, the wind solution through $r_c^{in}$ 
can make
a shock transition to join
its counter wind solution passing through $r_c^{out}$,
and thereby
increase the entropy accretion rate by the amount
${\Delta}{\dot \Xi}={\dot \Xi}(r_c^{out})-{\dot \Xi}(r_c^{in})$.
Here the numerical values of ${\cal S}_h$ along the
wind  solution passing through the inner critical point
are compared with the numerical values of ${\cal S}_h$
along the wind solution passing through the outer
critical point, and the shock locations  S1
and  S2 for the wind are found accordingly.
Here also, two theoretical shock locations are
obtained, which are shown by dot dashed azure vertical 
lines marked by S1 (at 649.41 $r_g$) and 
S2 (at 6.42 $r_g$), out of which only one is stable.
The shock strength corresponding to the stable outer shock 
can be calculated to be 20.24. Hence 
{\it extremely} strong shocks are formed for multi-transonic wind in general.
A part of the region $\left[{\cal E},\lambda,\gamma,a\right]_{\bf W}$
thus corresponds to  {\it mono-transonic} accretion
solutions  with
multi-transonic wind solutions with a shock.

Besides  $\gamma=4/3$ and $a=0.3$, for which Fig. 3.  has been drawn, one can perform 
a similar classification for any astrophysically relevant value of
$\gamma$ and $a$ as well. Some characteristic features of $\ptw$ would
be changed as $\gamma$ is being varied. For example, if
${\cal E}_{\rm max}$ is the maximum value
of the energy and if $\lambda_{\rm max}$ and $\lambda_{\rm min}$ are the maximum and
the minimum values of the angular momentum, respectively,
for $\ptw_{\bf A}$ for a fixed value of
$\gamma$,
then
$\left[{\cal E}_{\rm max},\lambda_{\rm max},\lambda_{\rm min}\right]$
anti-correlates with $\gamma$.
Hence,  as the flow makes a transition
from its ultra-relativistic
to its purely non-relativistic limit,
the area representing $\ptw_{\bf A}$
decreases.
%\begin{figure}
%\vbox{
%\vskip -0.0cm
%\centerline{
%\psfig{file=figure6.ps,height=11cm,width=16.2cm,angle=270.0}}
%\vskip -0.0cm
%{\bf Figure 6:} Solution topology for multi-transonic wind
%in Kerr geometry for a specific set of \eker as shown in the figure.
%See text for detail.}
%\end{figure}
\begin{center}
\begin{figure}[h]
\includegraphics[scale=0.6,angle=270.0]{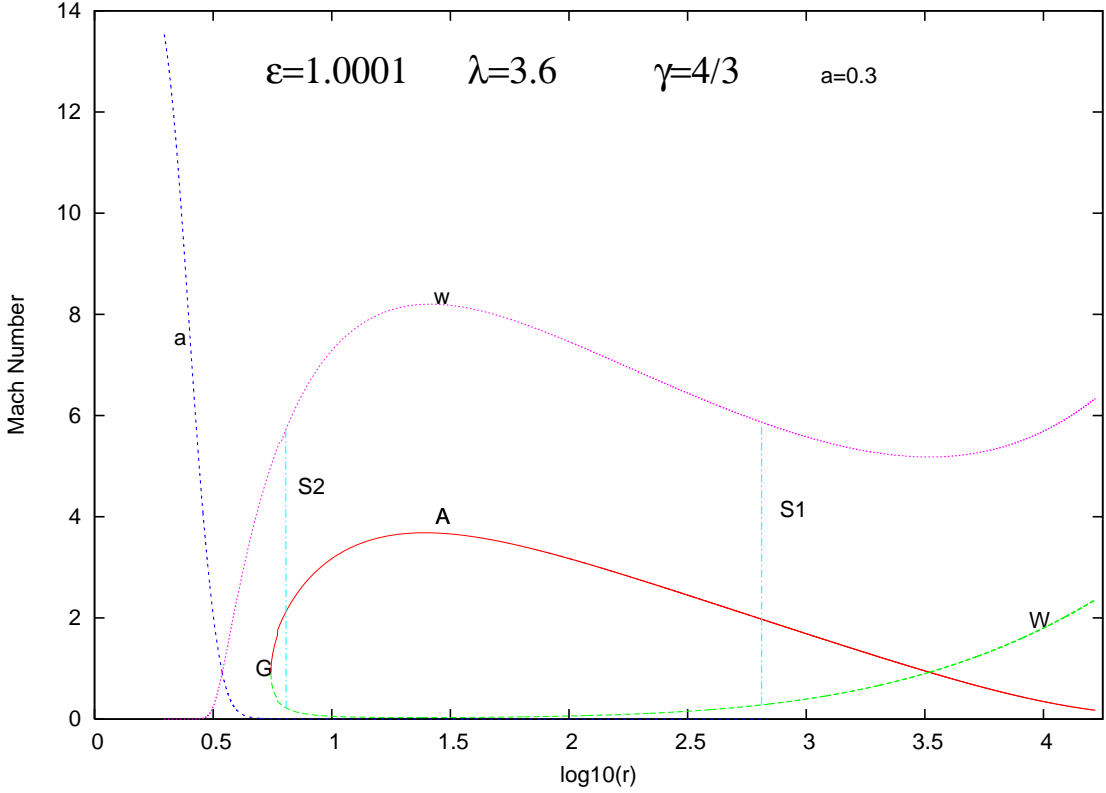}
\caption[]{Solution topology for multi-transonic wind
in Kerr geometry for a specific set of \eker as shown in the figure.
See text for detail.}
\label{fig6}
\end{figure}
\end{center}
\subsection{Dependence of Shock Location on Accretion Parameters}
\noindent
One finds that the shock location  correlates with
$\lambda$.
This is obvious because the higher the flow
angular momentum, the greater the rotational energy content
of the flow. As a consequence, the strength of the centrifugal
barrier which is responsible to break the incoming flow by forming a shock
will be higher and
the location of such a barrier will be farther away from the
event horizon.
However, the shock location
anti-correlates with ${\cal E}$ and $\gamma$.
This  means that for the same ${\cal E}$ and $\lambda$, in the purely
non-relativistic flow the shock
will form closer to the black hole compared with
the ultra-relativistic flow. Besides, we find that the shock strength
${\cal S}_i$ anti-correlates with the shock location $r_{sh}$,
which indicates that
the
closer to the black hole the shock forms , the higher the strength ${\cal S}_i$
and the entropy enhancement ratio $\Theta$ are. 
The ultra-relativistic flows
are supposed to
produce the strongest shocks.
The reason behind this is also easy to understand. The closer to the black hole the shock
forms, the higher the available gravitational
potential energy must be released, and the radial
advective velocity required to have a more vigorous shock jump will be larger.
Besides we note  that as  the flow gradually approaches its purely
non-relativistic limit,
the shock may form for lower and lower angular momentum,
which indicates that for purely non-relativistic
accretion, the shock formation may take place even for a quasi-spherical flow.
However, it is important to mention that
a shock formation will be allowed
not for every 
$\left[{\cal E},\lambda,\gamma,a\right]{\in}
\left[{\cal E},\lambda,\gamma,a\right]_{\bf A}$,
Equation (\ref{eqa57}) will be satisfied
only
for a specific subset of $\left[{\cal E},\lambda,\gamma,a\right]_{\cal A}$,
for which a steady, standing shock solution
will be found.

\subsection{Analogue Temperature}
The surface gravity is defined according to (\ref{eq33}).
For axisymmetric accretion described in the above sections, one can 
calculate that (Abraham, Bili\'c \& Das 2006; Das, Bili\'c \& Dasgupta 2006)
\begin{equation}
{\sqrt{-\chi^{\mu}\chi_{\mu}}}=
\frac
{r\sqrt{\Delta{B}}}
{r^3+a^2r+2a^2-2\lambda{a}}
\label{eq118}
\end{equation}
where $B$ can be defined as
\begin{equation}
B=g_{\phi\phi}+2\lambda{g_{t\phi}}+\lambda^2{g_{tt}}
\label{eq119}
\end{equation}

Since 
\begin{equation}
\frac{\partial}{\partial{\eta}}{\equiv}{\eta^\mu}{\partial_{\mu}}
=\frac{1}{\sqrt{g_{rr}}},
\label{eqan1}
\end{equation}
the expression for the analogue temperature can be calculated as
\begin{equation}
T_{AH}=\frac{\hbar}{2{\pi}{\kappa_B}}
\sqrt{1-\frac{2}{r_h}+\left(\frac{a}{r_h}\right)^2}
\frac{r_h{\zeta_1}\left(r_h,a,\lambda\right)}{{\zeta_1}\left(r_h,a,\lambda\right)}
\left|\frac{1}{1-c_s^2}\frac{d}{dr}\left(u-c_s\right)\right|_{\rm r=r_h}
\label{eqan2}
\end{equation}
where
\begin{eqnarray}
{\zeta_1}\left(r_h,a,\lambda\right)=
\sqrt{\frac{{\zeta_{11}}\left(r_h,a,\lambda\right)}{{\zeta_{12}}\left(r_h,a,\lambda\right)}},~
{\zeta_2}\left(r_h,a,\lambda\right)=r_h^3+a^2r_h+2a^2-2\lambda{a},~
& & \nonumber \\
{\zeta_{12}}\left(r_h,a,\lambda\right)=r_h^4+r_h^2a^2+2r_ha^2,
& & \nonumber \\
{\zeta_{11}}\left(r_h,a,\lambda\right)=
\left(r_h^2-2r_h+a^2\right)
[r_h^6+r_h^5\left(2a^2-\lambda\right)+2r_h^4\left(2a^2-2\lambda{a}+\lambda\right)+
& & \nonumber \\
r_h^3\left(a^4-\lambda{a^2}\right)+
%& & \nonumber \\
2r_h^2a\left(a-2\lambda+1\right)+
4r_h\left(a^2-2\lambda{a}+\lambda\right)]
\label{eqan3}
\end{eqnarray}

Using (\ref{eq109}-\ref{eq112}, \ref{eq113}-\ref{eq113a}), along with the expression for 
$(dc_s/dr)$ at $r_c$, one can calculate the location of the 
acoustic horizon (the flow {\it sonic} point), and the value of $c_s,du/dr$ and $dc_s/dr$
at the acoustic horizon, by integrating the flow from the 
critical point upto the acoustic horizon (sonic point). Such 
values can be implemented in the expression for $T_{AH}$ in
(\ref{eqan2}-\ref{eqan3}) to 
calculate the analogue temperature. 
The ratio $\tau=T_{AH}/T_H$ can also be calculated accordingly.

One can calculate the analogue temperature for the following five different categories of 
{\it accretion} flow all together, since we are not interested at this moment to study the
analogue effects in wind solutions:
\begin{enumerate}
\item Mono-transonic flow passing through the
single inner type critical/sonic point.
The range of \eker used
to obtain the result for this region corresponds to the 
region of Fig. 3  marked by {\bf I}.
\item Mono-transonic flow passing through the
single outer type critical/sonic point. 
The range of \eker used
to obtain the result for this region corresponds to the
region of Fig. 3  marked by {\bf O}.
\item Multi-transonic accretion passing through the
{\it inner} critical/sonic point. The range of \eker used
to obtain the result for this region corresponds to the
region of Fig. 3  marked by {\bf A}.
\item Multi-transonic accretion passing through the
{\it outer} critical/sonic point. The range of \eker used
to obtain the result for this region corresponds to the
region of Fig. 3  marked by {\bf A}.
\item Mono-transonic accretion passing through the inner 
critical/sonic point for the multi-transonic {\it wind} zone.
The range of \eker used
to obtain the result for this region corresponds to the
region of Fig. 3  marked by {\bf W}.
\end{enumerate}

In this section we would mainly like to concentrate 
to the study the dependence of $T_{AH}$ on the Kerr parameter $a$, also, we would
like to demonstrate that for some values of \eker, the analogue temperature may 
be comparable to the actual Hawking temperature. Hence we are interested in the 
region of \eker for which $\tau$ can have a value as large as possible.
We found that large value of $\tau$ can be 
obtained only for very high energy flow with large 
value of the adiabatic index. Such an almost purely nonrelativistic hot 
accretion does {\it not} produce multi-transonicity, it produce only 
mono-transonic flow passing through the inner type critical/sonic point.
Hence in the figure 7, we show the variation of $\tau$ with $a$ for 
a specific value of $\left[{\cal E},\lambda,\gamma\right]$ (as shown 
in the figure) for which 
$\left[{\cal E},\lambda,\gamma\right]{\in}\left[{\cal E},\lambda,\gamma\right]_{\bf I}$.
However, same $\tau-a$ figures can be drawn for any \eker taking for any of the 
other four categories of accretion mentioned above. 

In figure 7, the ratio of the analogue to the actual Hawking temperature $\tau$ 
has been plotted along the $Y$ axis, while the black hole spin parameter (the 
Kerr parameter $a$) has been plotted along the $X$ axis. 
It is obvious from the figure that there exists a preferred value of the
black hole spin parameter for which the acoustic surface gravity attains its
maximum value. Location of such a peak in the 
$\tau - a$ graph, i.e., the preferred value of the Kerr parameter which 
maximizes the surface gravity, sensitively depends on ${\cal E},\lambda$ 
and $\gamma$,
see Barai \& Das 2007 for further 
details. 
This is an extremely important finding 
since it manifest the fact that {\it the black hole spin angular 
momentum does influence the analogue gravity effect}, and tells 
how the background (fluid) metric influences the 
perturbative (acoustic) metric. 
Note that $\tau>1$ is possible to obtain for an extremely large value of 
${\cal E}$ having the adiabatic index almost equal to its purely 
non-relativistic limit ($\gamma=5/3$).
%\begin{figure}
%\vbox{
%\vskip -0.5cm
%\centerline{
%\psfig{file=figure7.ps,height=11cm,width=16.2cm,angle=270.0}}
%\vskip -0.0cm
%{\bf Figure 7:} Variation of the ratio of analogue to the actual Hawking temperature
%$\tau$ with the black hole spin angular momentum (the Kerr parameter $a$).} 
%\end{figure}
\begin{center}
\begin{figure}[h]
\includegraphics[scale=0.6,angle=270.0]{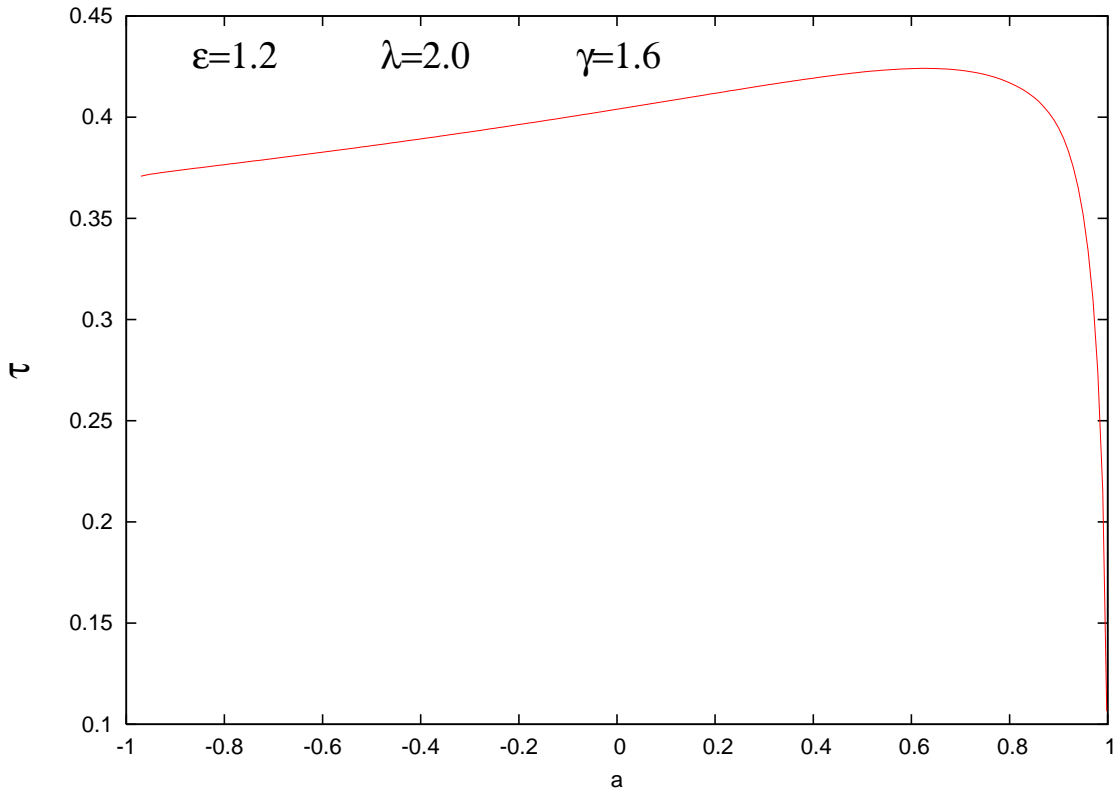}
\caption[]{Variation of the ratio of analogue to the actual Hawking temperature
$\tau$ with the black hole spin angular momentum (the Kerr parameter $a$).}
\label{fig7}
\end{figure}
\end{center}

As mentioned earlier, the discriminant ${\cal D}$ of the corresponding 
acoustic metric changes sign whenever the state of transonicity of the
flow flips from sub- to super-sonicity or vice versa. For multi-transonic 
shocked accretion flow, such state flipping occurs three times, first, from 
a sub-sonic to a supersonic state at the outer sonic point (the outer 
acoustic horizon $r_h^{out}$), then from the supersonic to the subsonic state
at the shock location through the Rankine-Hugoniot shock, and then from the 
subsonic to the supersonic state again at the inner sonic point (the inner 
acoustic horizon $r_h^{in}$). A transition from ${\cal D}<0$ (subsonic
flow) to ${\cal D}>0$ (supersonic flow) produces an acoustic black hole, 
while the reverse transition (${\cal D}>0{\longrightarrow}{\cal D}<0$) 
produces an acoustic {\it white hole} (Bercelo, Liberati, Sonego \& Visser 2004;
Abraham, Bili\'c \& Das 2006). It is thus obvious that for multi-transonic accretion 
encountering a stable shock, two acoustic black holes are formed, one at the inner and the 
other at the outer acoustic horizons (the inner and outer sonic points), and an 
acoustic {\it white hole} is produced at the shock. For relativistic 
accretion disc with constant thickness,
this has formally been demonstrated (Abraham, Bili\'c \& Das 2006)
by computing the value of ${\cal D}$ for the whole span of $r$ ranging from infinity to 
the event horizon to demonstrate that the ${\cal D}<0{\longrightarrow}{\cal D}>0$ 
transition indeed takes place at $r_s^{in}$ and at $r_s^{out}$, and 
${\cal D}>0{\longrightarrow}{\cal D}<0$ transition takes place at the shock location.
Similar calculation can also be performed for the disc geometry 
with the specific form of disc height (\ref{eq105}) used in this work.
\section{Black Hole Accretion in `Modified' Newtonian Potentials}
\noindent
Rigorous investigation of the complete general relativistic
multi-transonic black hole accretion disc
structure is extremely complicated. 
At the same time it is
understood that, as relativistic effects play an important role in the
regions close to the accreting black hole (where most of the
gravitational potential energy is released), purely Newtonian gravitational
potential 
cannot be a realistic choice to describe
transonic black hole accretion in general. To compromise between the ease of
handling of a
Newtonian description of gravity and the realistic situations
described by complicated general relativistic calculations, a series of
`modified' Newtonian potentials have been introduced
to describe the general relativistic effects that are
most important for accretion disk structure around Schwarzschild and Kerr
black holes (see Artemova, Bj\"{o}rnsson \& Novikov 1996;
Das 2002, and references therein for further discussion).

Introduction of such potentials allows one to investigate the
complicated physical processes taking place in disc accretion in a
semi-Newtonian framework by avoiding pure general relativistic calculations
so that
most of the features of spacetime around a compact object are retained and
some crucial properties of the analogous relativistic
solutions of disc structure could be reproduced with high accuracy.
Hence, those potentials might be designated as `pseudo-Kerr' or `pseudo-
Schwarzschild' potentials, depending on whether they are used to mimic the
space time around a rapidly rotating or non rotating/ slowly rotating
(Kerr parameter $a\sim0$) black
holes respectively.
Below we describe four such pseudo Schwarzschild potentials on
which we will concentrate in this article.
In this section, as well as in the following sections, we will 
use the value of $r_g$ to be equal to $2GM_{BH}/c^2$.

It is important to note that 
as long as one is not
interested in astrophysical processes extremely close
(within $1-2~r_g$) to a black hole horizon, one may safely
use the following black hole potentials to study
accretion on to a Schwarzschild
black hole with the advantage that use of these
potentials would simplify calculations by allowing one
to use some basic features of flat geometry
(additivity of energy or de-coupling of various
energy components, i.e., thermal ($\frac{c_s^2}{\gamma-1}$),
Kinetic ($\frac{u^2}{2}$) or
gravitational ($\Phi$) etc., see subsequent discussions)
which is not possible for
calculations in a purely Schwarzschild or a Kerr metric.
Also, one
can study more complex many body problems such as
accretion from an ensemble of companions or overall
efficiency  of accretion onto an ensemble of black holes
in a galaxy  or  for studying numerical hydrodynamic accretion flows
around a black hole etc. as simply as can be done in a
Newtonian framework, but with far better
accuracy. So a comparative study of multi-transonic 
accretion flow using all these 
potentials might be quite useful in understanding some 
important features of the analogue properties of astrophysical
accretion.

Also, one of the main `charms' of the classical 
analogue gravity formalism is that even if the governing equations for
fluid flow is completely non-relativistic (Newtonian), the
propagation of acoustic fluctuations embedded into it are 
described by a curved pseudo-Riemannian geometry. In connection to 
astrophysical accretion, one of the
best ways to manifest such interesting effect 
would be to study the analogue effects in the Newtonian and post-Newtonian accretion 
flow.
However, one should be careful in using these
potentials because none of these potentials discussed in the subsequent paragraphs
are `exact' in a sense that they are not directly
derivable from the Einstein equations.
These potentials
could only be used to obtain more
accurate correction terms over and above the pure
Newtonian results and any `radically' new results
obtained using these potentials should be cross-checked
very carefully with the exact general relativistic theory.

Paczy\'nski and Wiita (1980) proposed a pseudo-schwarzschild
potential of the form
\begin{equation}
\Phi_{1}=-\frac{1}{2(r-1)}
\label{tt1}
\end{equation}
which accurately reproduces the positions of the marginally stable orbit $r_s$ 
and the marginally bound orbit $r_b$, 
and provides the value 
of efficiency to be $-0.0625$, which is in close agreement 
with the value obtained in full general relativistic calculations. 
Also the Keplarian distribution of angular 
momentum obtained using this potential is exactly same as 
that obtained in pure 
Schwarzschild geometry. 
It is worth mentioning here that this potential 
was first introduced to study a thick accretion disc with super Eddington 
Luminosity. Also,
it is interesting to note that although it had been thought of
in 
terms of disc accretion, $\Phi_1$ 
is spherically symmetric with a scale shift of 
$r_g$.

To analyze the normal modes of acoustic oscillations within a 
thin accretion 
disc around a compact object (slowly rotating black hole or weakly 
magnetized neutron star), Nowak and Wagoner (1991) approximated some of the 
dominant relativistic effects of the accreting 
black hole (slowly rotating or 
non-rotating) via a modified Newtonian potential of the form
\begin{equation}
\Phi_{2}=-\frac{1}{2r}\left[1-\frac{3}{2r}+12{\left(\frac{1}{2r}\right)}^2\right
]
\label{tt2}
\end{equation}
$\Phi_2$ has correct form of $r_s$ as in the Schwarzschild case
but is unable to 
reproduce the value of $r_b$.
 This potential has the correct general relativistic value of the
angular velocity $\Omega_s$ 
at $r_s$. Also it reproduces the
radial epicyclic frequency $i\nu_\kappa$ (for $r>r_s$) close to its value obtained
from general relativistic calculations, 
and among all black hole potentials, $\Phi_2$ provides the best approximation for
$\Omega_s$ and $\nu_\kappa$.
However, this potential gives the
value of efficiency as $-0.064$ which is larger than that produced by 
$\Phi_1$, hence the disc spectrum computed using $\Phi_2$ would be more 
luminous compared to a disc structure studied using $\Phi_1$.

Considering the fact that the free-fall acceleration plays a very crucial 
role in Newtonian gravity, Artemova, Bj\"{o}rnsson \& Novikov (1996)
proposed two different 
black hole potentials to study disc accretion around a non-rotating black hole.
The first potential proposed by them produces exactly the
same value of the free-fall
acceleration of a test particle at a given value of $r$ as is obtained
for a test particle at rest with respect to the Schwarzschild reference
frame, and is given by
\begin{equation}
\Phi_{3}=-1+{\left(1-\frac{1}{r}\right)}^{\frac{1}{2}}
\label{tt3}
\end{equation}
The second one gives the value of the free fall acceleration that is equal 
to the value of the covariant component of the three dimensional free-fall 
acceleration vector of a test particle that is at rest in the Schwarzschild 
reference frame and is given by
\begin{equation}
\Phi_{4}=\frac{1}{2}ln{\left(1-\frac{1}{r}\right)}
\label{tt4}
\end{equation}
Efficiencies produced by $\Phi_3$ and $\Phi_4$ are $-0.081$ and $-0.078$ 
respectively.The magnitude of efficiency produced by $\Phi_3$
being 
maximum,calculation of disc structure using $\Phi_3$
will give  the maximum 
amount of energy dissipation and the corresponding spectrum would be the 
most luminous one. 
Hereafter we will refer to 
all these four potentials by $\Phi_i$ in 
general, where $\left\{i=1,2,3,4\right\}$ would correspond to $\Phi_1$
(\ref{tt1}), $\Phi_2$ (\ref{tt2}), $\Phi_3$ (\ref{tt3}) and $\Phi_4$ (\ref{tt4})
respectively.
One should notice that while all other $\Phi_i$ have
singularity at $r=r_g$, only $\Phi_2$ has a singularity at $r=0$.
%\begin{figure}
%\vbox{
%\vskip -5.8cm
%\centerline{
%\psfig{file=figure8.eps,height=15cm,width=15cm}}}
%\noindent {{\bf Fig. 8:}
%Newtonian potential and other
%pseudo-potentials $\Phi_i(r)$ ($i=1,2,3,4$) are plotted as a function of
%the logarithmic radial distance from the accreting black hole. This figure is 
%reproduced from Das \& Sarkar 2001.
%}
%\end{figure}
%
\vskip -3.0cm
\begin{center}
%\vskip -3.0cm
\begin{figure}[h]
\includegraphics[scale=0.8,angle=0.0]{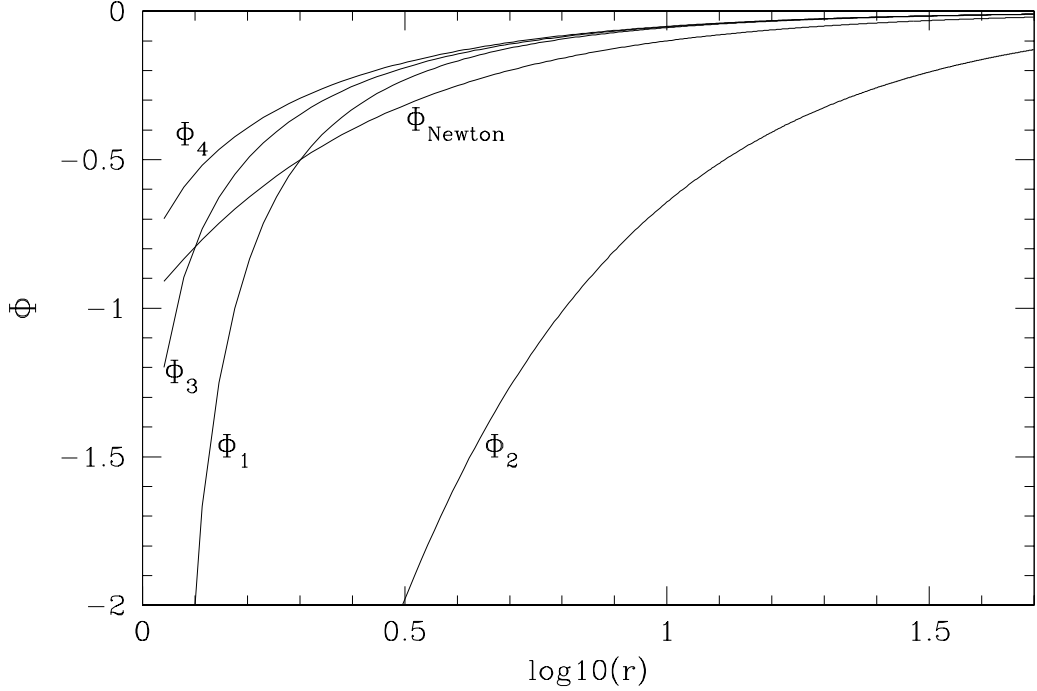}
\caption[]{Newtonian potential and other
pseudo-potentials $\Phi_i(r)$ ($i=1,2,3,4$) are plotted as a function of
the logarithmic radial distance from the accreting black hole. This figure is
reproduced from Das \& Sarkar 2001.}
\label{fig8}
\end{figure}
\end{center}

In figure 8 (reproduced from Das \& Sarkar 2001), 
we plot various $\Phi_i$ as a function of the radial distance
measured from the accreting black hole in units of $r_g$. Also in the same plot,
purely Newtonian potential is plotted.
If we now define a quantity ${\bf S}_i$ to be the `relative stiffness'
of a potential $\Phi_i$ as:
$$
{\bf S}_i=\frac{\Phi_i}{r}
$$
(that is, ${\bf S}_i$ is a measure of the numerical value of any $i$th
potential at a radial distance $r$), we find that for $r~>~2r_g$,
$$
{\bf S}_2~<~{\bf S}_{\rm N}~<~{\bf S}_1~<~{\bf S}_3~<~{\bf S}_4,
$$
which indicates that while $\Phi_2$ is a `flatter' potential compared to the
pure Newtonian potential $\Phi_{\rm N}$, all other pseudo potentials are
`steeper' to  $\Phi_{\rm N}$ for $r~>~2r_g$. \\
\noindent
One can write the modulus of free fall
acceleration obtained from all `pseudo' potentials except for $\Phi_2$
in a compact form as
\begin{equation}
\left|{{{{{\Phi}^{'}}_{i}}}}\right|=\frac{1}{2{r^{2-{\delta}_{i}}
{\left(r-1\right)}^{\delta_{i}}}}
\label{tt5}
\end{equation}
where ${\delta_{1}}=2$, $\delta_3=\frac{1}{2}$ and $\delta_4=1$.
$\left|{{{{{\Phi}^{'}}_{i}}}}\right|$
denotes the absolute value of the
space derivative of $\Phi_i$, i.e.,
$$
\left|{{{{{\Phi}^{'}}_{i}}}}\right|=\left|{\frac{d{\Phi_i}}{dr}}\right|
$$
whereas acceleration produced by $\Phi_2$ can be computed as,
\begin{equation}
{\Phi_2}^{'}=\frac{1}{2r^2}\left(1-\frac{3}{r}+\frac{9}{2r^2}\right)
\label{tt6}
\end{equation}

For axisymmetric accretion,
at any radial distance $r$ measured from the accretor,
one can define the effective potential $\Phi_i^{eff}(r)$
to be the summation of the gravitational
potential and the centrifugal potential for matter
accreting under the influence of $i$th pseudo
potential. $\Phi_i^{eff}(r)$ can be expressed as:
\begin{equation}
\Phi_i^{eff}(r)=\Phi_i(r)+\frac{\lambda^2(r)}{2r^2}
\label{tt7}
\end{equation}
where $\lambda(r)$ is the non-constant distance dependent
specific angular momentum of accreting material. One
then easily shows that $\lambda(r)$ may have an upper limit:
\begin{equation}
\lambda^{up}_i(r)=r^{\frac{3}{2}}\sqrt{\Phi^{'}_i(r)}
\label{tt8}
\end{equation}
where $\Phi^{'}_i(r)$ represents the derivative of $\Phi_i(r)$ with
respect to $r$.
For weakly viscous or inviscid flow, angular
momentum can be taken as a constant parameter ($\lambda$) and (\ref{tt7})
can be approximated as:
\begin{equation}
\Phi_i^{eff}(r)=\Phi_i(r)+\frac{\lambda^2}{2r^2}
\label{tt9}
\end{equation}
For general relativistic treatment of accretion, the
effective potential can not be decoupled in to its
gravitational and centrifugal components. 
The  general
relativistic effective potential $\Phi^{eff}_{GR}(r)$ (excluding
the rest
mass) experienced by the fluid accreting on to a Schwarzschild black hole 
can be expressed as:
\begin{equation}
\Phi^{eff}_{GR}(r)=r\sqrt{\frac{r-1}{r^3-{\lambda}^2\left(1+r\right)}}-1
\label{tt10}
\end{equation}
One can understand that the effective potentials in
general relativity cannot be obtained by linearly combining its
gravitational and rotational contributions because
various energies in general relativity are combined together to produce
non-linearly coupled new terms.\\
%\begin{figure}
%\vbox{
%\vskip -5.8cm
%\centerline{
%\psfig{file=figure9.eps,height=15cm,width=15cm}}}
%\noindent {{\bf Fig. 9:}
% Effective black hole  potentials for general relativistic
%($\Phi_{BH}^{eff}(r)$) as well as for pseudo-general
%relativistic($\Phi^{eff}_i(r)$)
% accretion discs as a function of the distance
%(measured from the event horizon in units or $r_g$) plotted
%in logarithmic scale. The specific angular momentum  is   chosen to
%be 2 in geometric units. The figure is reproduced from Das 2002.}
%\end{figure}
\begin{center}
\begin{figure}[h]
\includegraphics[scale=0.9,angle=0.0]{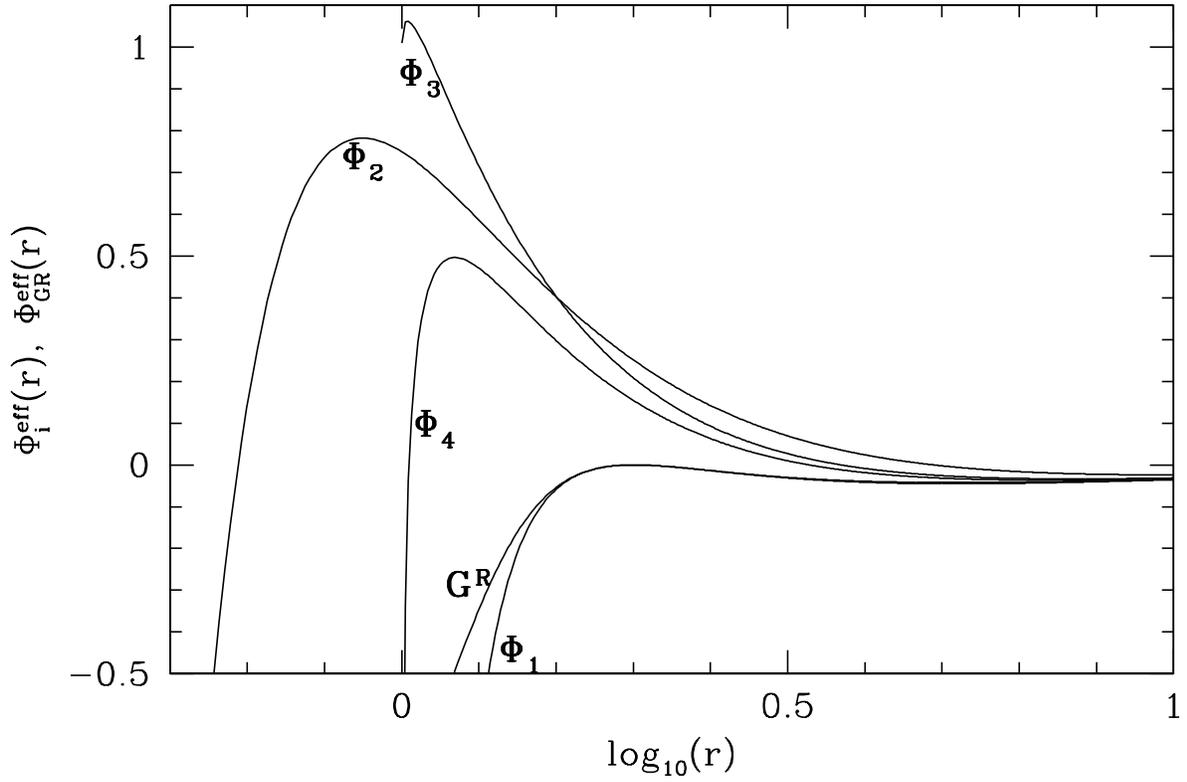}
\caption[]{% Effective black hole  potentials for general relativistic
($\Phi_{BH}^{eff}(r)$) as well as for pseudo-general
relativistic($\Phi^{eff}_i(r)$)
 accretion discs as a function of the distance
(measured from the event horizon in units or $r_g$) plotted
in logarithmic scale. The specific angular momentum  is   chosen to
be 2 in geometric units. The figure is reproduced from Das 2002.}
\label{fig9}
\end{figure}
\end{center}
\noindent
In Fig. 9. (reproduced from Das 2002), we plot $\Phi_i^{eff}(r)$ (obtained from (\ref{tt9})) and 
$\Phi^{eff}_{GR}(r)$ as a function of $r$ in logarithmic scale. The value of
$\lambda$ is taken to be 2
in units of $2GM/c$. $\Phi^{eff}$ curves for different $\Phi_i$
are marked exclusively in the
figure and the curve marked by ${\bf G^R}$ represents the
variation of $\Phi^{eff}_{GR}(r)$ with $r$. 
One can observe that $\Phi^{eff}_1(r)$ is in
excellent agreement with $\Phi^{eff}_{GR}(r)$.
Only for a very small
value of $r$ ($r{\rightarrow}r_g$),
$\Phi^{eff}_1$
starts deviating from $\Phi^{eff}_{GR}(r)$ and this deviation keeps
increasing as matter approaches closer and closer to
the event horizon. All other $\Phi^{eff}_i(r)$s
approaches to $\Phi^{eff}_{GR}(r)$ at a
radial distance (measured from the black hole) considerably
larger compared to the case for $\Phi^{eff}_1(r)$. If one defines 
${\Delta}_i^{eff}(r)$
to be the measure of the deviation of $\Phi^{eff}_i(r)$ with
$\Phi^{eff}_{GR}(r)$
at any
point $r$, 
$$
{\Delta}_i^{eff}(r)=\Phi^{eff}_i(r)-\Phi^{eff}_{GR}(r)
$$
One observes that ${\Delta}_i^{eff}(r)$ is always negative for 
$\Phi^{eff}_1(r)$,
but for other $\Phi^{eff}_i(r)$, it normally remains positive for
low values of $\lambda$ but may become negative for a very
high value of $\lambda$. If 
%${\Bigg{\vert}}{\Delta}^i_{eff}(r){\Bigg{\vert}}$
${{\vert}}{\Delta}_i^{eff}(r){{\vert}}$
be the modules or the absolute
value of ${\Delta}_i^{eff}(r)$, one can also see that, although only for a
very small range of radial distance very close to the event horizon, 
${\Delta}_3^{eff}(r)$ is maximum,
for the whole range of distance scale while $\Phi_1$ is the
best approximation of general relativistic space time,
$\Phi_2$ is the worst approximation and $\Phi_4$ and $\Phi_3$ are the
second and the third best approximation as long as the
total effective potential experienced by the accreting
fluid is concerned. It can be shown that 
${{\vert}}{\Delta}_i^{eff}(r){{\vert}}$ nonlinearly
anti-correlates with $\lambda$. The reason behind this is
understandable. As $\lambda$ decreases, rotational mass as
well as its coupling term with gravitational mass
decreases for general relativistic accretion
material while for accretion in any $\Phi_i$, centrifugal
force becomes weak and gravity dominates; hence
deviation from general relativistic case will be more
prominent because general relativity is basically a
manifestation of strong gravity close to the compact
objects.\\
\noindent
From the figure it is clear that for $\Phi^{eff}_{GR}(r)$
as well as for
all $\Phi^{eff}_i(r)$, a peak appears close to the horizon. The
height of these peaks may roughly be considered as the
measure of the strength of the centrifugal barrier
encountered by the accreting material for respective
cases. The deliberate use of the word `roughly' instead of
`exactly' is due to the fact that here we are dealing
with fluid accretion, and unlike particle dynamics, the
distance at which the strength of the centrifugal
barrier is maximum, is located further away from the
peak of the effective potential because here the total
pressure contains the contribution due to fluid or
`ram' pressure also. Naturally the peak height for  $\Phi^{eff}_{GR}(r)$
as well as for $\Phi^{eff}_i(r)$ increases with increase of $\lambda$ and
the location of this barrier moves away from the black hole 
with higher values of angular momentum. If the
specific angular momentum of accreting material lies
between the marginally bound and marginally stable
value, an accretion disc is formed. For inviscid or
weakly viscous flow, the higher will be the value of $\lambda$,
the higher will be the strength of the centrifugal
barrier and the more will be the amount of radial
velocity or the thermal energy that the accreting material 
must have to begin with so that it can be made to accrete
on to the black hole. In this connection it is important to
observe from the figure that accretion under $\Phi_1(r)$ will
encounter a centrifugal barrier farthest away from the
black hole compared to other $\Phi_i$.  For accretion under all $\Phi_i$s
except $\Phi_1$,the strength of centrifugal barrier at a
particular distance will be more compared to its value
for full general relativistic accretion.

In subsequent sections, we will use the above mentioned potentials
to study the analogue effects in spherically symmetric and in 
axisymmetric black hole accretion.
\section{Newtonian and Post-Newtonian Spherical Accretion as
an Analogue Model}
In this section,
we  study the analogue gravity phenomena in the spherical
accretion onto astrophysical black holes under the influence of
Newtonian as well as
 various post-Newtonian
pseudo-Schwarzschild potentials described above.
We  use the expressions `post-Newtonian' and `pseudo-Schwarzschild'
synonymously.
Our main goal is to provide a self-consistent calculation of
the analogue horizon temperature $T_{\rm AH}$
 in terms of the minimum number of
physical accretion parameters, and to study the dependence of $T_{\rm AH}$
on various flow properties. This section is largely based on Dasgupta,
Bili\'c \& Das 2005.
\subsection{Equation of Motion}
\noindent
The non-relativistic equation of motion  for spherically accreting matter
in a gravitational potential denoted by $\Phi$
 may be written as
\begin{equation}
\frac{{\partial{u}}}{{\partial{t}}}+u\frac{{\partial{u}}}{{\partial{r}}}+\frac{1}{\rho}
\frac{{\partial}p}{{\partial}r}+\frac{{\partial}\Phi}{\partial{r}}=0    ,
\label{eq48}
\end{equation}
The first term in (\ref{eq48}) is the Eulerian time derivative of the
dynamical velocity, the second term
is the `advective' term, the third term
is the
momentum deposition due to the pressure gradient and the
last term  is the gravitational force.
Another equation necessary to describe
the motion of the fluid is
the continuity
equation
\begin{equation}
\frac{{\partial}{\rho}}{{\partial}t}+\frac{1}{r^2}\frac{{\partial}}{{\partial}r}\left({\rho}ur^2\right)=0 .
\label{eq49}
\end{equation}
To integrate the above set of equations, one also needs the
equation of state that specifies the intrinsic properties of the fluid.
We will study  accretion described by either a polytropic
or an isothermal equation of state.
\subsection{Sonic Quantities}
\subsubsection{Polytropic Accretion}
\noindent
We employ
a polytropic equation of state of the form
$p=K{\rho}^\gamma$.
The  sound speed $c_s$ is defined by
\begin{equation}
c_s^2\equiv\left. \frac{\partial p}{\partial\rho}\right|_{\rm constant~~entropy}=\gamma\frac{p}{\rho},
\label{eq52}
\end{equation}

Assuming stationarity of the flow, we find the following conservation equations:

\noindent
1) Conservation of energy  implies  constancy
of the specific energy ${\cal E}$
\begin{equation}
{\cal E}=\frac{u^2}{2}+\frac{c_s^2}{{\gamma}-1}+\Phi .
\label{eq53}
\end{equation}

\noindent
2) Conservation of the baryon number implies constancy of the accretion rate ${\dot M}$
\begin{equation}
{\dot M}=4{\pi}{\rho}ur^2 .
\label{eq54}
\end{equation}
Equation (\ref{eq53}) is obtained  from (\ref{eq48}),
and (\ref{eq54}) follows directly from (\ref{eq49}).

Substituting $\rho$ in terms of $c_s$ and
differentiating (\ref{eq54}) with respect to $r$,
 we obtain
\begin{equation}
c_s'=\frac{c_s (1-\gamma)}{2}
\left(\frac{u'}{u}+\frac{2}{r}\right) ,
\label{eq55}
\end{equation}
where $'$ denotes the derivative with respect to $r$.
Next we  differentiate (\ref{eq53}) and eliminating $c_s'$ with the help
of (\ref{eq55}) we obtain
\begin{equation}
u'=\frac{2c_s^2/r-
\Phi'}{u-c_s^2/u} \, .
\label{eq56}
\end{equation}
One thus finds the critical point conditions
as
\begin{equation}
{u}_{\rm r=r_c}={c_{s}}_{\rm r=r_c}=\sqrt{\left|\frac{r\Phi'}{2}\right|_{\rm r=r_c}} \, ,
\label{eq57}
\end{equation}
As described in section 12.2, here also the critical point and the sonic points are 
equivalent, and the
location of the sonic point is identical to the location of the
acoustic horizon due to the assumption of stationarity and spherical symmetry.
Thus, hereafter we denote  
$r_h$ as the sonic point and the sphere of radius $r_h$ as the
acoustic horizon.
Hereafter, the subscript $h$ indicates that a particular
quantity is evaluated at $r_h$.
The location of the acoustic horizon
is obtained
by solving the algebraic equation
\begin{equation}
{\cal E}-\frac{1}{4}\left(\frac{\gamma+1}{\gamma-1}\right)r_h
{\Phi'_h}
-{\Phi_h}=0 .
\label{eq58}
\end{equation}
The derivative  $u'_h$
at the
corresponding sonic point is obtained by
solving the quadratic equation
\begin{eqnarray}
\left(1+\gamma\right)\left(u'_h\right)^2+
2\left(\gamma-1\right)\sqrt{\frac{2\Phi'_h}{r_h}}\,
u'_h & &
\nonumber \\
+\left(2{\gamma}-1\right)
\frac{\Phi'_h}
{r_h}
+{\Phi''_h} &= & 0,
\label{eq59}
\end{eqnarray}
which follows from (\ref{eq56}) in  the limit
$r{\rightarrow}r_h$
 evaluated with the help of l'Hospital's rule.

Finally, the gradient of the sound speed
at the acoustic horizon is obtained
by substituting  $u'_h$ obtained from (\ref{eq59})
into equation (\ref{eq55}) at the acoustic horizon
\begin{equation}
\left. c_s'\right|_h=\left(\frac{1-\gamma}{2}\right)
\left(u'_h+\sqrt{\frac{2\Phi'_h}{r_h}}\right) .
\label{eq60}
\end{equation}
\subsubsection{Isothermal Accretion}
\noindent
We employ the isothermal equation of state of the form
\begin{equation}
p=\frac{RT}{\mu}\rho=c_s^2{\rho}\, ,
\label{eq61}
\end{equation}
where $T$ is the
temperature,
$R$ and $\mu$ are the universal gas constant and the mean molecular weight, respectively.
 The quantity  $c_s$ is the isothermal sound speed defined by
\begin{equation}
c_s^2=\left.\frac{\partial p}{\partial \rho}\right|_T
={\Theta}T \, ,
\label{eq62}
\end{equation}
where the derivative is taken at fixed  temperature and the constant
$\Theta=\kappa_B/(\mu m_H)$ with $m_H \simeq m_p$ being the
mass of the hydrogen atom.
In our model we assume that the accreting matter is predominantly hydrogen,
hence $\mu \simeq 1$.
Now, the specific energy equation takes the form
\begin{equation}
{\cal E}=\frac{u^2}{2}+{\Theta}T\ln \rho+\Phi \, ,
\label{eq63}
\end{equation}
whereas the accretion rate is given by (\ref{eq54}) as before.

The radial change rate of the dynamical velocity
is again given
 by (\ref{eq56}). From (\ref{eq56})  and with (\ref{eq62})
we find
the sonic point condition as
\begin{equation}
u_h=\sqrt{\frac{r_h\Phi'_h}{2}}
=c_s=\sqrt{{\Theta}T} \, .
\label{eq64}
\end{equation}
since $c_s$ does not depend on $r$.
 The derivative of $u$ at $r_h$ is obtained
from (\ref{eq56})
by making use of l'Hospital's rule as before. We find
\begin{equation}
u'_h=-
\sqrt{-\frac{1}{2}\left(\frac{1}{r_h}\Phi'_h+\Phi''_h\right)}\, ,
\label{eq65}
\end{equation}
where the minus sign in front of the square root
indicates accretion (the plus would
correspond to a wind solution).
Note that the quantities in equations (\ref{eq64}) and (\ref{eq65}) are functions of
the fluid
temperature $T$  only. Hence the isothermal spherical accretion can be
essentially described as a one-parameter solution of the
hydrodynamical equations, parameterized by $T$.
\subsection{Analogue Temperature}
\noindent
From (\ref{eq33}) in Newtonian limit, i.e.,
\begin{equation}
|\chi^2|=g_{00}\rightarrow{\left(1+\frac{\Phi}{2c^2}\right)}
\label{eq66}
\end{equation}
gives a general expression for the
temperature of the analogue Hawking radiation in a spherically
accreting fluid
 in the Newtonian as well as in any
pseudo-Schwarzschild gravitational potential
\begin{equation}
T_{\rm AH}=\frac{\hbar}{2{\pi}\kappa_b}
\sqrt{\frac{2c^2+\Phi_h}{2c^2}}
\left[\frac{1}{1-c_s^2}\left|\frac{d}{dr}
\left(c_s-u\right)\right|\right]_{\rm r=r_h} \, .
\label{eq67}
\end{equation}
The quantities required to calculate the analogue temperature (\ref{eq67}) are
obtained using the formalism presented in section 15.2.
For polytropic accretion,
using equations (\ref{eq55})-(\ref{eq60}) one finds
\begin{eqnarray}
\tau\equiv \frac{T_{\rm AH}}{T_{\rm H}} =
4\sqrt{\frac{2+\Phi_h}{2}}\left(\frac{2}{2-r_h\Phi_h}\right)
\left(\frac{\gamma+1}{2}\right)
\nonumber  \\
\sqrt{\frac{\Phi'_h}{r_h}{\bf f}(\gamma)-
\left(1+\gamma\right)\Phi''_h}
\label{eq68}
\end{eqnarray}
where ${\bf f}(\gamma)=\left(0.00075\gamma^2-5.0015{\gamma}+3.00075\right)$.
The quantities
$\Phi_h$, $\Phi'_h$, and $\Phi''_h$
are obtained by calculating the values of
various potentials at $r_h$,
and  $r_h$ is calculated from (\ref{eq58}) for an
astrophysically
relevant choice of $\{ {\cal E}, \gamma\}$.

Note that if $(c_s' - u')_h$ is negative, one obtains
an acoustic  {\it white-hole} solution.
Hence the condition for the existence of the acoustic white hole is
\begin{equation}
\left(\frac{\gamma+1}{2}\right)
\sqrt{\frac{\Phi'_h}{r_h}{\bf f}(\gamma)-
\left(1+\gamma\right)\Phi''_h}\: < 0 .
\label{eq69}
\end{equation}
Since $\gamma$ and $r_h$ can never be negative, and since
$\Phi''_h$ and $\Phi''_h$ are
always real for the preferred domain of $\{ {\cal E}, \gamma\}$,
unlike general relativistic spherical accretion,
{\it acoustic white-hole
solutions are excluded in the astrophysical accretion governed by the Newtonian or
post-Newtonian potentials}.

For an isothermal flow, the quantity
$c_s'$ is zero and
using (\ref{eq65}) we find
\begin{equation}
\tau=4\sqrt{2}\left(\frac{1}{2-r_h\Phi'_h}\right)
\sqrt{-\left(1+\frac{\Phi_h}{2}\right)
\left(\Phi''_h+\frac{\Phi'_h}{r_h}\right)} \, ,
\label{eq70}
\end{equation}
where $r_h$ should be evaluated using (\ref{eq64}).
Clearly,  the fluid temperature $T$ completely determines
the analogue Hawking temperature. Hence, {\it a spherical
isothermally accreting astrophysical black hole provides
a simple system where  analogue gravity can be theoretically
studied using only one free parameter}.

For both polytropic and isothermal accretion,
for certain range of the parameter space, the analogue
Hawking temperature $T_{AH}$ may become {\it higher} than
the actual Hawking temperature $T_H$,
%For details of the variation of $T_{AH}$ on various accretion parameter,
see Dasgupta, Bili\'c \& Das 2005 for further details.
\section{Post-Newtonian Multi-transonic Accretion Disc as Analogue Model}
\noindent
In this section, we will study the analogue gravity phenomena for 
polytropic (adiabatic)
and isothermal rotating, advective, multi-transonic accretion disc in various 
pseudo-Schwarzschild potentials described in section 14.
\subsection{Flow Dynamics and Accretion Variables at the Critical Point}
\subsubsection{Polytropic Accretion}
\noindent
The local half-thickness,
$h_i(r)$ of the disc for any $\Phi_i$ can be obtained by balancing the
gravitational force by pressure gradient and can be expressed as:
\begin{equation}
h_i(r)=c_s\sqrt{{r}/\left({\gamma}{\Phi_i^{\prime}}\right)}
\label{kk1}
\end{equation}
where $\Phi_i^{\prime}=d\Phi_i/dr$.
For a non-viscous flow obeying the polytropic equation of state
$p=K{\rho}^{\gamma}$,
integration of radial momentum
equation:
\begin{equation}
u\frac{{d{u}}}{{d{r}}}+\frac{1}{\rho}
\frac{{d}P}{{d}r}+\frac{d}{dr}\left(\Phi^{eff}_{i}(r)\right)=0
\label{kk2}
\end{equation}
leads to the following energy conservation equation (on the 
equatorial plane of the disc) in steady state:
\begin{equation}
{\cal E}=\frac{1}{2}u^2+\frac{c_s^2}{\gamma - 1}
+\frac{{\lambda}^2}{2r^2}+\Phi_i
\label{kk3}
\end{equation}
and the continuity equation:
\begin{equation}
\frac{{d}}{{d}r}
\left[u{\rho}rh_i(r)\right]=0
\label{kk4}
\end{equation}
can be integrated to obtain the baryon number conservation equation:
\begin{equation}
{\dot M}=\sqrt{\frac{1}{\gamma}}uc_s{\rho}r^{\frac{3}{2}}
\left({\Phi_i^{\prime}}\right)^{-\frac{1}{2}}.
\label{kk5}
\end{equation}
The entropy accretion rate ${\dot \Xi}$ can be expressed as:
\begin{equation}
{\dot {\Xi}}=
\sqrt{\frac{1}{\gamma}}u
c_s^{\left({\frac{\gamma+1}{\gamma-1}}\right)}
r^{\frac{3}{2}}\left({\Phi_i^{\prime}}\right)^{-\frac{1}{2}}
\label{kk6}
\end{equation}
One can simultaneously solve (\ref{kk3}) and (\ref{kk6}) 
for any particular $\Phi_i$ and for a
particular set of values of $\left\{{\cal E}, \lambda, \gamma\right\}$.
For a particular value of $\left\{{\cal E}, \lambda, \gamma\right\}$,
it is now quite
straight-forward to derive the space gradient of the
acoustic velocity $\left(\frac{dc_s}{dr}\right)_i$ and the
dynamical flow velocity
$\left(\frac{du}{dr}\right)_i$ for flow in any particular
$i$th black hole  potential $\Phi_i$:
\begin{equation}
\left(\frac{dc_s}{dr}\right)_i=c_s\left(\frac{\gamma-1}{\gamma+1}\right)
\left(\frac{1}{2}\frac{{\Phi_i}^{{\prime}{\prime}}}{{\Phi_i}^{{\prime}}}
-\frac{3}{2r}-\frac{1}{u}\frac{du}{dr}\right)
\label{kk7}
\end{equation}
and,
\begin{equation}
\left(\frac{du}{dr}\right)_i=
\frac{
\left(\frac{\lambda^2}{r^3}+\Phi^{'}_i(r)\right)-
\frac{c_s^2}{\gamma+1}\left(\frac{3}{r}+
\frac{\Phi^{''}_i(r)}{\Phi^{'}_i(r)}\right)
}
{u-\frac{2c_s^2}{u\left(\gamma+1\right)}
}
\label{kk8}
\end{equation}
where
${\Phi_i}^{{\prime}{\prime}}$ represents the derivative
of ${\Phi_i}^{{\prime}}$.
Hence the critical point condition comes out to be:
\begin{equation}
\left[c_s^i\right]_{\rm r=r_c}=\sqrt{\frac{1+\gamma}{2}}\left[u^i\right]_{\rm r=r_c}=
\left[
\frac{\Phi^{'}_i(r)+{\gamma}\Phi^{'}_i(r)}{r^2}
\left(
\frac{\lambda^2+r^3\Phi^{'}_i(r)}{3\Phi^{'}_i(r)+r\Phi^{''}_i(r)}
\right)
\right]_{\rm r=r_c}
\label{kk9}
\end{equation}
Note that the Mach number $M_c$ at the critical point is {\it not} 
equal to unity, rather:
\begin{equation}
M_c=\sqrt{\frac{2}{\gamma+1}}
\label{kk10}
\end{equation}
Hence, the critical points and the sonic points are {\it not equivalent}. One needs to calculate the
sonic point, which is the location of the acoustic horizon, following the procedure as described
in section 13.5.

For any fixed set of $\left\{{\cal E}, \lambda, \gamma\right\}$, the critical points can 
be obtained by solving the following polynomial of $r$:
\begin{equation}
{\cal E}-{\left[\frac{\lambda^2}{2r^2}+\Phi_i\right]}_{\rm r=r_c}-\frac{2\gamma}{\gamma^2-1}
\left[
\frac{\Phi^{'}_i(r)+{\gamma}\Phi^{'}_i(r)}{r^2}
\left(
\frac{\lambda^2+r^3\Phi^{'}_i(r)}{3\Phi^{'}_i(r)+r\Phi^{''}_i(r)}
\right)
\right]_{\rm r=r_c}
=0.
\label{kk11}
\end{equation}
The dynamical velocity gradient at the critical point can be obtained by 
solving the following equation for $(du/dr)_{\rm r=r_c}$:
\begin{eqnarray}
\frac{4{\gamma}}{\gamma+1}\left(\frac{du}{dr}\right)^2_{c,i}
-2\left|u\right|_{\rm r=r_c}\frac{\gamma-1}{\gamma+1}
\left[\frac{3}{r}+\frac{\Phi^{''}_i(r)}{\Phi^{'}_i(r)}\right]_{\rm r=r_c}
\left(\frac{du}{dr}\right)_{c,i}
& & \nonumber \\
+\left|c^2_s\left[\frac{\Phi^{'''}_i(r)}{\Phi^{'}_i(r)}
-\frac{2\gamma}{\left(1+{\gamma}\right)^2}
\left(\frac{\Phi^{''}_i(r)}{\Phi^{'}_i(r)}\right)^2
+\frac{6\left(\gamma-1\right)}{\gamma{\left(\gamma+1\right)^2}}
\left(\frac{\Phi^{''}_i(r)}{\Phi^{'}_i(r)}\right)
-\frac{6\left(2\gamma-1\right)}{\gamma^2{\left(\gamma+1\right)^2}}
\right]\right|_{\rm r=r_c}
& & \nonumber \\
+
\Phi^{''}_i{\Bigg{\vert}}_{\rm r=r_c}-
\frac{3\lambda^2}{r^4_c}=0
\label{kk12}
\end{eqnarray}
Where the subscript $(c,i)$ indicates that the corresponding
quantities for any $i$th potential is being measured at its
corresponding critical point and $\Phi^{'''}_i=\frac{d^3\Phi_i}{dr^3}$.
\subsubsection{Isothermal Accretion}
\noindent
The isothermal sound speed is defined as:
\begin{equation}
c_{s}={\Theta}T^{\frac{1}{2}}
\label{kk13}
\end{equation}
where $\Theta=\sqrt{\frac{\kappa_B{\mu}}{m_{H}}}$ is a constant, ${m_{H}}{\sim}{m_{P}}$
being the mass of the hydrogen atom and $\kappa_B$ is Boltzmann's constant.
The local half-thickness $h_{i}(r)$  of the disc for any $\Phi_{i}(r)$ can be obtained by
balancing the gravitational force by pressure gradient and can be expressed as
\begin{equation}
h_{i}(r)=\Theta{\sqrt{\frac{rT}{\Phi_{i}^{\prime}{}}}}
\label{kk14}
\end{equation}
Solution of the radial momentum conservation equation and the continuity equation
provides the following two integral of motion on the equatorial 
plane of the isothermal accretion disc
\begin{equation}
\frac{u^2(r)}{2}+{\Theta}Tln{\rho(r)}+\frac{\lambda^2}{2r^2}+\Phi_i={\rm Constant}
\label{kk15}
\end{equation}
and
\begin{equation}
{\dot M}={\Theta}{\rho}(r)u(r)r^{\frac{3}{2}}\sqrt{\frac{T}{\Phi_i^{\prime}}}
\label{kk16}
\end{equation}
The dynamical flow velocity for a particular value of $\left\{{\cal E},\lambda\right\}$
can be expressed as
\begin{equation}
\frac{du}{dr} = \left[ \frac{\left(\frac{3\Theta^{2}T}{2r}+\frac{\lambda^{2}}{r^{3}} \right)
- \left(\frac{1}{2}\Theta^{2}T\frac{\Phi_{i}^{\prime\prime}(r)}{\Phi_{i}^{\prime}(r)}
+ \Phi_{i}^{\prime}\right)}{\left(u-\frac{\Theta^{2}T}{u} \right)}\right]
\label{kk17}
\end{equation}
where $\Phi_{i}^{\prime\prime}=\frac{d^2\Phi_i}{dr^2}$.
Since the flow is isothermal, $dc_s/dr=0$ everywhere identically.

The critical point condition can be expressed as:
\begin{equation}
\left|u\right|_{\rm r=r_c}=\Theta{T^{\frac{1}{2}}}=
\sqrt{
\frac{
{\Phi_i^{\prime}}{\Bigg{\vert}}_{\rm r=r_c}-\frac{\lambda^2}{r_c^3}}
{\frac{3}{2r_c}-\frac{1}{2}\left(\frac{{\Phi_i}^{\prime\prime}}{\Phi_{i}^{\prime}}\right)_{\rm r=r_c}}}
\label{kk18}
\end{equation}
Note that the Mach number at the critical point is exactly equal to unity, hence {\it the 
critical points and the sonic points are identical for isothermal accretion disc}. Therefore,
$r_c$ is actually the location of the acoustic event horizon $r_h$, and for a specific value of
$\left\{{\cal E},\lambda\right\}$, $r_h$ can be computed by solving the following equation for 
$r_h$:
\begin{equation}
\Phi_{i}^{\prime\prime}{\Bigg{\vert}}_{\rm r=r_h} + \frac{2}{\Theta^{2}T}\left(\Phi_{i}^{\prime}
\right)_{\rm r=r_h}^{2} - \left[\frac{3}{r_{h}\Theta} +
\frac{2{\lambda}^{2}}{T{\Theta}^{2}r_{h}^{3}}\right]\Phi_{i}^{\prime}{\Bigg{\vert}}_{\rm r=r_h} = 0
\label{kk19}
\end{equation}
The dynamical velocity gradient at the acoustic horizon can be obtained as:
\begin{eqnarray}
\left(\frac{du}{dr}\right)_{h,i} = \pm\frac{1}{\sqrt{2}} \left\{ \frac{1}{2}\Theta^
{2}T {\Bigg{[}} \left(\frac{\Phi_{i}^{\prime\prime}}{\Phi_{i}^{\prime}} \right)_{\rm r=r_h}^{2}
- \left(\frac{\Phi_{i}^{\prime\prime\prime}}{\Phi_{i}^{\prime}} \right)_{\rm r=r_h} \right]
& & \nonumber \\
- \left(\Phi_{i}^{\prime\prime}\Big{\vert}_{\rm r=r_h}
+\frac{3\Theta^{2}T}{2r_{h}^{2}}+\frac{3\lambda^{2}}{r_{h}^{4}} \right)
 {\Bigg{\}}}^\frac{1}{2}
\label{kk20}
\end{eqnarray}
\subsection{Multi-transonicity and Shock Formation}
\noindent
As in the case of general relativistic accretion disc, axisymmetric accretion 
under the influence of a generalized pseudo-Schwarzschild potential $\Phi_i$ 
also produces multiple critical/sonic points, both for polytropic as well as for 
the isothermal flow. For polytropic flow, (\ref{kk11}) can be solved to obtain 
various critical points, and the flow equations can be integrated from such 
critical points to find the corresponding sonic points. 

For accretion/wind solutions under the influence of various $\Phi_i$, one can define 
the square of the eigenvalue $\bar{\Omega}$ in the following way (Chaudhury, Ray \& Das 2006):
\begin{eqnarray}
{\bar{\Omega}}^2 = \frac{4 r_{\mathrm{c}}
\Phi^{\prime}(r_{\mathrm{c}})|c_{\mathrm{s}}^2|_{\rm r=r_c}}{(\gamma + 1)^2}
 \left[ \left(\gamma - 1 \right)
{\mathcal A} - 2 \gamma \left(1 + {\mathcal C} \right) + 2 \gamma
\frac{\mathcal{BC}}{\mathcal A} \right]  && \nonumber \\
-\frac{\lambda^2}
{\lambda_{\mathrm K}^2(r_{\mathrm{c}})}
\left[4 \gamma + \left(\gamma - 1 \right)
{\mathcal A} + 2 \gamma \frac{\mathcal{BC}}{\mathcal A} \right]
\label{multi1}
\end{eqnarray}
where
\begin{equation}
{\cal A} = r_{\rm c}\frac{\Phi^{\prime \prime}(r_{\rm c})}
{\Phi^{\prime}(r_{\rm c})} - 3, ~
{\cal B} = 1 + r_{\rm c}
\frac{\Phi^{\prime \prime \prime}(r_{\rm c})}
{\Phi^{\prime \prime}(r_{\rm c})}
- r_{\rm c}\frac{\Phi^{\prime \prime}(r_{\rm c})}
{\Phi^{\prime}(r_{\rm c})}, ~
{\cal C} = {\cal A} + 3, ~
\lambda_{\rm K}^2(r) = r^3 \Phi^{\prime}(r)
\label{multi2}
\end{equation}
For isothermal flows, a similar
expression for the related eigenvalues may likewise be derived. The
algebra in this case is much simpler and it is an easy
exercise to assure oneself that for isothermal flows one simply needs
to set $\gamma = 1$ in (\ref{multi1}), to arrive at a corresponding
relation for ${\bar{\Omega}}^2$.

A generic conclusion that can be drawn about the critical points from
the form of ${\bar{\Omega}}^2$ in (\ref{multi1}), is that for a conserved
pseudo-Schwarzschild axisymmetric flow driven by any potential, the only
admissible critical points will be saddle points and centre-type points.
For a saddle point, ${\bar{\Omega}}^2 > 0$, while for a centre-type point,
${\bar{\Omega}}^2 < 0$. Once the behaviour of all the physically relevant
critical points has been understood in this way, a complete qualitative
picture of the flow solutions passing through these points (if they
are saddle points), or in the neighbourhood of these points (if they
are centre-type points), can be constructed, along with an impression
of the direction that these solutions can have in the phase portrait
of the flow, see Chaudhury, Ray \& Das (2006) for further detail.

Application of the above mentioned methodology 
for finding out the nature of the critical point leads to the 
conclusion that for multi-transonic accretion and wind, 
the inner critical point $r_c^{in}$ and the outer critical 
point $r_c^{out}$ are of saddle type (`X' type), whereas the 
middle critical point $r_c^{mid}$ is of centre type (`O' type).
For mono-transonic accretion, the critical point will 
{\it always} be of saddle type and will be located either 
quite close to the event horizon (mono-transonic accretion 
passing through the `inner type' critical point) or quite
far away from the black hole (mono-transonic accretion 
passing through the `outer type' critical point).

Hereafter we will use the notation $\left[{\cal P}_i\right]$ for a set of
values of $\left[{\cal E}, \lambda, \gamma\right]$ for 
polytropic accretion in any particular
$\Phi_i$. 
For all $\Phi_i$, one finds a significant region of
parameter space spanned by $\left[{\cal P}_i\right]$ which allows
the multiplicity of
critical points for accretion as well as for wind
where two real physical inner and outer (with respect to
the location of the black hole event horizon) 
saddle type critical points $r_c^{in}$ and $r_c^{out}$ encompass
one centre type unphysical middle sonic point $r_c^{mid}$ in between.
For a particular
$\Phi_i$, if
${\cal A}_i\left[{\cal P}_i\right]$ denotes the universal set
representing the entire parameter space covering all
values of $\left[{\cal P}_i\right]$, and if
${\cal B}_i\left[{\cal P}_i\right]$ represents one particular  subset
of
${\cal A}_i\left[{\cal P}_i\right]$
which contains  only the
particular values of $\left[{\cal P}_i\right]$ for which the above mentioned
three critical points are obtained, then ${\cal B}_i\left[{\cal P}_i\right]$
can further be decomposed into two subsets ${\cal C}_i\left[{\cal P}_i\right]$
and ${\cal D}_i\left[{\cal P}_i\right]$ such that:
\begin{eqnarray}
{\cal C}_i\left[{\cal P}_i\right]~\subseteq~
{\cal B}_i\left[{\cal P}_i\right]~~~
{\rm \underline{only~for}}~~~
{\dot {\Xi}}\left(r_c^{in}\right)>
{\dot {\Xi}}\left(r_c^{out}\right)
& & \nonumber \\
{\cal D}_i\left[{\cal P}_i\right]~\subseteq~
{\cal B}_i\left[{\cal P}_i\right]~~~
{\rm \underline{only~for}}~~~
{\dot {\Xi}}\left(r_c^{in}\right)<
{\dot {\Xi}}\left(r_c^{out}\right)
\label{multi3}
\end{eqnarray}
then for $\left[{\cal P}_i\right] \in {\cal C}_i\left[{\cal P}_i\right]$,
we get multi-transonic {\it accretion} and for
$\left[{\cal P}_i\right] \in {\cal D}_i\left[{\cal P}_i\right]$
one obtains  multi-transonic {\it wind}.
%\begin{figure}
%\vbox{
%\vskip -5.8cm
%\centerline{
%\psfig{file=figure10.ps,height=15cm,width=15cm}}}
%\noindent {{\bf Fig. 10:}
%The complete classification of $\left[{\cal E},\lambda\right]$ for polytropic 
%accretion in the Paczy\'nski \& Wiita
%(1980) potential $\Phi_1$. The value of $\gamma$ is taken to be equal to $4/3$. Mono-transonic
%regions are marked by {\bf I} (accretion through the inner sonic point only)
%and {\bf O} (accretion through the outer sonic point only). The regions marked by ${\cal A}$ 
%and ${\cal W}$ represents the multi-transonic accretion and wind, respectively. The shaded 
%region represents the collection of $\left[{\cal E},\lambda\right]$ (for $\gamma=4/3$) for 
%which the stable Rankine-Hugoniot shock solutions are obtained.}
%\end{figure}
\begin{center}
\begin{figure}[h]
\includegraphics[scale=0.9,angle=0.0]{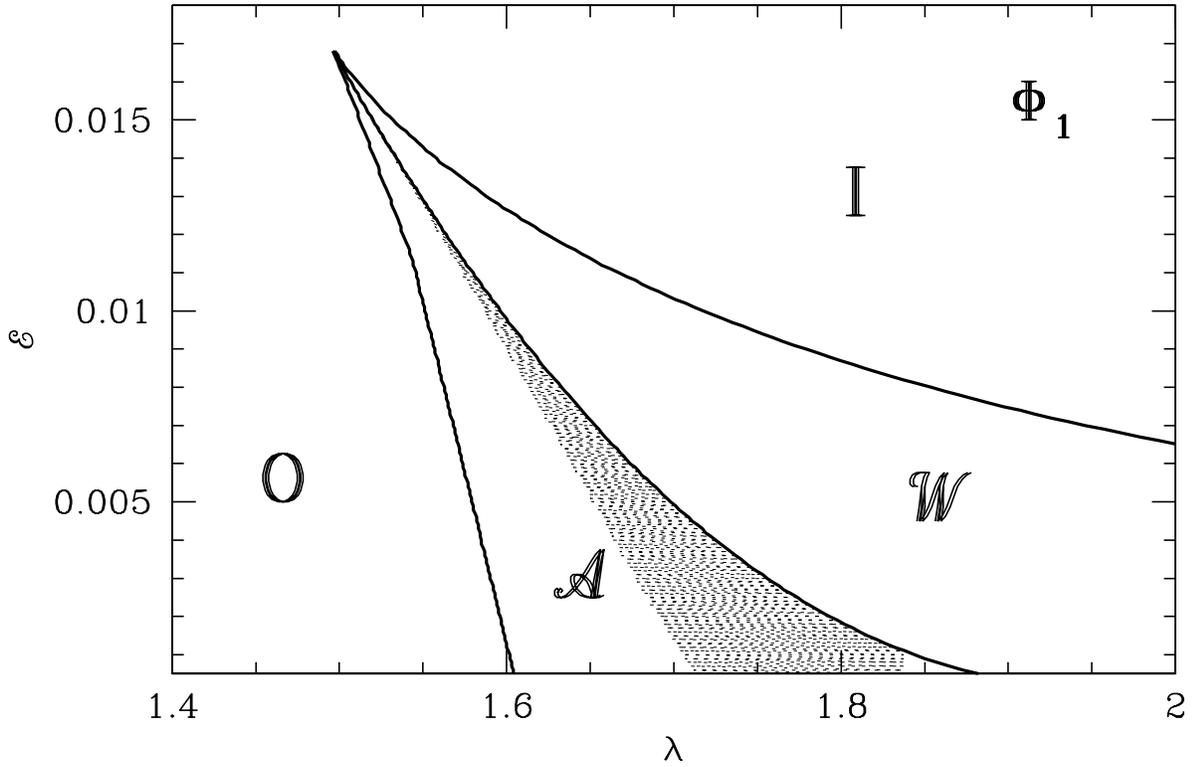}
\caption[]{The complete classification of $\left[{\cal E},\lambda\right]$ for polytropic
accretion in the Paczy\'nski \& Wiita
(1980) potential $\Phi_1$. The value of $\gamma$ is taken to be equal to $4/3$. Mono-transonic
regions are marked by {\bf I} (accretion through the inner sonic point only)
and {\bf O} (accretion through the outer sonic point only). The regions marked by ${\cal A}$
and ${\cal W}$ represents the multi-transonic accretion and wind, respectively. The shaded
region represents the collection of $\left[{\cal E},\lambda\right]$ (for $\gamma=4/3$) for
which the stable Rankine-Hugoniot shock solutions are obtained.}
\label{fig10}
\end{figure}
\end{center}

For the Paczy\'nski \& Wiita
(1980) potential $\Phi_1$, in figure 10 we classify the whole $\left[{\cal E},\lambda\right]$ 
parameter space for a fixed value of $\gamma=4/3$. The region marked by
{\bf I} represents the values of $\left[{\cal E},\lambda\right]$ for which 
accretion will be mono-transonic and will pass through the saddle type
inner critical point, whereas the region marked by
{\bf O} represents the values of $\left[{\cal E},\lambda\right]$ for which
accretion will be mono-transonic and will pass through the saddle type
outer critical point. The wedge shaped region bounded by heavy solid 
lines and marked by ${\cal A}$ (including the shaded region) represents 
the multi-transonic {\it accretion} zone for which 
$\left({\cal E}_i,\lambda_i\right) \in \left[{\cal P}_i\right]
\in {\cal C}_i\left[{\cal P}_i\right]~\subseteq~
{\cal B}_i\left[{\cal P}_i\right]$, whereas the wedge shaped region 
bounded by the heavy solid line and marked by ${\cal W}$ represents
the multi-transonic {\it wind} and mono-transonic accretion zone for 
which 
$\left({\cal E}_i,\lambda_i\right) \in \left[{\cal P}_i\right]
\in {\cal D}_i\left[{\cal P}_i\right]~\subseteq~
{\cal B}_i\left[{\cal P}_i\right]$.
A similar kind of parameter space division can easily be obtained for 
other $\Phi_i$ as well, see Das 2002 and Chaudhury, Ray \& Das 2006
for further detail.

If shock forms in accretion, then $\left[{\cal P}_i\right]$ responsible
for shock formation must be somewhere from the region for which
$\left[{\cal P}_i\right] \in {\cal C}_i\left[{\cal P}_i\right]$, though
not all $\left[{\cal P}_i\right] \in {\cal C}_i\left[{\cal P}_i\right]$
will allow shock transition. One can derive (see Das 2002 for further detail)
the Rankine-Hugoniot shock condition for the generalized potential $\Phi_i$
in the following form which will be satisfied only at the shock location
\begin{equation}
\left(1-\gamma\right)\left(\frac{{\rho_{-}}{{\dot {\Xi}}_{-}}}{\dot M}
\right)^{log_{\Gamma}^{1-\beta_1}}
{\cal E}_{{\left(ki+th\right)}}
-{\beta_1}{\left(1+\beta_1-{\rho}_{comp}\right)}^{-1}
+\left(1+\beta_1\right)^{-1}
=0
\label{multi4}
\end{equation}
where ${\dot M}$ is the mass accretion rate as defined in (\ref{kk5}),
${\cal E}_{{\left(ki+th\right)}}$ is the total specific thermal plus
mechanical energy of the accreting fluid:
$$
{\cal E}_{{\left(ki+th\right)}}=\left[{\cal E}-
\left(\frac{\lambda^2}{2r^2}+\Phi_i\right)\right],
$$
${\rho}_{comp}$ and $\Theta$ are
the density compression and entropy enhancement ratio respectively, defined
as
$\rho_{comp}=\left({\rho_{+}}/{\rho_{-}}\right)$ and
$\Theta=\left({\dot {\Xi}}_{+}/{\dot {\Xi}}_{-}\right)$
respectively; $\beta_1=1-\Gamma^{\left(1-{\gamma}\right)}$ and $\Gamma={\Theta}
{\rho_{comp}}$, ``$+$'' and ``$\_$'' refer to the post- and
pre-shock quantities.
The shock
strength ${\cal S}_i$ (ratio of the pre- to post-shock Mach number of the
flow) can be calculated as:
\begin{equation}
{\cal S}_i=\rho_{comp}\left(1+\beta_1\right)
\label{multi5}
\end{equation}
Equations (\ref{multi4}) and (\ref{multi5})
cannot be solved
analytically because they are non-linearly coupled. However,
one can solve the above set of equations 
using iterative numerical
techniques. An efficient numerical
code has been developed in Das 2002, which takes
$\left[{\cal P}_i\right]$ and $\Phi_i$ as its input and can calculate the 
shock location $r_{sh}$
along with
any sonic or shock quantity as a function of
$\left[{\cal P}_i\right]$. One obtains a two-fold degeneracy 
for $r_{sh}$, and the local stability analysis ensures that
the shock which forms in
between the {\it sonic} points $r_s^{out}$ and $r_s^{mid}$ is
stable for {\it all} $\Phi_i$. Hereafter, we will be interested only in
such stable shocks and related quantities.

If $\left[{\cal P}_i\right]
\in {\cal F}_i\left[{\cal P}_i\right]~\subseteq~
{\cal C}_i\left[{\cal P}_i\right]$
represents the region of parameter space for which
multi-transonic supersonic
flows is expected to
encounter a Rankine-Hugoniot shock at $r_{sh}$, where they
become hotter, shock compressed and subsonic
and will again become supersonic only after passing through $r_{in}$ before
ultimately crossing the event horizon, then one can also define
$\left[{\cal P}_i\right]
\in {\cal G}_i\left[{\cal P}_i\right]$ which is complement
of ${\cal F}_i\left[{\cal P}_i\right]$ related to
${\cal C}_i\left[{\cal P}_i\right]$ so that for:
\begin{equation}
\left\{{\cal G}_i\left[{\cal P}_i\right]\Bigg{\vert}
\left[{\cal P}_i\right]\in{\cal C}_i\left[{\cal P}_i\right]
~{\rm and}~
\left[{\cal P}_i\right]\notin{\cal F}_i\left[{\cal P}_i\right]
\right\},
\label{multi6}
\end{equation}
the shock location becomes imaginary in
${\cal G}_i\left[{\cal P}_i\right]$,
hence no stable shock forms in that region.
Numerical simulation shows that (Molteni, Sponholz \& Chakrabarti 1996)
the shock keeps oscillating back and forth in this region. One anticipates that
${\cal G}_i\left[{\cal P}_i\right]$ is also an important zone which might be
responsible for the Quasi-Periodic Oscillation (QPO) of the black hole 
candidates, and the frequency for such QPO can be computed for 
all pseudo-Schwarzschild potentials (see Das 2003 for further 
details).

The wedge shaped shaded region in figure 10 represents the $\left[{\cal P}_i\right]
\in {\cal F}_i\left[{\cal P}_i\right]~\subseteq~
{\cal C}_i\left[{\cal P}_i\right]$ zone, for which steady standing stable
Rankine-Hugoniot shock forms, while the white region of 
the multi-transonic accretion (marked by ${\cal A}$) represents 
the $\left\{{\cal G}_i\left[{\cal P}_i\right]\Bigg{\vert}
\left[{\cal P}_i\right]\in{\cal C}_i\left[{\cal P}_i\right]
~{\rm and}~
\left[{\cal P}_i\right]\notin{\cal F}_i\left[{\cal P}_i\right]
\right\}$ zone.

Similarly, solution of (\ref{kk19}) provides the multi-transonic accretion 
and wind regions for the isothermal accretion in various $\Phi_i$. The 
corresponding shock conditions can also be constructed and can be solved 
for a particular value of $\left[T,\lambda\right]$ to find the 
region of parameter space responsible for the formation for stable shock solutions.
See Das, Pendharkar \& Mitra 2003 for details about the multi-transonicty and
shock formation in isothermal accretion disc around astrophysical black holes.
\subsection{Analogue Temperature}
\noindent
For axisymmetric accretion in Newtonian limit, one obtains (Bili\'c, Das \& Roy 2007)
from (\ref{eq33}) 
\begin{equation}
\left|{\chi}^2\right|=
\sqrt{\chi^{\mu}\chi_{\mu}}=
\sqrt{(1+2\Phi)\left(1-\frac{\lambda^2}{r^2}-2\Phi\frac{\lambda^2}{r^2}\right)}
\label{eqeq1}
\end{equation}
Hence the analogue temperature for the pseudo-Schwarzschild, axisymmetric, transonic 
accretion with space dependent acoustic velocity would be (Bili\'c, Das \& Roy 2007):
\begin{equation}
T_{\rm AH}=\frac{\hbar}{2{\pi}\kappa_B}
\sqrt{\left|(1+2\Phi)\left(1-\frac{\lambda^2}{r^2}-2\Phi\frac{\lambda^2}{r^2}\right)\right|_{\rm r=r_h}}
\left[\frac{1}{1-c_s^2}\left|\frac{d}{dr}
\left(c_s-u\right)\right|\right]_{\rm r=r_h}
\label{eqeq2}
\end{equation}
As discussed earlier, once the critical points are found by solving (\ref{kk11}), one can 
integrate the flow equations to find the sonic point $r_s$, which actually is the location 
of the acoustic horizon $r_h$. One then finds the value of $\left(du/dr\right)_{\rm r=r_h}$ 
and $\left(dc_s/dr\right)_{\rm r=r_h}$. Thus once a specific set of values
for $\left[{\cal E},\lambda,\gamma\right]$ for polytropic accretion is provided, all the corresponding terms 
in (\ref{eqeq2}) could readily be known and one thus comes up with an accurate 
estimation of $T_{AH}$, as well as $\tau$, the ratio of the analogue to the actual Hawking
temperature, as a function of $\left[{\cal E},\lambda,\gamma\right]$. 

In figure 11, we demonstrate the variation of $\tau$ (plotted along the $Z$ axis) on 
$\left[{\cal E},\lambda\right]$ (for a fixed value of $\gamma=4/3$) for multi-transonic
shocked accretion flow in Paczy\'nski \& Wiita
(1980) potential $\Phi_1$. $\left[{\cal E},\lambda\right]$ used to obtain such result,
corresponds to the shaded region of figure 10 (for which stable Rankine-Hugoniot 
shock forms in polytropic accretion). As discussed in section 13.9, two acoustic black holes are formed 
at the inner and the outer sonic points, and an acoustic {\it white hole} is formed at the
shock location. The analogue temperature corresponding to the white hole in not defined. The 
red surface in the figure corresponds to the variation of $\tau$ with $\left[{\cal E},\lambda\right]$
for the outer acoustic horizon (the outer sonic points) and the blue surface corresponds to 
the variation of $\tau$ with $\left[{\cal E},\lambda\right]$ for the inner acoustic horizons (the 
inner sonic points). It is observed that for a fixed value of $\left[{\cal E},\lambda,\gamma\right]$,
$\tau_{r_h^{in}}>\tau_{r_h^{out}}$. 

Although the above figure has been obtained for a fixed value of $\gamma$(=4/3), one can 
obtain the same $\left[\tau-{\cal E}-\lambda\right]$ variation for any value of $\gamma$
producing the multi-transonic shocked accretion flow. In general, $\tau$ co-relates with $\gamma$.
$\left[\tau-{\cal E}-\lambda\right]$ variation can also be studied for mono-transonic accretion passing 
through the inner or the outer sonic point only, and for mono-transonic accretion flow
in multi-transonic wind region (flow described by $\left[{\cal E},\lambda\right]$ obtained 
from the ${\cal W}$ region of the figure 10). 

All the above mentioned variation can also be studied for all other $\Phi_i$, see Bili\'c, Das \& 
Roy (2007) for further detail.  
%\begin{figure}
%\vbox{
%\vskip -0.0cm
%\centerline{
%\psfig{file=figure11.ps,height=11cm,width=16.2cm,angle=270.0}}
%\vskip -0.0cm
%{\bf Figure 11:} Variation of $\tau$ on ${\cal E}$ and $\lambda$ for 
%multi-transonic shocked accretion in the Paczy\'nski \& Wiita
%(1980) potential $\Phi_1$. The
%red surface in the figure corresponds to the variation of $\tau$ with $\left[{\cal E},\lambda\right]$
%for the outer acoustic horizons and the blue surface corresponds to
%the variation of $\tau$ with $\left[{\cal E},\lambda\right]$ for the inner acoustic horizons. 
%This figure has been reproduced from Bili\'c, Das \& Roy (2007)}
%\end{figure}
%
\begin{center}
\begin{figure}[h]
\includegraphics[scale=0.4,angle=270.0]{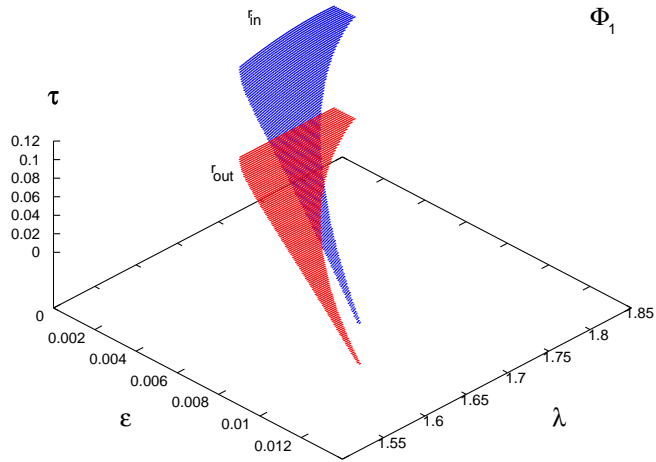}
\caption[]{Variation of $\tau$ on ${\cal E}$ and $\lambda$ for
multi-transonic shocked accretion in the Paczy\'nski \& Wiita
(1980) potential $\Phi_1$. The
red surface in the figure corresponds to the variation of $\tau$ with $\left[{\cal E},\lambda\right]$
for the outer acoustic horizons and the blue surface corresponds to
the variation of $\tau$ with $\left[{\cal E},\lambda\right]$ for the inner acoustic horizons.
This figure has been reproduced from Bili\'c, Das \& Roy (2007)}
\label{fig11}
\end{figure}
\end{center}

It is now easy to calculate the analogue temperature 
for isothermal axisymmetric accretion in pseudo-Schwarzschild 
potential. Since $c_s$ is a function of the 
bulk temperature of the flow $T$, and since for isothermal accretion $T$ 
is constant throughout, the space derivative of the acoustic velocity $(dc_s/dr)$
is identically zero everywhere for any potential $\Phi_i$. Hence the expression for
the analogue temperature can be obtained by setting $(dc_s/dr)=0$ in (\ref{eqeq2}). 
The dependence of $T_{AH}$ on $\left[T,\lambda\right]$ has been 
discussed in Bili\'c, Das \& Roy (2007) in detail.

\section{Epilogue}
\noindent
The primary motivation of this review article is to demonstrate the following:

One can propose that the general relativistic as well as the Newtonian/post-Newtonian
accretion flow around an astrophysical black hole can be considered as an
example of classical analogue gravity model realized in nature. To accomplish this task,
one first formulates and solves the equations describing
the accretion processes around black holes, and then provides the arguments that 
such accretion is transonic in general \footnote{Except for 
a very few special cases. For 
example, if
infalling matter is supplied from the supersonic stellar wind, accretion may
not be transonic if there is no shock formation at a length scale reasonable
well away from the event horizon.} and the accreting material 
must encounter a sonic point at some specific length scale 
determined by various accretion parameters. The collection of such sonic points
forms a null hypersurface, generators of which are the acoustic
null geodesics, i.e. the phonon trajectories. Such a surface can be shown
to be identical with  an acoustic event horizon. The acoustic surface
gravity and the corresponding analogue horizon temperature $T_{AH}$ at
the acoustic horizon are then computed in terms of fundamental accretion
parameters. Physically, the analogue temperature is associated with the
thermal phonon radiation analogous to the Hawking radiation of the
black-hole horizon. Acoustic {\it white holes} can also be generated if the
accretion flow is multi-transonic and if such 
multi-transonic black-hole accretion encounters a stable shock. Such a white hole,
produced at the shock, is always flanked by two acoustic black holes
generated at the inner and the outer sonic points.

At this point one might as a crucial question: Does the accretion processes {\it only} 
around a black hole represents an analogue system, or any kind of astrophysical 
accretion exhibits the analogue gravity phenomena {\it in general}? From the discussions 
presented in this article, one understands that two of the essential requirements 
for a physical system to manifest the classical analogue gravity effects are the 
following:
\begin{enumerate}
\item The system should consists of transonic, barotropic fluid, and the fluid should, 
preferably, be inviscid in order not to violate the Lorentzian invariance.
\item An acoustic perturbation (or equivalent perturbation, 
a surface gravity wave for example, see, e.g., 
Sch$\ddot{\rm u}$tzhold \& Unruh 2002) should propagate 
within such fluid for which a space time metric can be constructed. Such metric should 
incorporate a singularity (not always in a formal sense though), from which one can come 
up with the notion of the acoustic horizon.
\end{enumerate}
Hence, it is obvious that hydrodynamic, non-dissipative accretion onto {\it any} 
astrophysical object {\it should} manifest the analogue gravity phenomena, if such 
accretion exhibits transonic properties, 
and if such accreting fluid configuration possesses a 
specific well defined symmetry (spherically symmetric or axisymmetric flow, for 
example). Hence, hydrodynamic, transonic, astrophysical accretion possessing a suitable symmetric
geometrical configuration may exhibit the analogue properties {\it in general}, where the 
accretor resembles the sink.

{\it Transonic accretion in astrophysics can be conceived to constitute an 
extremely important class of 
classical analogue gravity model}. Among all the classical analogue systems studied in the
literature so far, {\it only} an accreting astrophysical object incorporates {\it gravity} 
(through the general body force term in the Euler's equation, even if the accretion is studied within the 
framework of the Newtonian space-time) in the analogue model. Also, the simplest possible 
analogue model may be constructed for such objects. For example, the spherically symmetric 
accretion of isothermal fluid onto a Newtonian/semi-Newtonian gravitating mass constitutes an 
analogue system which can be {\it completely} determined using a {\it single} parameter, the bulk
flow temperature of the infalling material (see section 15.3). 

However, among all the accreting astrophysical systems capable of manifesting the 
classical analogue effect, black hole accretion process deserves a very special status. The
accreting astrophysical black holes are the {\it only}
real physical candidates for which both the black-hole event horizon 
and the analogue sonic horizon may co-exist. Hence,
the application of the analogue Hawking effect to the theory of
transonic black hole accretion will be useful to compare
the properties of these two types of horizons.

Recently, the spacetime geometry on the
equatorial slice through a Kerr black hole has been shown to be
equivalent to the geometry experienced by phonons in a rotating
fluid vortex (Visser \& Weinfurtner 2005). Since many astrophysical
black holes are expected to possess non-zero spin (the Kerr
parameter $a$), a clear understanding of the influence of spin on
analogue models will be of great importance. Some important features
on the dependence of the analogue temperature on the black hole 
spin angular momentum of an astrophysical black hole has been
discussed in this article. 
In section 13.9 (Fig. 7
and related discussions),
it has been shown that the black hole spin
{\it does} influence the analogue gravity effect in a rotating relativistic 
fluid around it. Also the spin (of the black hole) - angular momentum (of the 
accreting material) coupling modulates such effect. Analogue effect 
is more prominent for retrograde (counter-rotating) flow, resulting a higher 
value of the corresponding analogue temperature.

In connection to the acoustic geometry, one can define
an `anti-trapped surface' to be a hypersurface in which
the fluid flow will be outward directed with the normal component
of the three-velocity greater than the local speed of sound. In stationary geometry, an anti-trapped surface will
naturally be constructed by the collection of sonic
points corresponding to a spherically symmetric or
axisymmetric transonic wind solution emanating out from an astrophysical
source.
Transonic outflow (wind) is ubiquitous in astrophysics,
spanning a wide range from solar/stellar winds to  large-scale
outflows from active galaxies, quasars, galactic micro-quasars
and energetic gamma ray bursts (GRB). In section 13.7,
it has been shown how
to identify the critical and the sonic points corresponding to the
wind solutions. Such a scheme can be useful in studying
the transonic properties of outflow from astrophysical sources.
Hence the formalism presented in this paper can be applied to study
the analogue effects in transonic winds as well. Recently
Kinoshita, Sendouda \& Takahashi (2004) performed the causality analysis of the
spherical GRB outflow using the concept of effective acoustic geometry.
Such an investigation can be extended into a more robust form by
incorporating the kind of work presented in this article,
to study the causal structure of the transonic GRB
outflows in axisymmetry, i.e. for energetic directed outflow
originating from a black-hole accretion disc system progenitor.

In connection to the study of accreting black hole system as 
a classical analogue gravity model, so far the analogy has been applied to
describe the classical perturbation of the fluid in terms of a
field satisfying the wave equation in an effective geometry.
Such works do not aim to provide a  formulation by which
the phonon field generated in this system could be quantized.
To
accomplish this task, one would need to show that the effective action for the
acoustic perturbation is equivalent to a field theoretical action
in curved space, and the corresponding commutation and dispersion
relations  should
directly follow (see, e.g., Unruh \& Sch$\ddot{\rm u}$tzhold 2003).
Such considerations are beyond the scope of
this article.

While describing the accretion disc dynamics, 
the viscous transport of the angular momentum is not
explicitly taken into account. Viscosity, however, is quite a subtle
issue in studying the analogue effects for disc accretion.
Even thirty three years after the discovery of
standard accretion disc theory (Shakura \& Sunyaev, 1973;
Novikov \& Thorne 1973), exact modeling of viscous
transonic black-hole accretion, including
proper heating and cooling mechanisms, is still quite an arduous task, even for a
Newtonian flow, let alone for general relativistic accretion.
On the other hand,
from the analogue model point of view, viscosity
is likely to destroy  Lorenz invariance, and hence the assumptions behind building up an
analogue model may not be quite  consistent.
Nevertheless, extremely large radial velocity
close to the black hole implies $\tau_{inf}\ll \tau_{visc}$, where $\tau_{inf}$ and
$\tau_{visc}$ are the infall and the viscous time scales, respectively.
Large radial velocities even at larger distances are due to the fact
that the angular momentum content of the accreting fluid
is relatively low (Beloborodov \& Illarionov 1991; 
Igumenshchev \& Beloborodov 1997;
Proga \& Begelman 2003).
Hence,
the assumption of inviscid flow is not unjustified from
an astrophysical point of view.
However,
one of the most significant effects of the introduction of viscosity
would be the reduction of the angular momentum.
It has been observed that the location of the sonic points
anti-correlates with $\lambda$, i.e. weakly rotating flow makes the
dynamical velocity gradient steeper, which indicates that for
viscous flow the acoustic horizons will be pushed further out and the flow would
become supersonic at a larger distance for the same set of other initial
boundary conditions. 

In section 13.2, while constructing the geometry of the general relativistic accretion disc,
the expression for the disc height has been derived using the prescription of 
Abramowicz, Lanza \& Percival (1997). However, a number of  other models  for the disc height exist in the literature
(Novikov \& Thorne 1973; Riffert \& Herold 1995;
Pariev 1996; Peitz \& Appl 1997; Lasota \& Abramowicz 1997)
The use of any other disc
height model would not alter our  conclusion
that black-hole accretion disc solutions form an important class
of analogue gravity models (see, e.g.,
Das 2004 for further details about the
investigation of the relativistic disc dynamics 
using the disc height proposed by Lasota \& Abramowicz (1997)). 
However, the numerical values of
$T_{AH}$ and other related quantities would  be different for
different disc heights.

For all  types of
accretion discussed here, the analogue temperature $T_{\rm AH}$
is many orders of magnitude  lower compared with the fluid temperature
of accreting matter. However, the study of analogue effects may be measurably 
significant for accretion onto primordial black holes
because the analogue as well as the 
actual Hawking temperature may be considerably high for such situations. There may be
a possibility that intense Hawking radiation may not allow any accretion due to the
domination of strong radiation pressure. However, the situation may be completely different
for Randall-Sundrum type II cosmology, where during the high energy regime of braneworld
cosmology, accretion may have significant effects on increasing the mass of the
primordial black holes (Guedens, Clancy \& Liddle 2002; Guedens, Clancy \& Liddle 2002a;
Majumdar 2003).
In braneworld scenario, the accretion onto the
primordial black holes from surrounding radiation bath may completely dominate over the
evaporation process as long as radiation dominations persists. It would be interesting 
to investigate 
the analogue effects in primordial black hole accretion in Randall-Sundrum type - II
cosmology, to study whether analogue radiation can really dominate over the accretion phase,
resulting the enhancement of the black hole evaporation process.
One may also like to investigate whether the first `black hole explosions' due to
Hawking radiation would be acoustic-mediated explosions of the medium
surrounding the primordial black holes.

In recent years, considerable attention has been focused on the study of
gravitational collapse of massive matter clump, in particular,
on
the investigation of the
final fate of such collapse (for a review see, e.g., Krolak 1999).
Goswami and Joshi (2004) have studied  the role of the equation of
state and initial data in determining the final fate of the continual spherical
collapse of barotropic fluid
in terms of naked singularities and the black-hole formation.
It is  tempting to study the analogue effects in such a collapse
model. Since at some stage the  velocity of the collapsing
fluid will exceed the velocity of local acoustic perturbation
one might encounter a
sonic horizons at the radial locations of the
corresponding transonic points in a stationary configuration.
One should, however, be careful about the issue that
many results in
analogue models are based on the assumption of a stationary flow,
whereas a  collapse scenario is a full time dependent  dynamical
process.

The correspondence between general relativity and analogue gravity has
so far been exploited only on a kinematical, i.e.
geometrical level. The analogue gravity systems lack a proper dynamical
scheme, such as Einstein's field equations in general relativity and
hence the analogy is not
complete.
A certain progress in this direction has recently been made
by Cadoni and Mignemi (Cadoni 2005; Cadoni \& Mignemi 2005),
who have established a dynamical correspondence between
analogue and dilaton gravity in 1+1 dimensions.
We believe that the approach presented in this article in which
an arbitrary background geometry serves as a source for
fluid dynamics may shed a new light towards a full analogy between
general relativity and analogue gravity.

\section*{Acknowledgments}
\noindent
During the process of understanding the general relativistic theory of black hole 
accretion, black hole thermodynamics, and the theory of analogue gravity
in last couple of years, I have been greatly benefited by insightful 
discussions with Marek A. Abramowicz, Narayan Banerjee, Peter Becker, Mitch Begelman,
Jacob Bekenstein, Neven Bili\'c, Roger Blandford,
Brandon Carter, Bozena Czerny, Andy Fabian, Juhan Frank, Werner Israel, Theodore A. (Ted) Jacobson,
Sayan Kar, Stefano Liberati, Parthasarathi Majumdar, John Miller, Mark Morris, 
Igor D. Novikov, John CB Papaloizou,
Tsvi Piran, A. R. Prasanna, Agatha Rozanska, Ralph Sch$\ddot{\rm u}$tzhold,
Nikola I. Shakura, Frank Shu, Ronald Taam, 
Gerard 't Hooft,
William (Bill) Unruh, Matt Visser, Robert Wagoner, Paul J. Wiita
and Kinwah Wu.
I gratefully acknowledge useful discussions with Jayanta K. Bhattacharjee 
and Arnab K. Ray regarding the 
dynamical systems approach to study the transonic behaviour of 
black hole accretion.

It is also a great pleasure to acknowledge the hospitality provided 
by the Theoretical Institute for Advanced Research in Astrophysics (TIARA)
(in the form of a visiting faculty position, Grant no. 94-2752-M-007-001-PAE), 
where a part of this work \footnote{The present one is a slightly modified version of the invited review article
published in the Indian Journal of Physics (September 2006, volume 80,
number 9, pg. 887 - 934), a special issue  dedicated to the revered memory
of the noted cosmologist
and relativist, late Professor Amal Kumar Raychaudhuri, well 
known for his profound contribution to the field of relativistic astrophysics
and cosmology,
especially for the famous Raychaudhur Equation.} has been carried out.
 
{}

\end{document}